\documentclass[10pt,a4paper]{article}

\usepackage[english]{babel}
\usepackage[utf8]{inputenc}
\usepackage{amsmath}
\usepackage{graphicx}
\usepackage{geometry}
\usepackage{wrapfig}
\usepackage{multicol}
\usepackage{hyperref}
\usepackage{float}
\usepackage{blindtext}
\usepackage[final]{pdfpages}
\usepackage[colorinlistoftodos]{todonotes}
\usepackage{subfigure}

\usepackage[round]{natbib}   % omit 'round' option if you prefer square brackets
\title{The impact of asteroid shapes and topographies on their reflectance spectroscopy}

\date{}

\begin{document}

\maketitle
S.M. Potin$^{1}$, S. Douté$^{2}$, B. Kugler$^{3}$, F. Forbes$^{3}$\\
{\scriptsize $^{1}$Laboratoire d'Etudes Spatiales et d'Instrumentation en Astrophysique (LESIA), Observatoire de Paris, Université PSL, CNRS, Sorbonne université, Université de Paris, 5 place Jules Janssen, 92195 Meudon, France. $^{2}$Université Grenoble Alpes, CNRS, Institut de Planétologie et d’Astrophysique de Grenoble (IPAG), Saint-Martin d'Hères, France. $^{3}$Université Grenoble Alpes, CNRS, Inria, Grenoble INP, Laboratoire Jean Kuntzmann (LJK), Grenoble, France\\
Corresponding author: sylvain.doute@univ-grenoble-alpes.fr}

\begin{abstract}
Here we report the comparison between unresolved reflectance spectroscopy of Solar System small bodies and laboratory measurements on reference surfaces. We measure the bidirectional reflectance spectroscopy of a powder of howardite and a sublimation residue composed of a Ceres analogue. The spectra are then inverted using the Hapke semi-empirical physical model and the MRTLS parametric model to be able to simulate the reflectance of the surfaces under any geometrical configuration needed. We note that both models enable an accurate rendering of the reflectance spectroscopy, but the MRTLS model adds less noise on the spectra compared to the Hapke model. Using the parameters resulting from the inversions, we simulate two spherical bodies and the small bodies (1)Ceres and (4)Vesta whose surfaces are homogeneously covered with the Ceres analogue and powder of howardite respectively. We then simulate various scenarios of illumination and spectroscopic observations, i.e. spot-pointing and fly-bys, of these small bodies for phases angles between 6° and 135°. The unresolved reflectance spectroscopy of the simulated bodies is retrieved from the resulting images, and compared to the reflectance spectroscopy of the reference surface measured in the laboratory. Our results show that the photometric phase curves of the simulated bodies are different from the reference surfaces because of the variations of the local incidence and emergence angles due to the shape and topography of the surface. At low phase angle, the simulated bodies are brighter than the reference surfaces, with lower spectral slope and shallower absorption bands. We observe the maximum differences at wide phase angles with the various simulated observations of (4)Vesta due to its high surface topography. Finally, we highlight the differences in the spectral parameters derived from the unresolved observations at 30° with laboratory measurements acquired under a single geometrical configuration.
\end{abstract}

Keywords:Spectroscopy; Asteroids, surface; Asteroid Ceres; Asteroid Vesta

\section{Introduction}
\hspace*{0.5cm}The Solar System small bodies can be observed from ground-based or Earth orbiting telescopes but only span over a few pixels on the images due to their distance to the observer, their small size, and atmospheric conditions blurring the images \citep{rivkin_2002, onaka_2008, muller_2013, moreno_2016, usui_2019}. Adaptive optics (AO) systems correct the atmospheric perturbations and increase the spatial resolution of the ground-based telescopes. The sizes and shapes of asteroids can be resolved using AO, though the surface topography is still unresolved \citep{marchis_2006, carry_2010, viikinkoski_2017}. The highest spatial resolution on the surface of the small bodies is achieved during in-situ observations by spacecrafts orbitting their targets. In these cases, the small scale topography, boulders and textural variations are observed \citep{fujiwara_2006, stern_2019, watanabe_2019, lauretta_2019}.\\
\hspace*{0.5cm}Moreover, an asteroid presents compositional and textural variations over its surface, and recent close encounters on Near-Earth Asteroids (NEAs) showed multiscales heterogeneities. The spacecraft Hayabusa2 from the Japa\-nese Aerospace eXploration Agency (JAXA) observed compositional heterogeneities on the scale of the asteroid surface \citep{kitazato_2019}, while its lander, the Mobile Asteroid Surface Scout (MASCOT) observed macroscopic inclusions on the surface of the asteroid \citep{MASCOT}. The Origins, Spectral Interpretation, Ressource Identification, and Security Regolith Explorer (OSIRIS-REx) spacecraft from the National Aeronautics and Space Administration (NASA) observed dark and bright boulders on the surface of the asteroid Bennu \citep{dellagiustina_2021}. In the case of unresolved spectroscopic observations of asteroids, all these heterogeneities and texture variations are averaged over the whole observed surface.\\
\hspace*{0.5cm}The dependance of the spectral reflectance on the direction of illumination and observation has been intensively studied. Variations of the albedo with increasing phase angle, i.e. photometric phase curves, were discovered on early observations \citep{gehrels_1970, lane_1973, blanco_1979}. The geometry of illumination and observation also impacts the spectral slope and generates the phase reddening, namely the increase of slope, so the increase of the measured reflectance at long wavelengths, with increasing phase angle \citep{schroder_2014, Potin_mukundpura, fornasier_2020}. The amplitude of the absorption bands that can be detected on the reflectance spectra of asteroids and meteorites is also impacted by the geometry of illumination and observation \citep{Potin_mukundpura, fornasier_2020}.\\
\hspace*{0.5cm}Reflectance spectroscopy of samples in the laboratory is generally performed under a single geometry, often with a phase angle of 30° with either a nadir illumination or observation \citep{bishop_1994, milliken_2005, pommerol_2008a, cloutis_2011, takir_2019, Potin_NEA, deangelis_2019}. Reflectance spectra in the laboratory can only be acquired under a finite number of geometries, depending on the type of instrumental setup and on the studied surface. Moreover, measuring the reflectance spectroscopy of a surface in the laboratory generally required samples with an homogeneous composition and a preliminary preparation of the surface. In order to be statistically relevant, a reflectance spectrometer field of view must take into account a number of facets, randomly oriented, high enough of the sample grains. It thus creates a limitation on the type of sample a spectrometer can analyse, either fine grained powder, or coarse grains depending on the respective size of the illumination and observation spots. The study of rock surface is possible assuming that the sample spans the whole extend of the illumination spot and is flat enough not to cast shadows on the studied surface, lowering the measured reflectance value. Some setups also necessitate the use of pressed pellets, or samples must be provided with a fix distribution of grain sizes. All these parameters restrict the samples availabilities in the laboratory.\\

We address the following scientific questions. Firstly, to which extent unresolved topography, roughness, and the overall shape of asteroids are going to control their reflectance spectra depending on the geometrical acquisition conditions? The study of the effects induced by other unresolved heterogeneities such as composition and texture will not be the subject of the paper. Then, under which conditions is it possible to interpret asteroid spectra using reference reflectance spectroscopy of samples in the laboratory? To answer these questions, we first propose a methodology (section 2) : we measure the spectral bidirectional reflectance distribution function (BRDF) of asteroidal analogue materials and we build models of this quantity for any geometry. We examine and discuss the lab measurements and their modeling (section 3). We then simulate non resolved spectral observations of spherical bodies, Ceres, and Vesta using the BRDF models and detailed shape models, for a series of acquisition geometries. We describe phase curves in reflectance, spectral slope and band depth obtained from the simulated observation (section 4). Finally, we discuss the effects of shape and topography by comparing the reference lab measurements with the simulated results (section 5).

\section{Methodology}
\hspace*{0.5cm}We propose here to compare for a series of geometries the unresolved spectroscopic observation of an asteroidal body to the reflectance spectroscopy of a flat surface constituted with the same material. To do so, we virtually apply the composition and texture of a surface studied in the laboratory on the shape model of an asteroidal body. The simulated astronomical target is then placed in the field of view of a camera and illuminated under a given direction to simulate sunlight. Each facet of the simulated target will scatter the illuminating light according to the laboratory BRDF, and integrating the contribution of each illuminated and observed facet will recreate the unresolved observation of the body.\\

In this section we describe in details each step of the laboratory measurements, modeling, and integration into the simulation of unresolved observations.

\subsection{Presentation of the surfaces}
\hspace*{0.5cm}Two different samples were used, a sample of howardite and a Ceres analogue. The sample of howardite consists of a fine powder of the meteorite Northwest Africa (NWA) 2060, resulting from the analysis from \cite{beck_2012}. The sample presents a broad distribution of grain sizes, between 50 and 75$\mu$m. The pyroxene-bearing composition of Vesta, and its relationship to howardite, eucrite and diogenite (HED) meteorites has been observed first using ground-based telescopic spectrometry and then confirmed during the flight of the Dawn spacecraft around the small body \citep{gaffey_1997, mcsween_2011, mcsween_2013, desanctis_2012, russell_2012, longobardo_2014}.\\
\hspace*{0.5cm}The second sample presents a composition similar to the dwarf planet Ceres, and originating from a sublimation experiment \citep{Stefan_Ceres}. This sample is mostly composed of magnetite and $NH_{4}$-bearing nontronite, respectively at 44 vol$\%$ (63 w$\%$) and 35 vol$\%$ (22 w$\%$), with antigorite and dolomite \citep{Stefan_Ceres, ferrari_2019, deangelis_2017}. The residue presents a highly porous texture, and has been studied in bidirectional reflectance spectroscopy after all sublimation experiments.

\subsection{Laboratory bidirectional reflectance spectroscopy}
\hspace*{0.5cm}The bidirectional reflectance spectroscopy of the selected surfaces is acquired with the spectro-gonio radiometer SHADOWS \citep{Potin_SHADOWS}. The spectra are measured from 340 nm to  4200 nm according to the configuration described in table \ref{configuration spectro}.

\begin{table}[H]
\begin{center}
\begin{tabular}{lll}
\hline
Range (nm) & Step (nm) & Resolution (nm)\\
\hline
340 - 679 & 20 & 4.85 - 4.75\\
680 - 1499 & 20 & 9.71 – 9.38\\
1500 - 2999 & 20 & 19.42 – 18.73\\
3000 - 4200 & 20 & 38.84 – 38.44\\
\hline
\end{tabular}
\caption{Spectroscopic configuration of SHADOWS for the bidirectional reflectance spectroscopy of the studied surfaces.}
\label{configuration spectro}
\end{center}
\end{table}

\begin{figure}[H]
\centering
    \includegraphics[width = 0.6\textwidth]{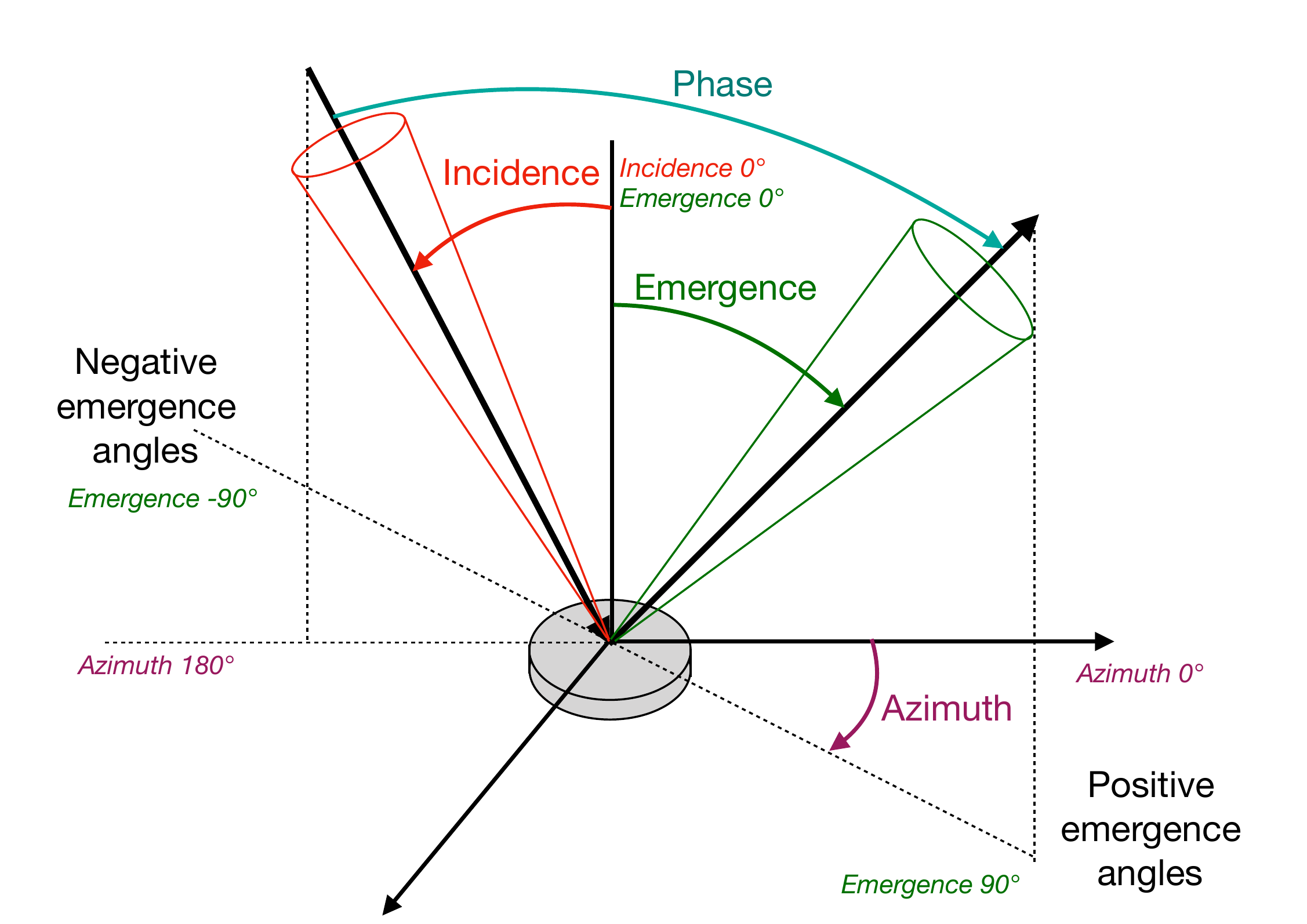}
    \caption{Schematic definition of the geometrical configuration of SHADOWS used in this study. Picture from \cite{Potin_SHADOWS}.}
    \label{schema geo shadows}
\end{figure}

The spectro-gonio radiometer illuminates the sample under a given incidence angle and the reflected light is measured from another fixed angle, defined as the emergence angle. The phase angle is considered as the angle formed by the direction of illumination and observation. The incidence and emergence angle are calculated with respect to the normal to the surface. In this configuration, a nadir illumination corresponds to an incidence angle of 0°. The measurements can be performed in the principal plane including the normal to the surface and the direction of incidence, or in a plane of any other azimuth angle. Figure \ref{schema geo shadows} shows the geometrical configuration of the spectro-gonio radiometer SHADOWS.

All surfaces are analyzed under the same list of geometrical configurations. The spectra are measured with an emergence angle from -70° to 70° with an angular step of 10°, for an incidence angle going from nadir (0°) to 60° with an angular step of 20° in the principal plane. One series of spectral measurements is also performed out of the principal plane, with an azimuth angle of 30°, an incidence angle of 20° and the emergence angle ranging from -70° to 70° with the same angular step of 10°. It is important to note that the goniometer cannot measure the reflectance of a surface at a phase angle lower than 5° because of the occultation of the detectors by the incidence mirror. This will have an impact on the inversion of the spectra and on the final simulation, which will be described in the related sections. The laboratory measurement results in a series of 71 different geometrical configurations, for which a single spectrum is measured. As the incidence angle is varying along with the emergence angle, several geometrical configurations result in the same phase angle.\\
\hspace*{0.5cm}The reflectance factor of the samples is calculated by dividing the measured signal for a given sample by the one of a Spectralon and an Infragold targets, acquired with a nadir illumination and an observation of 30°. \\
\hspace*{0.5cm}This measurement method results in a series of 71 spectra for each surface, each spectrum corresponding to a single geometrical configuration (incidence, emergence and azimuth angle). The bidirectional reflectance spectra of each composition and texture are presented in appendix. \\

\subsection{Spectral inversion}
\hspace*{0.5cm}Laboratory measurements on meteorites can only be acquired under a finite number of geometrical configurations. Inverting photometric models consists in estimating the best values of their parameters to reproduce the BRDFs sampled in the laboratory. It is then possible to generate the reflectance spectra that would have been measured under any geometrical configuration needed for the planetary simulation. \\
\hspace*{0.5cm}For the analysis of the measured BRDFs we consider two photometric models: the popular semi-empirical Hapke model based on physical first principles \citep{hapke_1981, hapke_1984, hapke_2008} and a modified version of the Ross-Thick Li-Sparse model \citep{Roujean1992, wenner_1995, Lucht2000}. The latter is a fully parametric model based on a linear combination of several scattering components, the so-called kernels, that are physically derived from simplified surface and volume scattering scenarii. The kernels provide the basic BRDF shapes for characterizing the heterogeneous scattering of soil-vegetation systems and their combination has proved to be accurate in recreating many types of natural surfaces  \citep{schlapfer_2015, jiao_2016, jiao_2019, matsuoka_2016}, including planetary surfaces \citep{Fernando2013}. The standard Ross-Thick Li-Sparse BRDF model calculates the reflectance as a weighted sum of three terms:
\begin{equation}
\begin{split}
R\left( {\theta,\vartheta,\varphi,\lambda} \right) = f_{\mathit{iso}}\left( \lambda \right) + f_{\mathit{vol}}\left( \lambda \right)K_{\mathit{vol}}\left( {\theta,\vartheta,\varphi} \right)
+ f_{\mathit{geo}}\left( \lambda \right)K_{\mathit{geo}}\left( {\theta,\vartheta,\varphi} \right)
\end{split}
\end{equation}

where the terms play the roles of Lambertian (iso), volumetric (vol), and geometric (geom) components respectively, with $\lambda$ the wavelength and the triplet $(\theta,\vartheta,\varphi)$ representing the incidence, emergence, and azimuth angles. Spectral weights $f_{\mathit{iso}}$, $f_{\mathit{vol}}$, and $f_{\mathit{geo}}$ determine the surface reflectance properties by controlling the respective contribution of the three components. Their values do not have a direct physical meaning. Note that the volumetric kernel contains a simplified treatment of the opposition effect derived by \citet{breon_2002} with fixed width and amplitude as this effect is poorly constrained by our measurements:
\begin{equation}
 (1+(1+\xi/\xi_0)^{-1})
\end{equation}

Here $\xi$ is the phase angle, and $\xi_0$ is a characteristic angle in relation to the ratio of scattering element size and the medium vertical density. We take $\xi_0$=1.5° that has been suggested as a typical value representing a wide range of surface conditions. \\
\hspace*{0.5cm}Following \citet{jiao_2019} we add a forward scattering kernel to the standard Ross-Thick Li-Sparse model for a better modeling of the BRDF of our samples:
\begin{equation}
\begin{split}
    K_{fwd}\left( {\theta,\vartheta,\varphi} \right) = R_{0}\left( {\theta,\vartheta,\varphi} \right)\left( {1‐\alpha \cdot \cos\xi \cdot \exp\left( {- \cos\xi} \right)} \right) + 0.4076\alpha ‐1.1081
    \end{split}
\end{equation}

The parameter $\alpha$ ensures that the forward scattering kernel is normalized to zero (i.e., $K_{fwd}\left( {\theta,\vartheta,\varphi} \right)=0$) when the view and sun are both in the nadir direction (i.e., $\theta=0$ and $\vartheta=0$).
The reflectance $R_0$ for a semi-infinite, layer of nonabsorbing, large, equant, and translucent particles was derived in an approximate version by \citet{kokhanovsky_2005}:
\begin{align}
  R_{0}\left( {\mu_{s},\mu_{v},\varphi} \right) = & \frac{K_{1} + K_{2} \cdot \left( {\mu_{s} + \mu_{v}} \right) + K_{3} \cdot \mu_{s}\mu_{v} + P\left( \xi \right)}{4 \cdot \left( {\mu_{s} + \mu_{v}} \right)}\\
  P\left( \xi \right) = & 11.1e^{‐0.087 \cdot {({180‐\xi})}} + 1.1e^{‐0.014 \cdot {({180‐\xi})}}
\end{align}

The three constants $K$ are determined by $K_1=1.247$, $K_2=1.186$, and $K_3=5.157$. The parameter $\xi$ is the phase angle (in degrees) and we define $\mu_s=\cos(\theta)$ and $\mu_c=\cos(\vartheta)$. 
As a result the modified Ross-Thick Li-Sparse (MRTLS) BRDF model calculates the reflectance as a weighted sum of four terms:
\begin{equation}
\begin{split}
R\left( {\theta,\vartheta,\varphi,\lambda} \right) = f_{\mathit{iso}}\left( \lambda \right) + f_{\mathit{vol}}\left( \lambda \right)K_{\mathit{vol}}\left( {\theta,\vartheta,\varphi} \right) 
+ f_{\mathit{geo}}\left( \lambda \right)K_{\mathit{geo}}\left( {\theta,\vartheta,\varphi} \right) + f_{\mathit{fwd}}\left( \lambda \right)K_{\mathit{fwd}}\left( {\theta,\vartheta,\varphi} \right)
\end{split}
\label{MRTLS}
\end{equation}

The inversion of the MRTLS model can be stated as finding the solution $\boldsymbol{f}_{sol}=(f_{\mathit{iso}},f_{\mathit{vol}},f_{\mathit{geom}},f_{\mathit{fwd}})$ such that $\left\|\boldsymbol{R}_{obs}-\mathrm{\mathbf{K}}\boldsymbol{f}\right\|^2$ is minimized, $\boldsymbol{R}_{obs}$ being the vector of reflectance measurements. We use the Maximum Likelihood Estimate to solve the linear regression problem adapted to the kernel formulation of the MRTLS model. For a set of $D$ measurement geometries, each parametrized by a triplet $(\theta,\vartheta,\varphi)$ of incidence, emergence, and azimuth angles, we can model the forward problem as:
\begin{equation}
\begin{split}
R\left( {\theta^d,\vartheta^d,\varphi^d} \right) = f_{\mathit{iso}} + f_{\mathit{vol}}K_{\mathit{vol}}\left( {\theta^d,\vartheta^d,\varphi^d} \right)+ 
f_{\mathit{geo}}K_{\mathit{geo}}\left( {\theta^d,\vartheta^d,\varphi^d} \right)+ f_{\mathit{fwd}}K_{\mathit{fwd}}\left( {\theta^d,\vartheta^d,\varphi^d} \right)\\
(d=1:D)
\end{split}
\end{equation}
where we omit the wavelength dependency for the sake of simplicity. This set of equations can be reformulated in a matrix form : $\boldsymbol{R}=\mathrm{\mathbf{K}}\boldsymbol{f}$. In this case the solution of the system of linear equations can be expressed using the pseudo inverse form of matrix $\mathrm{\mathbf{K}}$:
\begin{equation}
\boldsymbol{f}_{sol}=\mathrm{\mathbf{K}}^t\left(\mathrm{\mathbf{K}}\mathrm{\mathbf{K}}^t\right)^{-1} \boldsymbol{R}_{obs} 
\label{fsol}
\end{equation}
If the measurements are affected by some uncertainties expressed by the covariance matrix $\mathrm{\Sigma_{rr}}$, then these uncertainties can be propagated to the solution using the following equation:
\begin{equation}
\mathrm{\mathbf{\Sigma_{ff}}}=\left(\mathrm{\mathbf{K}}\mathrm{\mathbf{K}}^t\right)^{-1}\mathrm{\mathbf{K}}^t \mathrm{\mathbf{\Sigma_{rr}}} \mathrm{\mathbf{K}} \left(\mathrm{\mathbf{K}}\mathrm{\mathbf{K}}^t\right)^{-1}
\end{equation}

Finally the reflectance values for the measurement geometrical configurations $\left( {\theta,\vartheta,\varphi} \right)$ or any other setting can be computed by inserting the four components of the solution vector $\boldsymbol{f}_{sol}$ into equation \ref{MRTLS}. Note that the retrieval of the kernel weights and the modeled reflectance calculation are performed independently for all wavelengths of our measurements. \\

The Hapke model is inverted differently on our measurements because of its highly non-linearity regarding some of its parameters. For that purpose we operate an efficient method based on a learning approach in a Bayesian framework \citep{kugler_2021}. The latter offers a natural solution to propagate uncertainties on measures, while the variance of the likelihood assesses the uncertainty induced by the inversion. The high number of geometries and wavelengths to be inverted makes approaches based on Markov Chain Monte-Carlo simulations \citep{geyer_1992} unacceptably slow.\\

The method we propose considers the couple of random variables $(\boldsymbol{X},\boldsymbol{R})$, where $\boldsymbol{X}$ is the vector of photometric parameters $(\omega, \overline{\theta}, b, c)$, respectively the scattering albedo, macroscopic roughness, asymmetry parameter and backscattering fraction. The variable $\boldsymbol{X}$ has an uniform prior distribution on the physical parameters space. The variable $\boldsymbol{R}$ is the corresponding vector of reflectance values with $\boldsymbol{R}=F(\boldsymbol{X})+ \epsilon$ where $\mathrm{F}$ is the forward Hapke model and $\epsilon$ is a centered Gaussian noise accounting for measure and model uncertainties. \\

The main idea is to use a two steps approach. First, we build a parametric surrogate statistical model of $(\boldsymbol{X},\boldsymbol{R})$. We choose the family of so-called Gaussian Locally-Linear Mappings (GLLiM) \citep{deleforge_2015}. Such models are expressive enough to approximate highly non-linear, complex, forward models, while remaining tractable. The learning step is performed by generating a training dictionary composed of samples from $(\boldsymbol{X},\boldsymbol{R})$, on which the likelihood is maximized by a standard Expectation-Maximiza\-tion (EM) algorithm. Regarding the dimensionality of the problem, the number of statistical parameters is proportional to $D$, the number of measurement geometries, making this model suitable even for high dimensional observables like ours (71). Moreover, the computational cost of the learning phase is independent of the number of observations to be subsequently analyzed. \\

In a second step, we use the surrogate model to estimate the posterior probability distribution of the random variable $\boldsymbol{X}$ given $\boldsymbol{R}_{obs}$ ($p(\boldsymbol{X} \  | \  \boldsymbol{R}_{obs})$) independently for each observable $\boldsymbol{R}_{obs}$, i.e. each wavelength of our goniometric measurements. The inverse of the surrogate GLLiM model has an explicit formulation in the form of a Gaussian mixture. From this posterior density, the mean and variance are straightforward to compute and provide a first order solution. To refine the solution, the mixture can be further explored with mode search methods and can be exploited in order to perform importance sampling of the target posterior distribution. Among these alternative solutions (mean, modes, or importance sampling estimates), the best in terms of the reconstruction error $\boldsymbol{X}_{sol}=(\omega, \overline{\theta}, b, c)$ is retained and introduced into the Hapke direct formulation to compute the reflectance values for the measurement geometrical configurations $\left( {\theta,\vartheta,\varphi} \right)$ or for any other configuration.

\subsection{Spectroscopic observation of the target}
\hspace*{0.5cm}The measurement of the unresolved reflectance of planetary bodies integrates their spatial heterogeneities, as well as changes of observation geometries due to the large-scale structures of the surfaces (slopes, craters, ...) and due to the roughness at small scales. Our goal is to understand which factors control the spectrophotometric properties of the objet resulting from the aggregation of physical signals at different sub-pixel scales. For that purpose we use a planetary image simulation tool built upon the physically based renderer \href{http://www.mitsuba-renderer.org/index_old.html}{Mitsuba}\footnote{Last accessed October 27 2021} that computes light propagation from sources to a scene, and then to a sensor according to ray tracing and integration schemes. Several lines of development are considered for achieving a high degree of realism: high resolution shape models, description of material distribution, bidirectional reflectance measured in the laboratory as a function wavelength for analogue materials, mixing of spectral signatures at different scales, 3D radiative transfer of sunlight between the different facets of the object, and up to the sensor. A synthetic hyperspectral image is generated according to a given scene, sensor characteristics, and observation geometry (sub-solar and sub-sensor points, distance between the object and the observer). The sensor is modeled like a simple pinhole camera defined by its angular field of view, the number of columns, and the number of lines in the pixel matrix. The bidirectional reflectance of the surface materials is expressed either using the Hapke model or the Modified Ross-Thick Li-Sparse (MRTLS) model. The calculation of the spectral radiance at each pixel of the sensor is based on an inverse ray tracing algorithm and MonteCarlo integration scheme. In particular the algorithm addresses the direct illumination on a facet as a function of the local incidence and emergence angles, self-shadowing, occultation, as well as the calculation of the diffuse illumination (i.e. reflections of direct and diffuse irradiance from the neighborhood facets). \\
\hspace*{0.5cm}In the context of our experiments, we distribute our analogue materials uniformly over the whole shape model of Vesta \citep{russell_2011, preusker_2014} and Ceres \citep{preusker_2016, roatsch_2016}. Each interaction between a ray and a facet involves the calculation of the bidirectional reflectance according to the Hapke or MRTLS model fed by their respective spectro-photometric parameters described in section 3. Note that the computation is conducted for multiple wavelengths in parallel. The final image is expressed for each pixel in the form of a radiance normalized to a unit incoming flux. The shape models of the targets present a spatial resolution of 92m for Vesta and 137m for Ceres, and each pixel thus integrates the spectroscopic contribution of all facets contained in the projection of the pixel onto the surface. The spatial resolution of the resulting images is then fixed by the distance between the target and the observer. The unresolved reflectance spectra resulting from the simulations will not be impacted by a varying resolution as all illuminated and observed facets will contribute to the spectroscopy.\\
\hspace*{0.5cm}In the context of our experiments, we will consider two scenarii of observation with a series of varying acquisition geometries in both cases: a spot-pointing sequence and a fly-by sequence. In the first case the sensor is always facing the same hemisphere of the object, the sub-observer point is fixed on the surface, while the sub-solar point moves only in longitude. In the second case the sensor has a straight trajectory in the equatorial plane of the object with a "close encounter" and looks constantly towards the center of the object. Regardless of the scenario, we explore the same series of ten phase angles from 6° to 135°. 

\section{Results of the bidirectional reflectance measurement and inversion by the models}
\subsection{Presentation of the spectra and comparison between the two inversed models}
\hspace*{0.5cm}To reduce the risk of misinterpretation, it is important that the reflectance spectra resulting from the inversion of the models, and so applied on the surface of our simulated asteroid, are as close as possible to the reference laboratory spectra. We present here the bidirectional reflectance measurement performed on the powder of howardite and on the Ceres sublimation residue, and the spectra resulting from the inversion by the Hapke and MRTLS models. For clarity, we only present the spectra acquired under nadir illumination with an emergence angle of 30°, and spectra acquired with an illumination angle of 60° and an observation at 70°.

\begin{figure}[H]
\centering
\includegraphics[width = 0.6\textwidth]{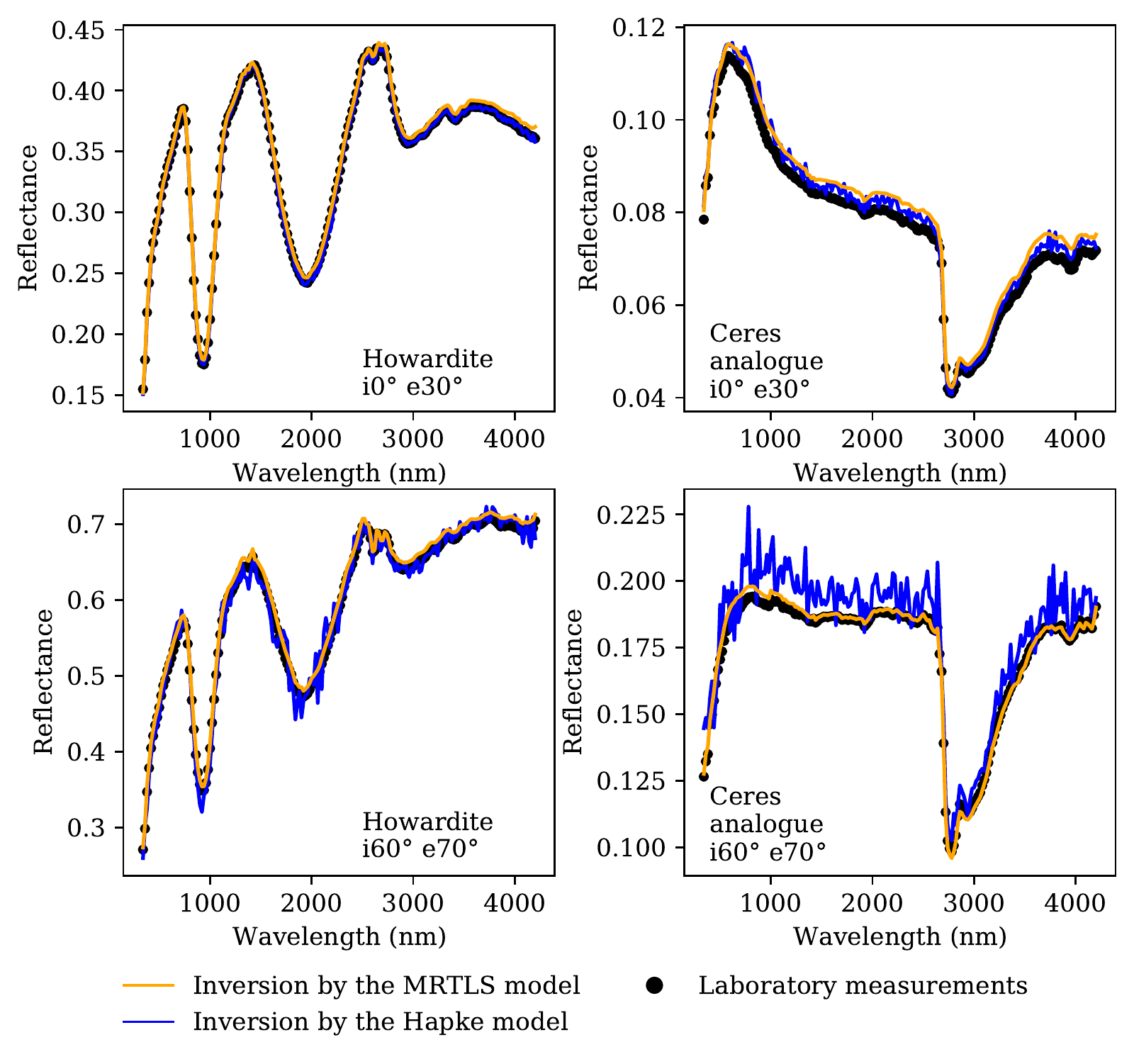}
\caption{Reflectance spectra of the howardite (left) and Ceres analogue (right) measured in the laboratory (black circles), and resulting from the inversion of the MRTLS model (orange) and the Hapke model (blue). The spectra shown here correspond to the reflectance with a nadir illumination and an emergence angle of 30° (top panels) and with an illumination angle of 60° and an emergence angle of 70°. (bottom panels)}
\label{compar model reflec}
\end{figure}

\begin{figure}[H]
\centering
\includegraphics[width = 0.5\textwidth]{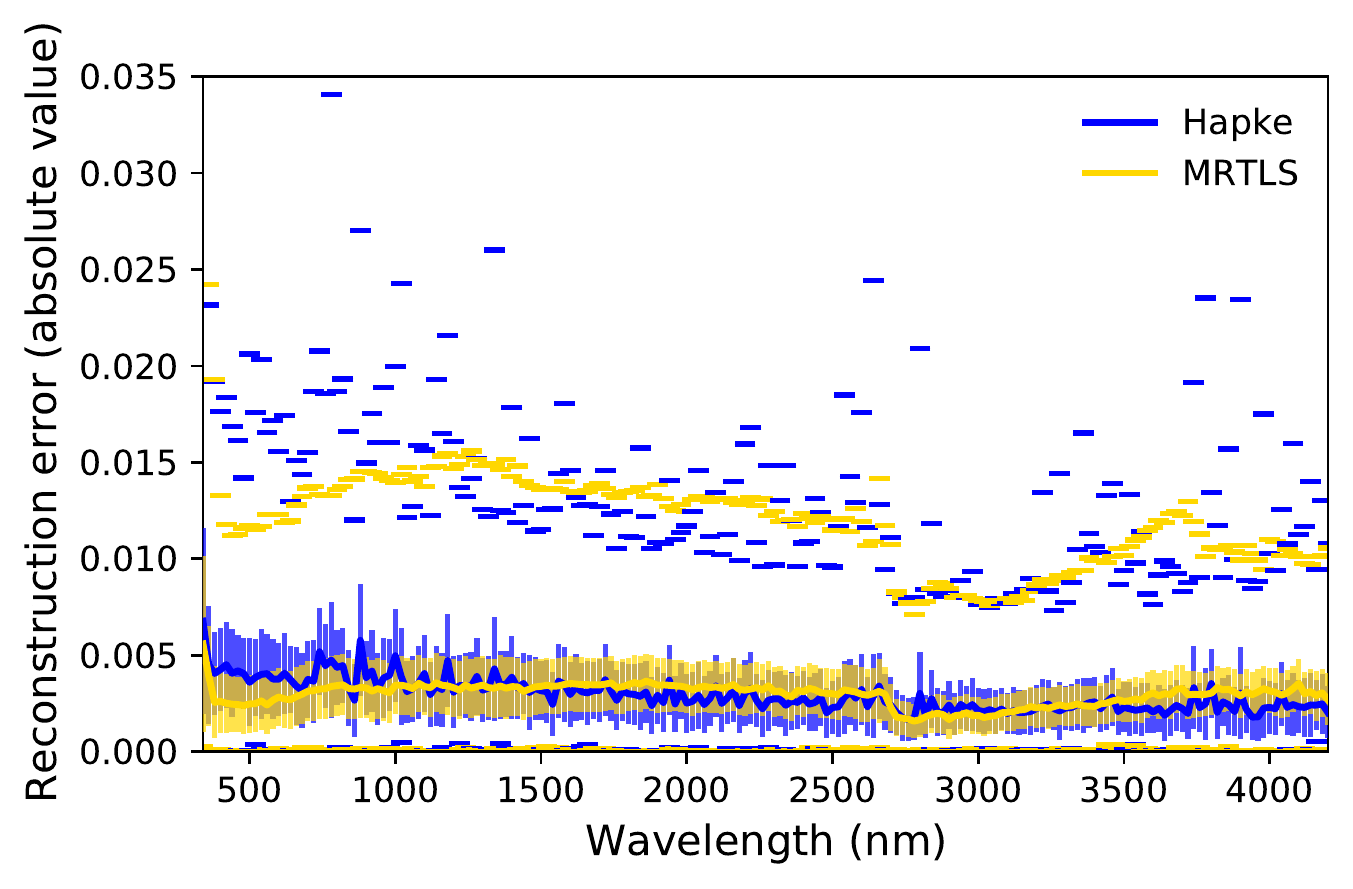}
\caption{Reconstruction errors of the Hapke (blue) and MRTLS (orange) models, considering all geometries at each wavelength. The solid lines represent the median values, the error bars represent the standard deviation around the median value, the markers are positionned at the minimum and maximum values of the reconstruction error.}
    \label{residue Hapke RTLSR}
\end{figure}

\hspace*{0.5cm}The reflectance spectrum of the howardite is red-sloped (i.e. increase of reflectance with increasing wavelength) and exhibits absorption features typical of HED meteorites and of their parent body (4)Vesta \citep{desanctis_2012}. The bands observed around 900 and 2000 nm trace the abundance of pyroxene in the sample, and their detected positions are linked to the pyroxene mineralogical composition \citep{burbine_2009, burbine_2018}. Finally, the band detected around 3000 nm indicates the presence of hydroxyl -OH groups linked to hydrated minerals and water molecules \citep{bishop_1994, frost_2000}.\\
\hspace*{0.5cm}The Ceres sublimation residue exhibits a broad absorption band centered around 1100 nm related to magnetite in its composition. A strong absorption feature around 2800 nm, and signs of NH4-bearing organics are detected around 3400 nm (\cite{poch_2020}). The strong blue slope (i.e. decrease of reflectance with increasing wavelength) results from the sublimation experiment giving the sample a highly porous foam-like texture.\\
\hspace*{0.5cm}We can note from figure \ref{compar model reflec} that the MRTLS and the Hapke models accurately reproduce the spectral features of the initial laboratory spectra. However, reflectance spectra reproduced by the Hapke model exhibit a higher noise compared to the spectra of the MRTLS model or the laboratory spectra, as it can be seen on the Ceres analogue spectrum. To evaluate the noise added to the spectroscopy by the inversions, the reconstruction error is calculated here as the absolute value of the difference between the reflectance value resulting from the inversion and the value measured in the laboratory used as reference. Figure \ref{residue Hapke RTLSR} presents the reconstruction errors of the Hapke and MRTLS models, considering all geometries at each wavelengths. \\

Figure \ref{residue Hapke RTLSR} shows that the two models result in similar median values of the reconstruction error. However, the standard deviation and maximum values are generally lower on the invesion by the MRTLS model than by the Hapke model. We make the hypothesis that this noise is due to the uncertainties on the Hapke parameters derived during the inversion, impacting the calculated reflectance afterwards.\\

\begin{figure}[H]
\centering
\includegraphics[width = 0.6\textwidth]{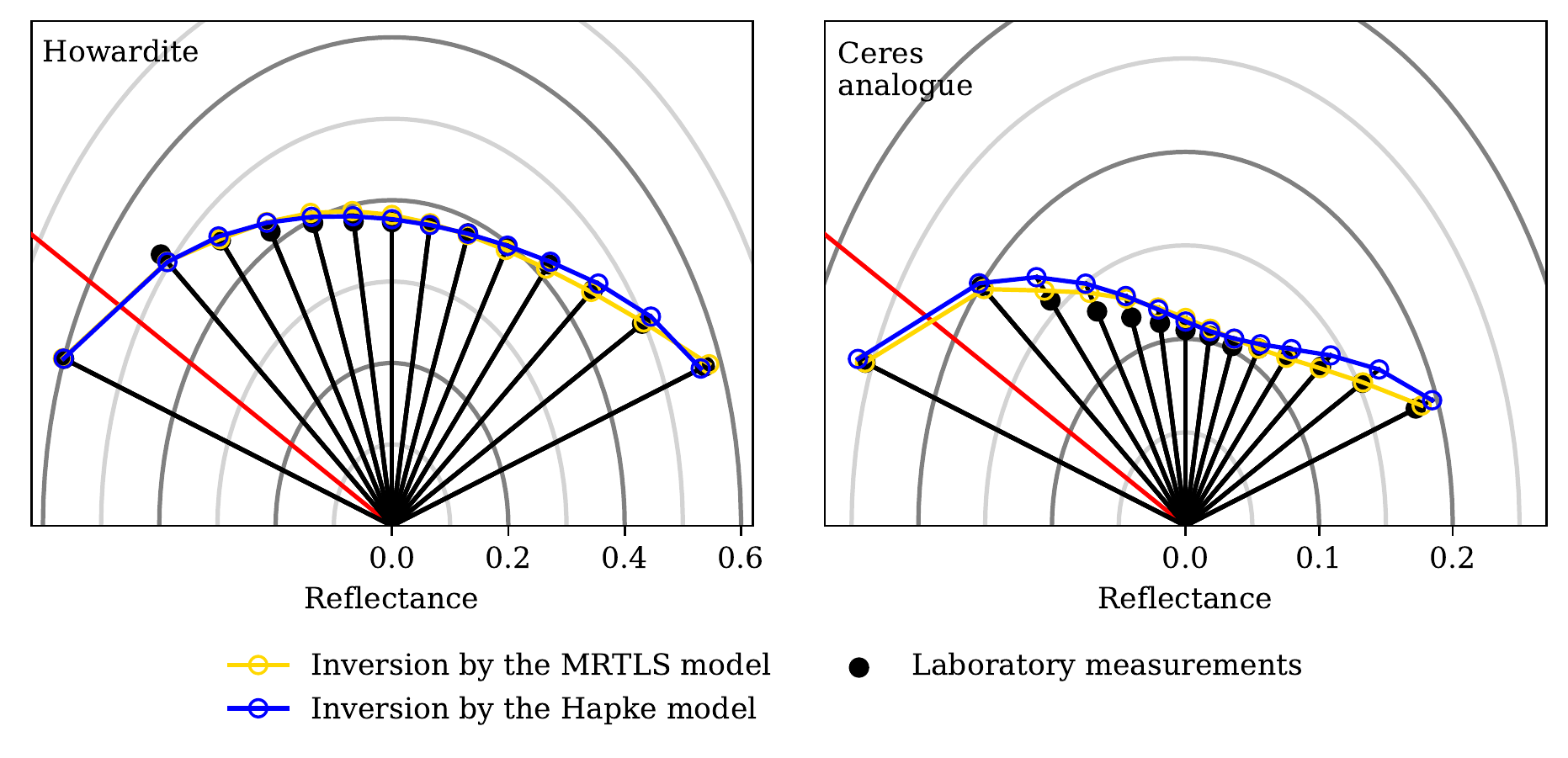}
\caption{Polar plot of the bidirectional reflectance measured at 560 nm on all surfaces (Left: howardite; Right: Ceres analogue). The reflectance values are taken from the spectra at incidence 60° measured in the laboratory (black), derived from the inversion by the MRTLS model (orange), and derived from the inversion by the Hapke model (blue).}
    \label{BRDF compar model}
\end{figure}

As the geometry of measurement impacts all spectral parameters, it is important to analyse the ability of the models to accurately recreate the spectra from under any given geometries. As an example, figure \ref{BRDF compar model} presents a polar plot of the reflectance of the surfaces at 560nm, with an incidence angle of 60°. On this representation, the direction of illumination is represented by the red line, and each marker represents a reflectance measurement. The angle from the vertical to the marker corresponds to the emergence angle (here from -70° to 70°), while the length between the marker and the origin corresponds to the reflectance value. In this representation, the BRDF of a perfectly lambertian surface reflecting the light homogeneously in every direction would be represented as a semi-circle.

A quantitative analysis of the differences between the laboratory values and those resulting from the inversion is presented further in this section. Both models recreate the bidirectional reflectance behaviour of all surfaces. However, as it can be seen for the howardite and Ceres analogue samples, the Hapke model returns a higher reflectance than the original data in the lobe at high phase angle tracing the forward scattering of the surface. In this region, the MRTLS model returns a closer value of reflectance compared to the Hapke model, and even the accurate value of reflectance in the case of the powdered sample.\\
\hspace*{0.5cm}Near opposition, where the incidence and emergence follow the same direction creating a phase angle of 0°, the reflectance of a surface increases. This opposition surge results from the combination of two phenomena, the Coherent Backscattering Opposition Effect (CBOE) and Shadow-Hiding backscatter Opposition Effect (SHOE) \citep{hapke_1986, hapke_2002}. Though the laboratory measurements show the beginning of the opposition surge, the spectra from both models do not show this increase of reflectance at low phase angle. This results from the fact that the reflectance is not measured at the opposition, thus the models are not constrained in this area. The spectra resulting from the inversions consider the reflectance as following the trend constrained by reflectance values around the opposition. This results in a method limitation: under a geometry close to opposition, the reflectance of a simulated body will be underestimated compared to the value of reflectance of the same real body. At phase angles wider than 10°, the models accurately reproduce the reflectance of the surface, thus the observation will not be biased by the unconstrained opposition surge.\\

Along with the value of reflectance, spectroscopic analysis of planetary surfaces is also based on several parameters. The spectral slope is calculated considering the reflectance measured at two separated wavelengths as follow:
\begin{equation*}
    Slope = \frac{R_{\lambda 2} - R_{\lambda 1}}{\lambda 2 - \lambda 1}
\end{equation*}
with $R_{\lambda}$ the reflectance measured at the wavelength $\lambda$. The band depth is calculated considering a linear continuum between the two inflexion points around the absorption feature:
\begin{equation*}
    Band Depth = 1 - \frac{R_{center}}{R_{continuum}}
\end{equation*}
with $R_{center}$ the reflectance at the center of the band, considered as the wavelength at which the reflectance is at its minimum value inside the absorption feature, and $R_{continuum}$ the reflectance of the continuum, calculated considering a linear continuum and at the same wavelength as the center of the band.\\
\hspace*{0.5cm}No apparent shift of the band center is detected on our data with varying illumination and observation geometry. However, the laboratory measurements were acquired with a spectral step of 20nm, so a shift lower than this value would not be detected by the goniometer. No shift of the band center is thus reproduced by the Hapke and MRTLS models, and so will not appear on the BRDFs resulting from the inversions and on the bodies simulations.\\

To avoid introducing a bias in their estimations, the reflectance and spectral slope must be calculated at wavelengths outside any absorption features. Table \ref{wavelength spectro} presents the wavelengths used for the spectral calculations for the analysis of the howardite and sublimation residue.\\

\begin{table}[H]
\begin{center}
\begin{tabular}{lll}
\hline
 & Howardite & Ceres analogue\\
\hline
Reflectance & 740 nm & 580 nm\\
Slope & 740 - 1380 nm & 580 - 2200 nm \\
Continuum & 2720 - 3760 nm & 2560 - 3720 nm \\
Band depth & 2960 nm & 2760 nm \\
\hline
\end{tabular}
\caption{Wavelengths used for the determination of the spectral parameters in the case of the analysis of the howardite and Vesta, and of the sublimation residue and Ceres.}
\label{wavelength spectro}
\end{center}
\end{table}

The analysis of the bidirectional behaviour of the reflectance measured on the reference surfaces can be found in appendix. Figure \ref{compar model parameters} presents the spectral parameters calculated on the laboratory spectra and the set of spectra derived from the Hapke and MRTLS models. All geometrical configurations are plotted on the figure. 

\begin{figure}[H]
\centering
\includegraphics[width = 0.6\textwidth]{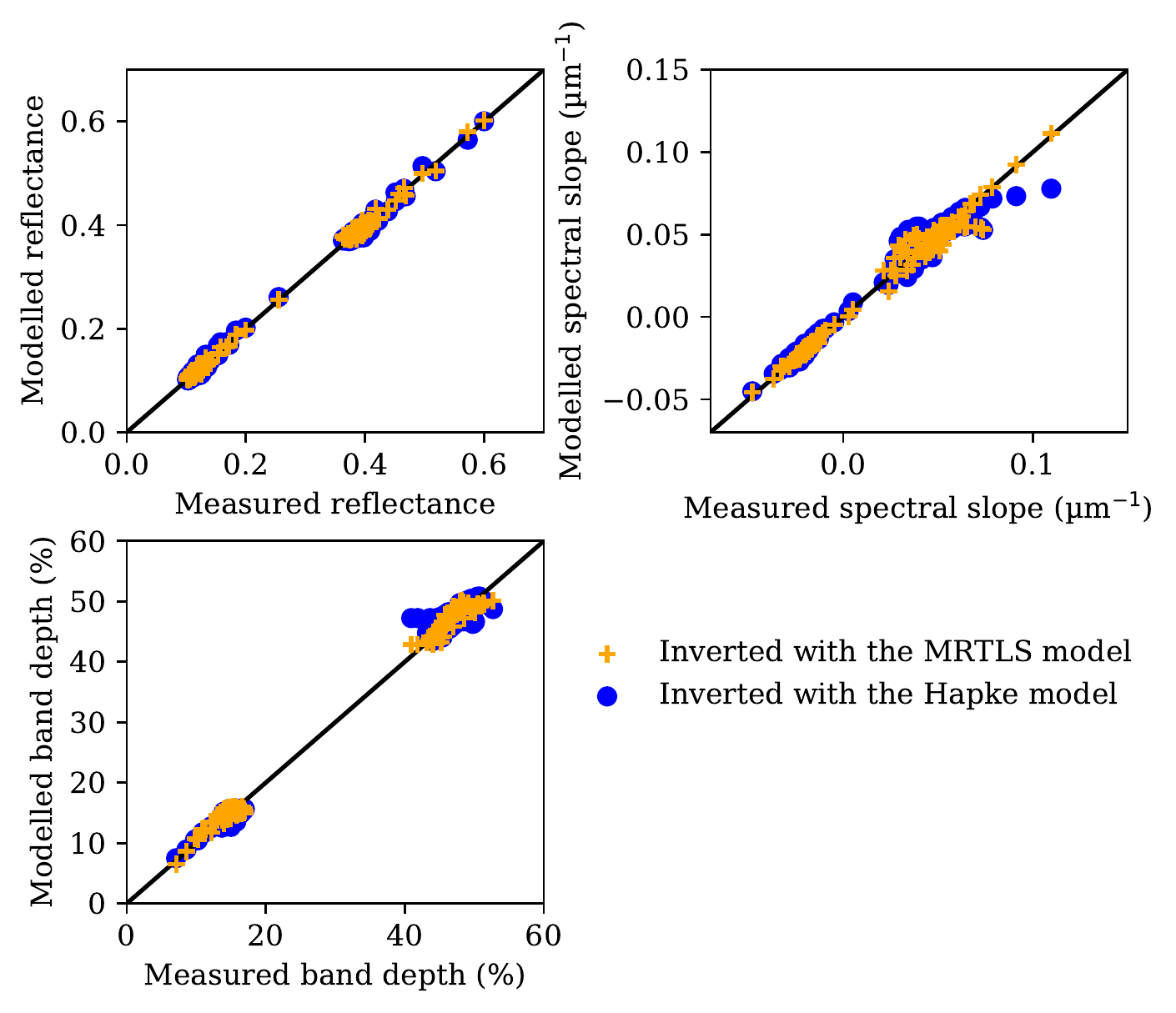}
\caption{Values of the spectral parameters derived from the spectra resulting from the inversion by the MRTLS model (orange) and the Hapke model (blue), with respect to the measurement values taken as reference (black line).}
\label{compar model parameters}
\end{figure}

The gap between the laboratory spectra and the inversed spectra is considered as the $\chi^2$ of the series, calculated as follow:
\begin{equation*}
    \chi^2 = \sum \frac{({{Spec}_{lab}}^{(i,e,a)} - {{Spec}_{model}}^{(i,e,a)})^2}{{\sigma}_{lab}^2}
\end{equation*}
with $Spec_{lab}$ and $Spec_{model}$ respectively the spectral parameters measured in the lab under the geometry (i,e,a) and the spectral parameters derived from the Hapke of MRTLS model under the same geometry, and $\sigma_{lab}$ the error on the spectral parameters derived from the laboratory spectra. The model which reproduces the most accurately the laboratory data will return the lowest $\chi^2$ values. The $\chi^2$ values calculated with the series of the spectra derived from the Hapke and MRTLS model are presented in table \ref{table_chi_2}.\\

\begin{table}[H]
\begin{center}
\begin{tabular}{lllll}
\hline
 & \multicolumn{2}{c}{Howardite} & \multicolumn{2}{c}{Ceres analogue}\\
\hline
 & MRTLS & Hapke & MRTLS & Hapke\\
\hline
Reflectance & 2.21 $10^6$& 2.34 $10^6$ & 3.53 $10^5$ & 6.30 $10^5$\\
Slope & 3.89 $10^4$ & 6.77 $10^4$ & 3.22 $10^3$ & 785 $10^3$\\
Band depth & 7.18 $10^4$ & 5.62 $10^4$& 6.30 $10^4$ & 2.43 $10^5$\\
\hline
Total & 2.32 $10^6$ & 2.47 $10^6$ & 4.19 $10^5$ & 8.81 $10^5$\\
\hline
\end{tabular}
\caption{$\chi^2$ values of the spectral inversion by the MRTLS and Hapke models.}
\label{table_chi_2}
\end{center}
\end{table}

Taking into account all values, the MRTLS model returns a $\chi^2$ value of 2.74 $10^6$ while the Hapke model returns a value of 3.35 $10^6$. This can come from the higher noise shown in the spectra resulting from the inversion by the Hapke model. Given the $\chi^2$ value and low noise spectroscopy, we will then use in priority the MRTLS model in our simulation.

\section{Results of the simulated scenes}
\hspace*{0.5cm}In this section, we present several simulated scenes. The first scenario presents the observation of a simple spherical and homogeneous body, the others present the observation of the targets (1)Ceres and (4)Vesta during spot-pointing protocoles and fly-bys. At each time, the target is simulated and observed according to the chosen conditions. The reflectance is averaged over the observed surface in order to obtain an unresolved spectroscopic observation of the body. \\

\subsection{Homogeneously covered sphere}
\hspace*{0.5cm}As a first test, we select the easiest shape possible for a planetary body. We thus virtually and homogeneously cover a sphere with one of the studied surface and simulate the unresolved spectroscopic observation of this body. \\

As this simulation does not require extensive calculation time, we can consider all combinations of surface composition and reflectance model type (MRTLS or Hapke) and, in each case, we cover the sphere uniformely with the corresponding reflectance. We thus create in total four different simulated spheres.

\subsubsection{Unresolved observation at fixed phase angle}
\begin{figure}[H]
\begin{center}
\includegraphics[width = 0.6\textwidth]{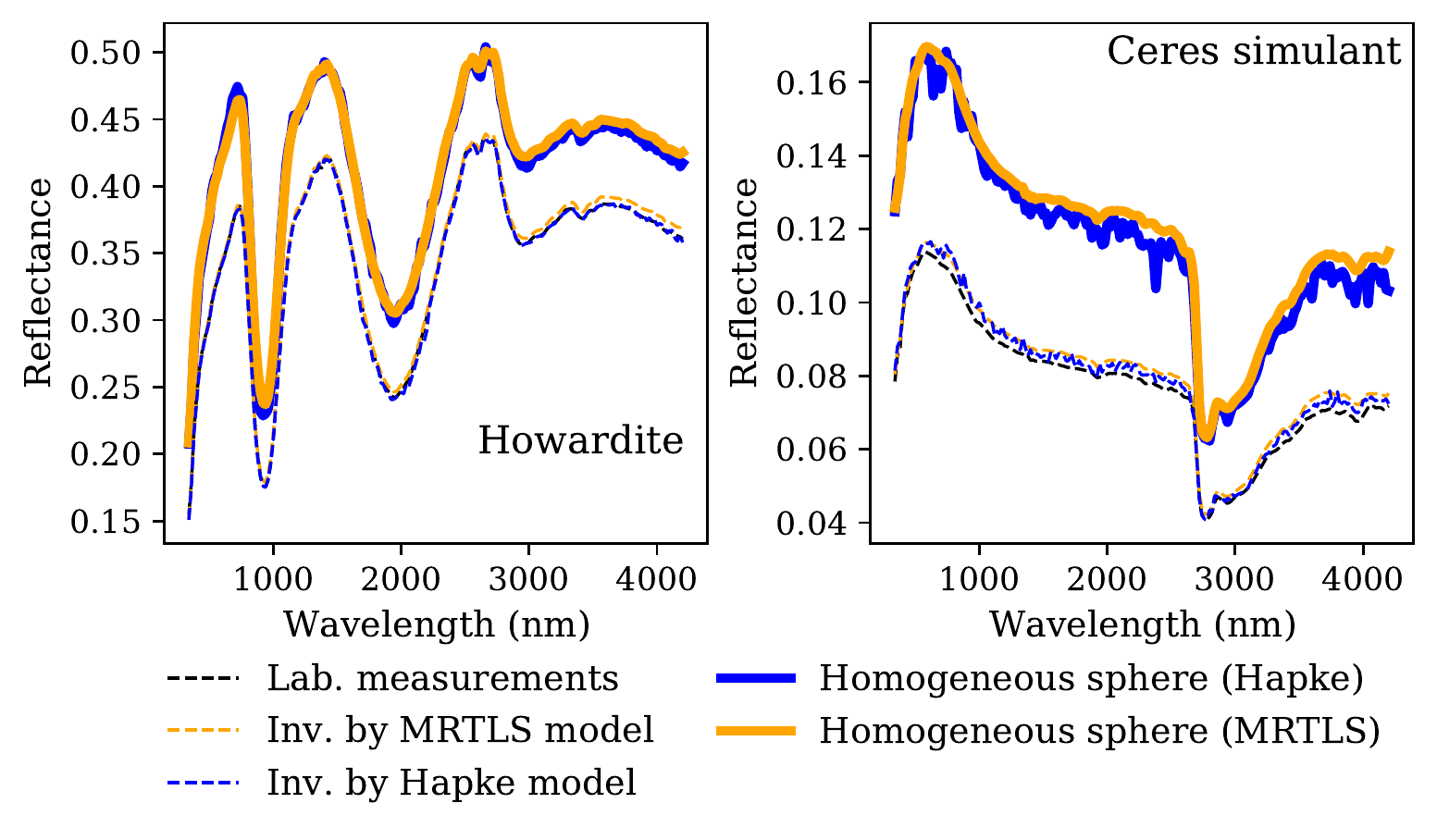}
\caption{Reflectance spectra at phase angle 30° of the simulated spheres, homogeneously covered with the howardite (left panel) and the Ceres simulant (right panel), and resulting from the inversion by the Hapke (blue) and MRTLS (orange) model. The reflectance spectra of the reference surfaces at incidence 0° and emergence 30° are also plotted in dotted lines (Black: Laboratory measurement, Orange: resulting from the inversion by the MRTLS model, Blue: resulting from the inversion by the Hapke model).}
\label{spectra_spheres}
\end{center}
\end{figure}

\hspace*{0.5cm}We first analyse the unresolved reflectance spectra of the simulated spheres at fixed geometry. Figure \ref{spectra_spheres} presents the reflectance spectra of the simulated spheres at phase angle 30°.

The unresolved simulated spheres present reflectance spectra similar to what is measured in the laboratory on the reference surfaces. The spectra of the simulated sphere covered with howardite presents the two pyroxene bands around 900 and 2000 nm and the hydration band around 3000nm. The simulated sphere covered with the Ceres simulant also presents the blue slope typical of its reference surface and the strong -OH band around 3000 nm. It has been shown in previous investigations that a strong slope can hide absorption bands \citep{Potin_mukundpura}. But in this case, the effect of shape on the integrated reflectance spectroscopy is not strong enough to alter the spectrum beyond recognition of its composition.\\
\hspace*{0.5cm}However, figure \ref{spectra_spheres} shows that the integrated reflectance spectra of the simulated sphere are significantly brighter than the reference surface at the same phase angle. For the howardite, the reference surface presents a reflectance value of 0.3851 at 740 nm, while the reflectance of the simulated spheres are of 0.4687 when inverted by the MRTLS model and 0.4737 if inverted using the Hapke model. In this case, the spherical shape of the simulated body increases the reflectance compared to the reference surface by roughly 0.11. The Ceres analogue reference surface presents a reflectance value at 580 nm of 0.1138, while the simulated spheres show a reflectance of 0.1726 when inverted with the MRTLS model and of 0.1726 when using the Hapke model. In this case, the spherical shape of the simulated body increases the reflectance value by 0.06.\\
\hspace*{0.5cm}The effect of the shape on the spectral slope and absorption band depth can hardly been seen on figure \ref{spectra_spheres} and will be further analyzed in the next section presenting the calculated spectral phase curves. One can expect stronger effects of the shape of the body on its integrated reflectance spectra in the case of a reference surface showing more drastic variations with the observation geometry than the surfaces presented here.

\subsubsection{Unresolved observation under several phase angles}
\begin{figure}[H]
\begin{center}
\includegraphics[width = 0.6\textwidth]{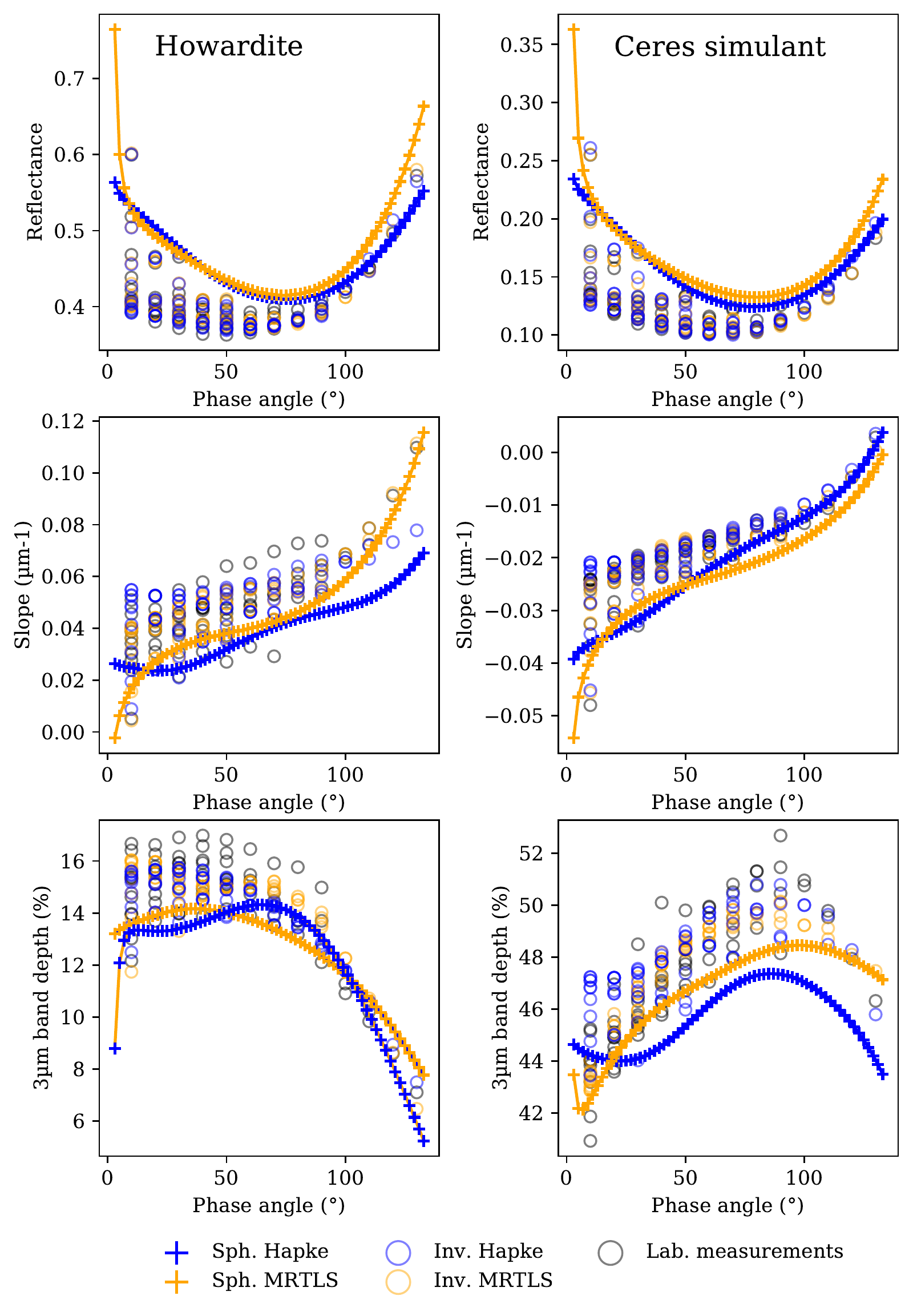}
\caption{Spectral parameters of the unresolved observation of the simulated spheres homogeneously covered with the howardite and the Ceres analogue. Sph. Hapke: homogeneous sphere covered with the surface derived from the inversion by the Hapke model. Sph. MRLSR: homogeneous sphere covered with the surface derived from the inversion by the MRTLS model. Inv. Hapke: spectral parameters calculated on the spectra resulting from the inversion by the Hapke model. Inv. MRTLS: spectral parameters calculated on the spectra resulting from the inversion by the MRTLS model.}
\label{spectro sphere homogenes}
\end{center}
\end{figure}

\hspace*{0.5cm}We now analyse the variations of the spectral parameters with increasing phase angle. The reflectance of the homogeneous spheres is calculated for scenes of varying illumination and observation geometry, generating phase angles from 1 to 140°. The phase curves, namely the variation of the spectral parameters with increasing phase angle, are compared to the ones measured in the laboratory and derived from the inversion by the Hapke and MRTLS models. Figure \ref{spectro sphere homogenes} compares the phase curves of the two simulated objects and their respective reference surfaces. The laboratory data and spectra derived from the inversion present several points for each value of phase angle, corresponding to the different incidence angles at which the spectra were acquired and derived.

We first notice on figure \ref{spectro sphere homogenes} a difference in the spectral components of the simulated observations with the model used for the inversion. Although the photometric phase curves (i.e. variations of the reflectance value with increasing phase angle) of the simulated objects are similar, the variations of the spectral slope of the sphere simulated by the Hapke model are steeper than the reference or the sphere simulated with the MRTLS model. The Hapke model returns variations of the band depth with increasing phase angle drastically different from the reference surface. While the band depths resulting from the inversion by the MRTLS model follow the convex shape of the reference surface curve, the Hapke model shows an inflexion of the band depth curve around a phase angle of 30°, followed by the global decrease of the band depth with increasing phase angle. These variations are due to the noise added to the spectra by the inversion models. As the MRTLS model has shown to return less noise on the reflectance spectra and the phase curves, we will apply the parameters derived from the MRTLS on the next simulated objects.\\
\hspace*{0.5cm}Independently on the inversion model used to simulate the surface, the phase curves of the simulated spheres are different from the phase curves derived from the laboratory and inverted spectra. The photometric phase curves of the simulated objects show higher reflectance than the references spectra on the whole range of phase angle. The spectral slope of the object simulated with the MRTLS model is globally lower than on the measurements, while the simulation with the Hapke model returns a spectral slope lower than the reference until a phase angle of 50°, then higher than the reference. Finally, the simulated objects show spectral features fainter than their reference surface.\\

The simulated object is homogeneously covered with a single reference surface, so no compositional heterogeneities are involved. Moreover, no terrain variation has been added to the simulation. This test shows that a spherical shape is enough to induce noticeable spectral variations between the integrated reflectance of the target and its surface itself. 

\subsection{Spot Pointing}
We now add the effect of irregular shape and surface topography to the analysis. Not to add the variations of observed area, we simulate a protocole of spot-pointing around the targets, where the position of the observer is fixed with respect to the surface, and the illumination is movable to change the phase angle. We used the small bodies (1)Ceres and (4)Vesta as simulated targets for this scenario and the next.\\
\hspace*{0.5cm}To limitate the computation time, we select 10 geometries starting from the lowest phase angle possible for the simulation, so 6°, to 120° in waxing. We also simulated the waning object from 45° to 6° of phase angle. The simulated images of both objects can be found in appendix.

\subsubsection{(1)Ceres}
\begin{figure}[H]
\begin{center}
\includegraphics[width = 0.6\textwidth]{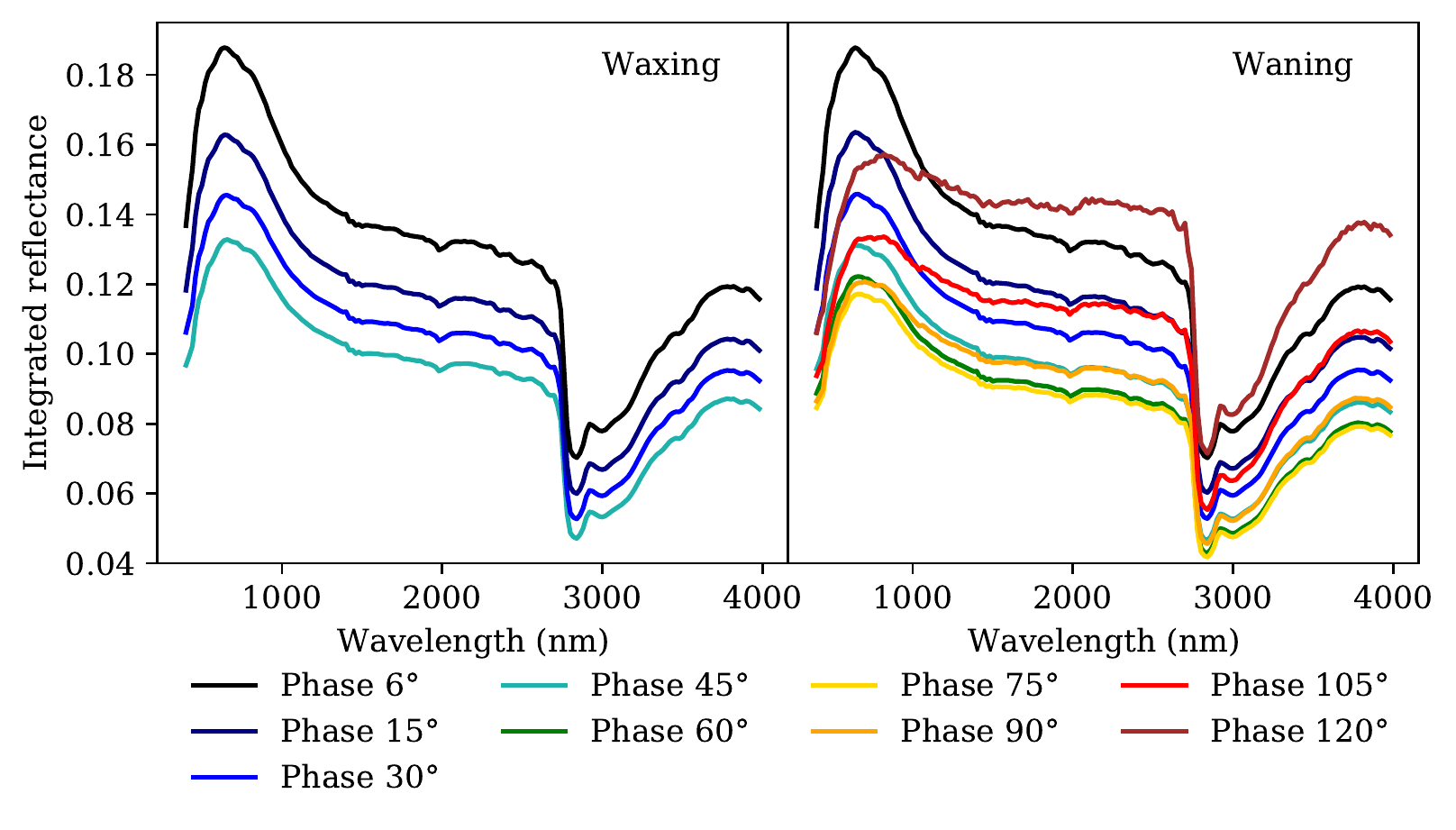}
\caption{Integrated reflectance spectra of the simulated Ceres observed in spot pointing during waxing (left) and waning (right) phase.}
\label{spectra SP Ceres}
\end{center}
\end{figure}

\hspace*{0.5cm}We first simulate the spot pointing around Ceres. The altitude of the observer is fixed at 24332 km above the surface. Figure \ref{spectra SP Ceres} presents the reflectance spectra of the unresolved simulated observations.\\

The spectra of the unresolved observations are again similar to the spectroscopy of the reference surface. It is interesting to note here that the spectra derived from the observation of the waxing and waning phases are identicals. In this case, the illumination condition has no effect on the reflectance spectroscopy of the object. However, figure \ref{spectra SP Ceres} shows that the reflectance of the simulated Ceres decreases for phases angle lower than 75°, then increases back to a value similar to what was observed at low phase angle. Moreover, it can be observed that the spectral slope of the integrated spectra increases, from a strong blue slope at low phase angle to an almost flat spectrum at the widest phase angle simulated here.\\

\begin{figure}[H]
\begin{center}
\includegraphics[width = 0.6\textwidth]{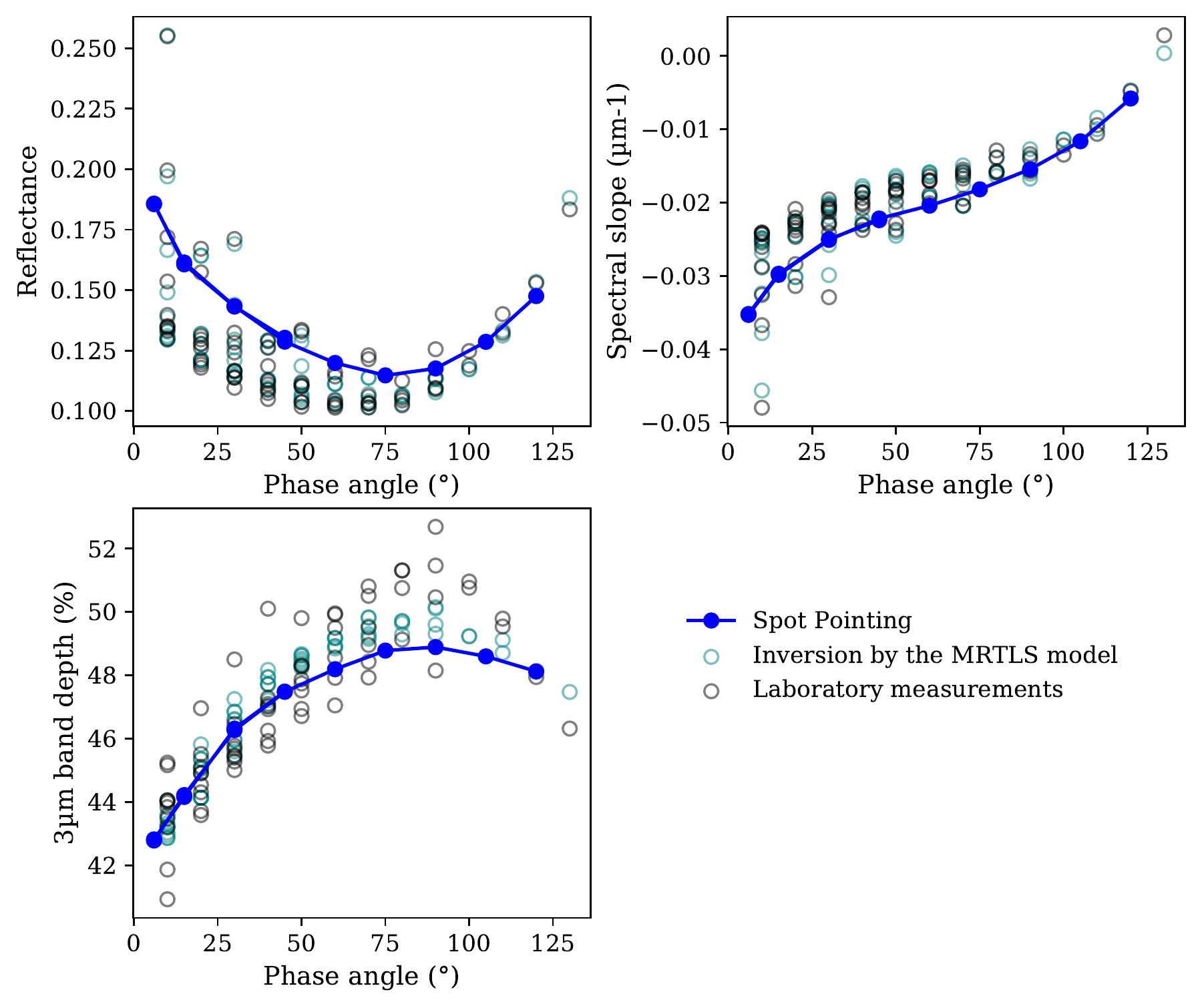}
\caption{Spectroscopic phase curves of the unresolved observations of Ceres in spot pointing, compared to the spectral parameters derived from the laboratory measurements and inversions by the MRTLS model.}
\label{SP Ceres}
\end{center}
\end{figure}

We now analyse the variations of the spectral parameters with increasing phase angle. Figure \ref{SP Ceres} presents the phase curves calculated from the simulated observations.\\

The phase curves of the simulated observations are similar to those of the reference surface, here the sublimation residue of the Ceres analogue. The photometric phase curve shows a concave shape similar to the phase curve of the reference surface, with however a reflectance value higher than what has been measured in the laboratory. The variations of the spectral slope and absorption band depths are also within the range of values of the reference surface. However, the spectral slope of the simulated body tends to be lower than the laboratory measurements. Ceres presents a spherical shape, with small terrain variations compared to its size, so results similar to the first test with the spherical body are expected. 

\subsubsection{(4)Vesta}
\begin{figure}[H]
\begin{center}
\includegraphics[width = 0.6\textwidth]{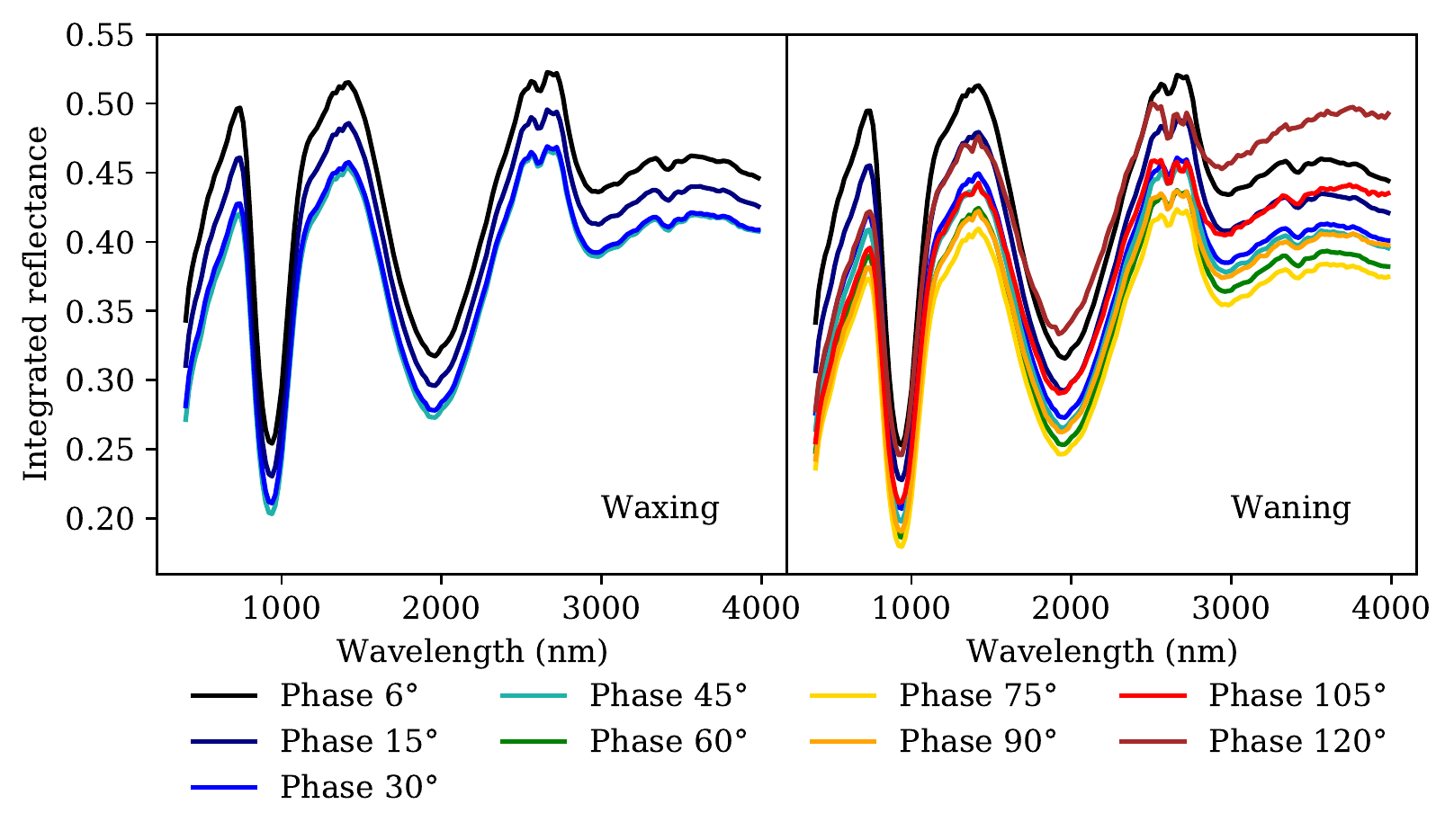}
\caption{Integrated reflectance spectra of the simulated Vesta observed in spot pointing during waxing (left) and waning (right) phase.}
\label{spectra SP Vesta}
\end{center}
\end{figure}

\hspace*{0.5cm}Similar to Ceres, we now simulate a spacecraft observing Vesta in spot spointing. The altitude of the observer is fixed at 12166 km above the surface. Figure \ref{spectra SP Vesta} presents the integrated reflectance spectroscopy derived from the simulated observations.\\

The integrated reflectance spectra of the simulated Vesta show the same pyroxene and hydration absorption features as the reference surface. The differences between the waxing and wining phases of the body are faint and difficult to detect on this figure. However, one can note the decrease of reflectance led by the increasing phase angle, coupled with the increase of the spectral slope.\\

\begin{figure}[H]
\begin{center}
\includegraphics[width = 0.6\textwidth]{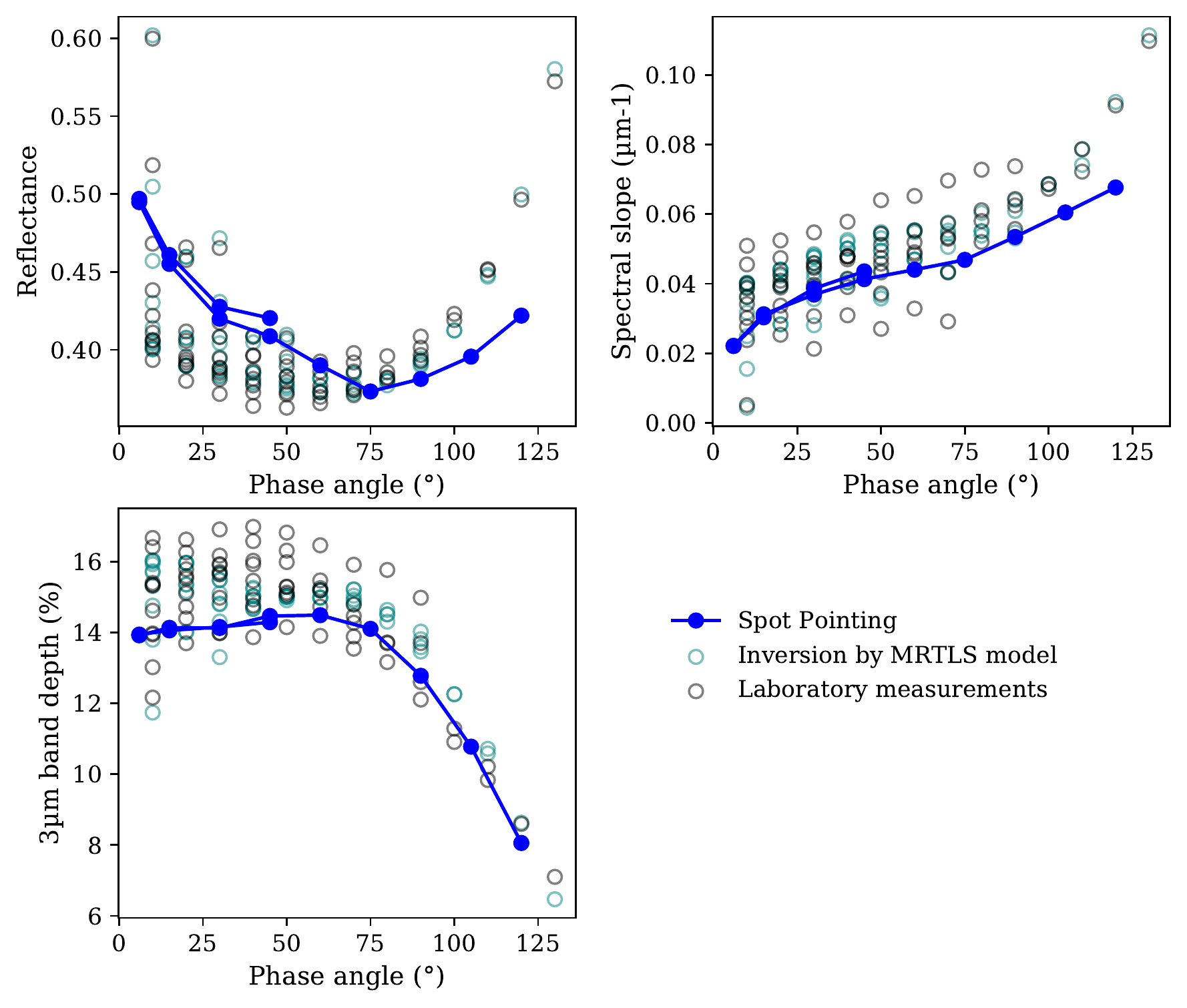}
\caption{Spectroscopic phase curves of the unresolved observations of Vesta in spot pointing, compared to the spectral parameters derived from the laboratory measurements and inversions by the MRTLS model.}
\label{SP Vesta}
\end{center}
\end{figure}

We now compare the spectral phase curves of Vesta derived from the simulated spot pointing observations. Figure \ref{SP Vesta} presents the spectral phase curves of Vesta calculated from the simulated observations.\\

The difference between the waxing and waning phases can only be seen on the photometric phase curve, where the reflectance measured in the waxing phase is slightly greater than in waning phase. At phase angle 45°, the reflectance of the simulated Vesta is of 0.420 in waxing phase for 0.408 in waning phase. The differences on the spectral slope and absorption band depth are too negligible in this simulation.\\
\hspace*{0.5cm}With increasing phase angle, we can observe on figure \ref{SP Vesta} that the simulated observation presents similar phase cuves than the reference surface. The  amount of reflectance shows a concave shape with increasing phase angle, with a minimum met at 75°. The unresolved spectra redden with increasing phase angle, and the detected band depth decreases. However, when compared to the reference surface, the simulated Vesta is brighter than what has been measured in the laboratory for phase angles lower than 60°. At wider phase angle, the simulated Vesta is darker than the reference surface. The two photometric curves begin to differ at phase angle 90°. The spectral slope derived from the simulated observations is globally lower than the reference surface. Similar to the photometric phase curve, the divergence from the reference is shown at phase angle 90°. Finally, the absorption bands calculated on the simulated observation are globally shallower than what has been measured in the laboratory, although the two phase curves do not differ from one another.\\

The simulation of spot pointing observation implies variations in the illumination conditions, such as shadows casted by the topography of the surface. However the surface observed on the simulated body is identical in all observations, as the sub-observer point is fixed on the simulated surface.

\subsection{Fly-by}
\hspace*{0.5cm}We now add the variation of the observed surface to the analysis. We thus simulate the fly-by of the observer near the small bodies Ceres and Vesta. The small bodies are homogeneously covered with their reference surface, namely the sublimation residue for Ceres and the howardite for Vesta. In this scenario, the illumination source and the asteroid are fixed, and the observer is following a straight line, going away from the light source, approaching and then passing by the small body, pointing toward the center of the target at all time. Observations are simulated at ten different geometries, ranging from phase angle 6° to 135°. The simulated images of both objects can be found in appendix.\\

To allow a comparison between the spot-pointing experiment and the fly-by, the altitude and position of the observer with respect to the surface are set identical for the phase angle of 30°.

\subsubsection{(1)Ceres}
\begin{figure}[H]
\begin{center}
\includegraphics[width = 0.45\textwidth]{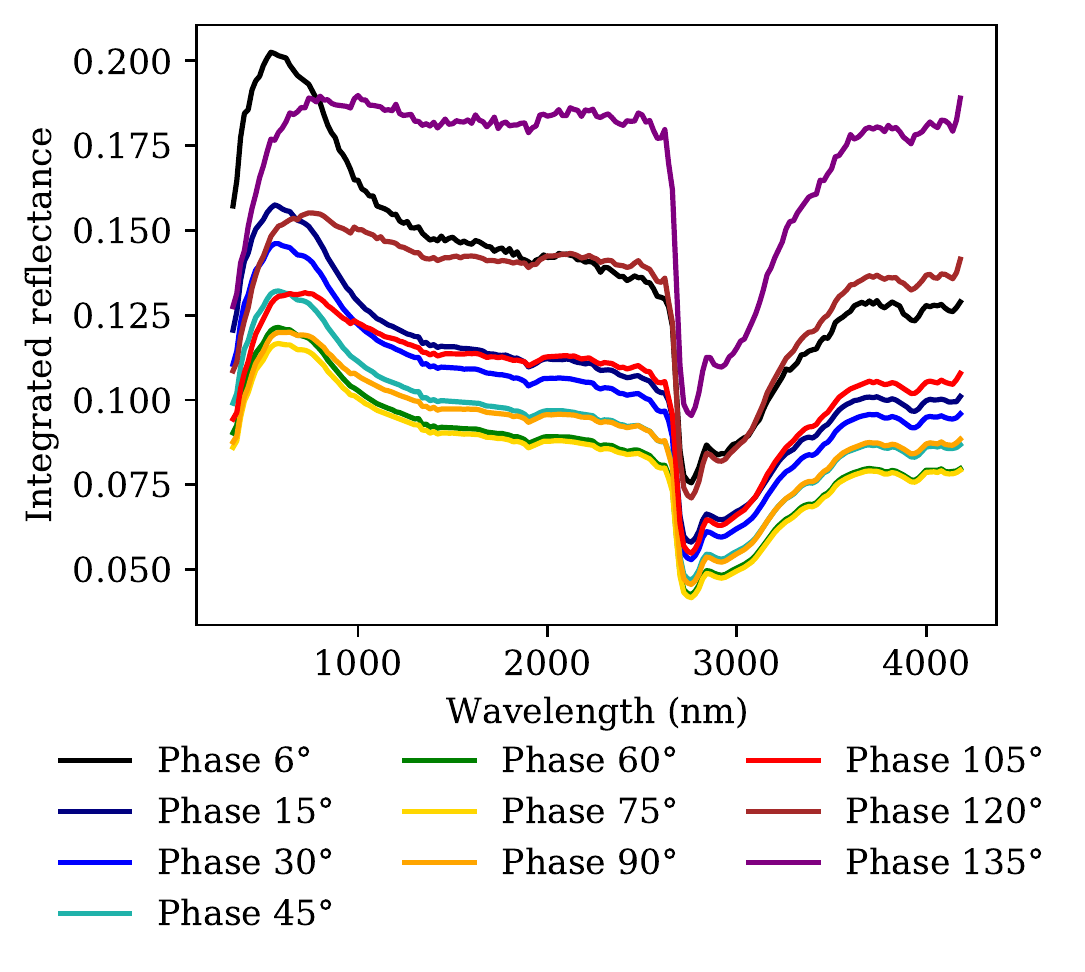}
\caption{Integrated reflectance spectra of Ceres resulting from the simulated fly-by. }
\label{spectra FB Ceres}
\end{center}
\end{figure}

\hspace*{0.5cm}We first analyse the simulated fly-by around Ceres. In this scenario, the simulated observer follows a trajectory along the equator of Ceres and is at an altitude of 12166 km above the surface at his closest approach. Figure \ref{spectra FB Ceres} presents the resulting reflectance spectra integrated over the whole surface of Ceres. \\

Similar to the results obtained for Vesta, the increasing phase angle induces strong variations of the measured reflectance spectroscopy. The reflectance value in the visible range is measured at 0.201 at phase angle 6°. This value decreases with increasing phase angle until reaching 0.116 at phase angle 75°. This value then increases until 0.178 at the widest simulated angle 135°. The spectra of the simulated Ceres present a strongly blue slope at low phase angle, reddening with increasing angle until a nearly flat spectrum at the widest phase angle simulated.\\

\begin{figure}[H]
\begin{center}
\includegraphics[width = 0.6\textwidth]{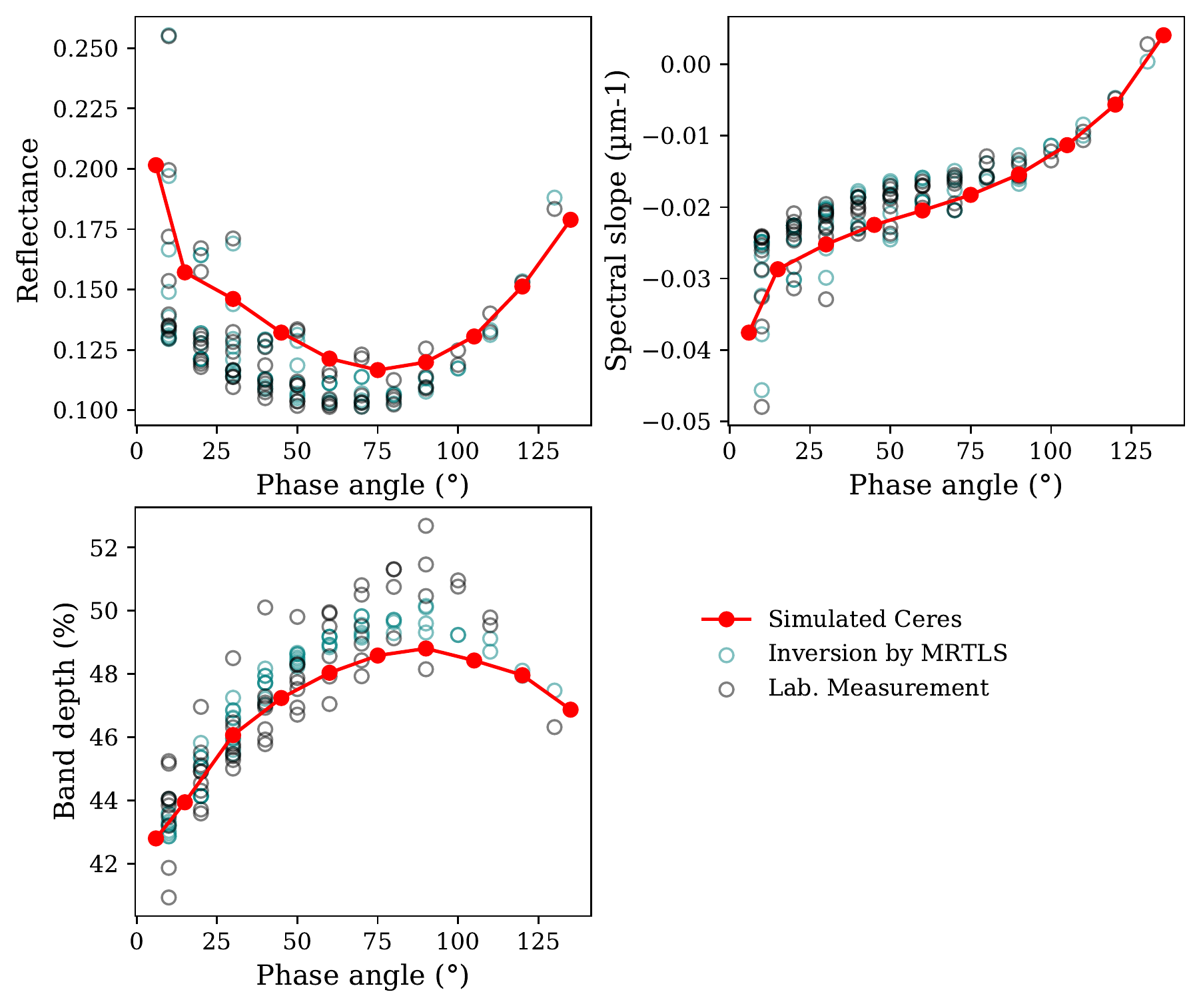}
\caption{Spectral phase curves resulting from the simulated fly-by around Ceres homogeneously covered with the Ceres analogue. The observations are compared to the spectral parameters derived from the reference surface.}
\label{PC FB Ceres}
\end{center}
\end{figure}

Figure \ref{PC FB Ceres} presents the spectral phase curves derived from the simulated fly-by around Ceres.

It can easily be seen from figure \ref{PC FB Ceres} that the spectroscopic phase curves of the simualted fly-by of Ceres present similar evolution with increasing phase angle when compared to the reference surface. This behaviour is similar to what has previously been observed on the homogeneous spheres. We note that the reflectance value of the simulated Ceres is generally higher than what has been measured in the laboratory on the reference surface and the spectral slope is lower than the reference surface except at phase angle wider than 100°. Finally, the absorption band depth of the simulated Ceres only becomes lower than most of the measured values at phase angles wider than 75°.

\subsubsection{Equatorial fly-by of(4)Vesta}
\begin{figure}[H]
\begin{center}
\includegraphics[width = 0.45\textwidth]{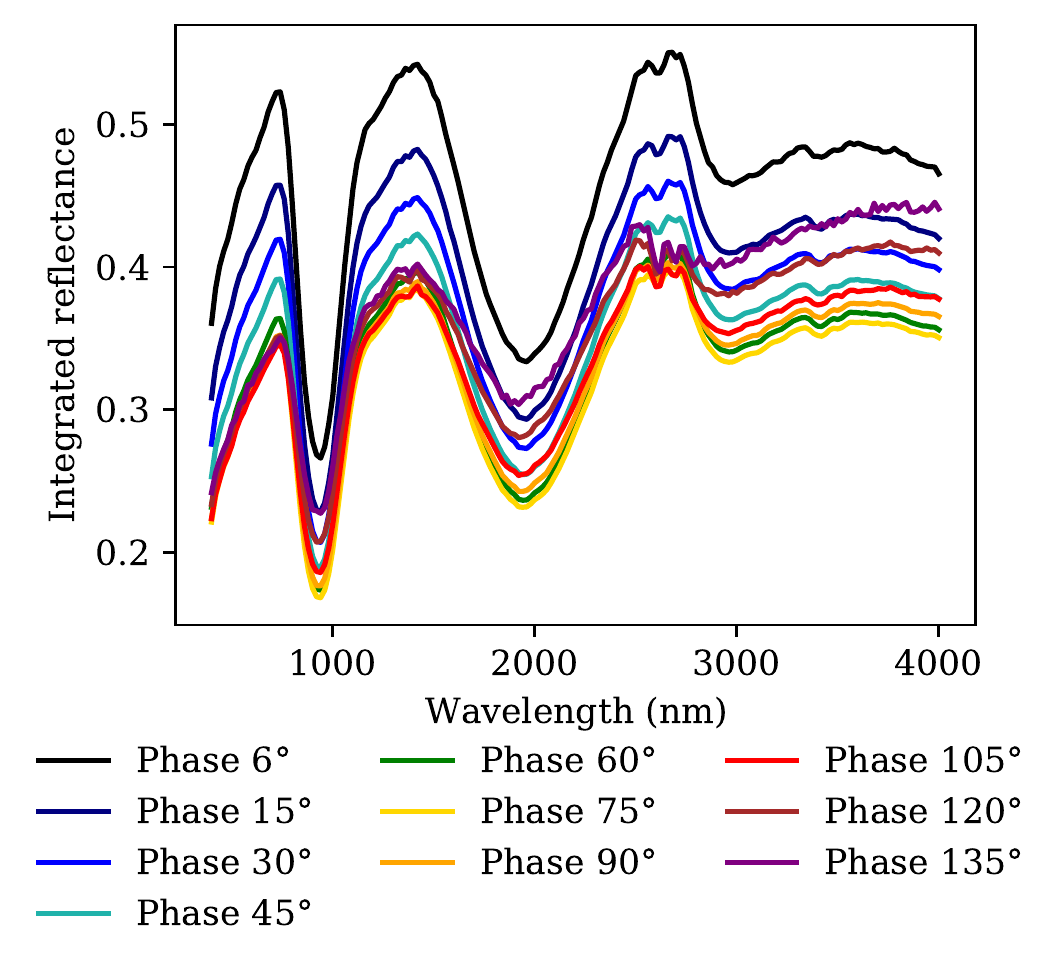}
\caption{Integrated reflectance spectra of Vesta resulting from the simulated fly-by.}
\label{spectra FB vesta}
\end{center}
\end{figure}

\hspace*{0.5cm}We now simulate the fly-by of the observer around Vesta. The observer trajectory, illumination conditions and observation geometries are similar to the previous fly-by of Ceres. However, the observer is at an altitude of 6083 km above the surface at his closest approach. Figure \ref{spectra FB vesta} presents the resulting reflectance spectra integrated over the whole surface of Vesta. \\

\begin{figure}[H]
\begin{center}
\includegraphics[width = 0.6\textwidth]{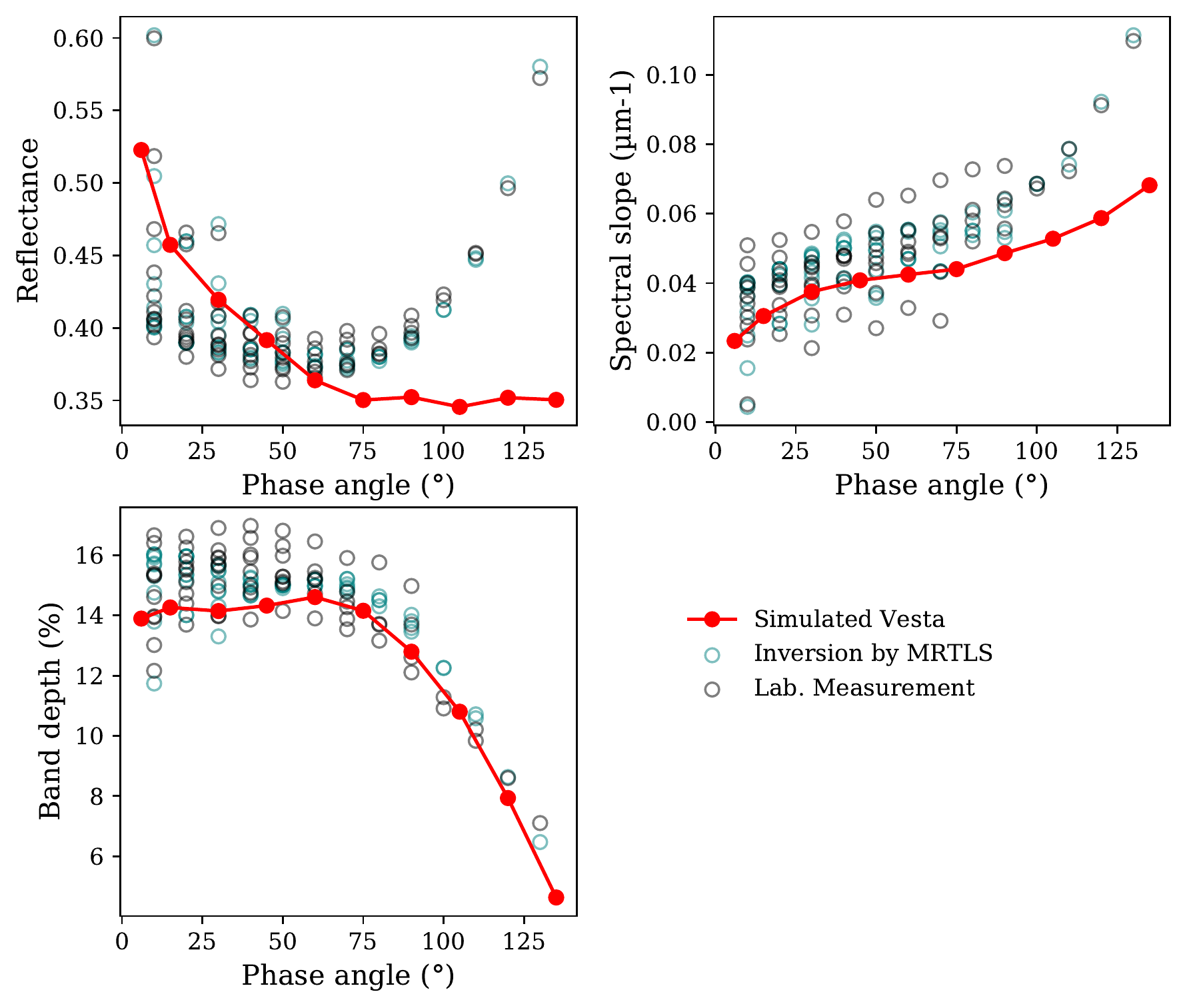}
\caption{Spectral phase curves resulting from the simulated fly-by around Vesta homogeneously covered with the howardite powder. The observations are compared to the spectral parameters derived from the reference surface.}
\label{PC FB Vesta}
\end{center}
\end{figure}

The reflectance spectroscopy of the simulated Vesta presents similar spectral parameters than the reference surface. All spectral signatures, pyroxenes and OH bands, are detected in the spectra throughout the simulation. The reflectance value in the visible range decreases with increasing phase angle, passing from 0.523 to 0.350. Unlike the simulated fly-by of Ceres, no increase of reflectance is observed at wide phase angle. Figure \ref{PC FB Vesta} presents the evolution of the spectral parameters with increasing phase angle measured on the simulated equatorial fly-by of Vesta.\\

We first notice on figure \ref{PC FB Vesta} the different behaviours shown by the photometric phase curve of the simulated fly-by. The reflectance decreases within the whole range of phase angles, while the reflectance of the reference surface presents a concave shape, increasing for phase angles wider than 60°. Differences are also present when comparing the spectral slope derived from the fly-by and the reference surface. The phase curve of the simulated object separates from the measurements values around a phase angle of 75°. However, the variations of the band depth with increasing phase angle of the simulated fly-by show the exact variations of the reference surface.\\
\hspace*{0.5cm}For phase angles lower than 45°, the unresolved reflectance of the simulated Vesta is greater than the reference values. The simulated body is then darker than the reference surface at wider phase angle, until reaching a reflectance of 0.35 at phase angle 135° where 0.57 was measured on the reference surface. The spectral slope resulting from the simulation is significantly lower than the reference surface, on the whole angular range. The greater difference is met at the widest simulated phase angle, where a slope of 0.068 is derived from the simulation for 0.110 on the reference surface. Finally, the absorption bands detected on the simulations are shallower than what had been calculated on the reference surface, over the whole angular range studied here. At phase angle 30°, the simulated Vesta presents a 3µm-band depth of 14.15$\%$, while the values derived from the reference surface span from 13.3 to 16.9$\%$ between the measurements and spectral inversion by the MRTLS model.\\
\hspace*{0.5cm}The surface parameters applied on the simulated Vesta are those resulting from the inversion by the MRTLS model, based on the laboratory measurements. The simulated surface of this Vesta is thus identical to the one measured in the laboratory. The body is homogeneously covered with the same composition and texture. The only differences between the laboratory measurements and simulations are the global shape of the object and the varying observed terrain on its surface. In this case, the fly-by itself, coupled with a non-spherical shape of the body and non-negligeable surface topography are strong enough to induce drastic differences between the phase curves of the target and its surface in itself.

\subsubsection{Polar fly-by of (4)Vesta}
\hspace*{0.5cm}In the previous fly-by, both Ceres and Vesta show unvarying shapes along the observations. Ceres is nearly spherical and the fly-by around Vesta was set along the equatorial ridge of the body, thus observing an oblong shape at each measurement. To add the variation of shape, we simulate a fly-by around Vesta, starting from its north pole to the south. During this fly-by, the observed shape of Vesta goes from a disk at the poles to an oblong shape near the equator.\\

\begin{figure}[H]
\begin{center}
\includegraphics[width = 0.45\textwidth]{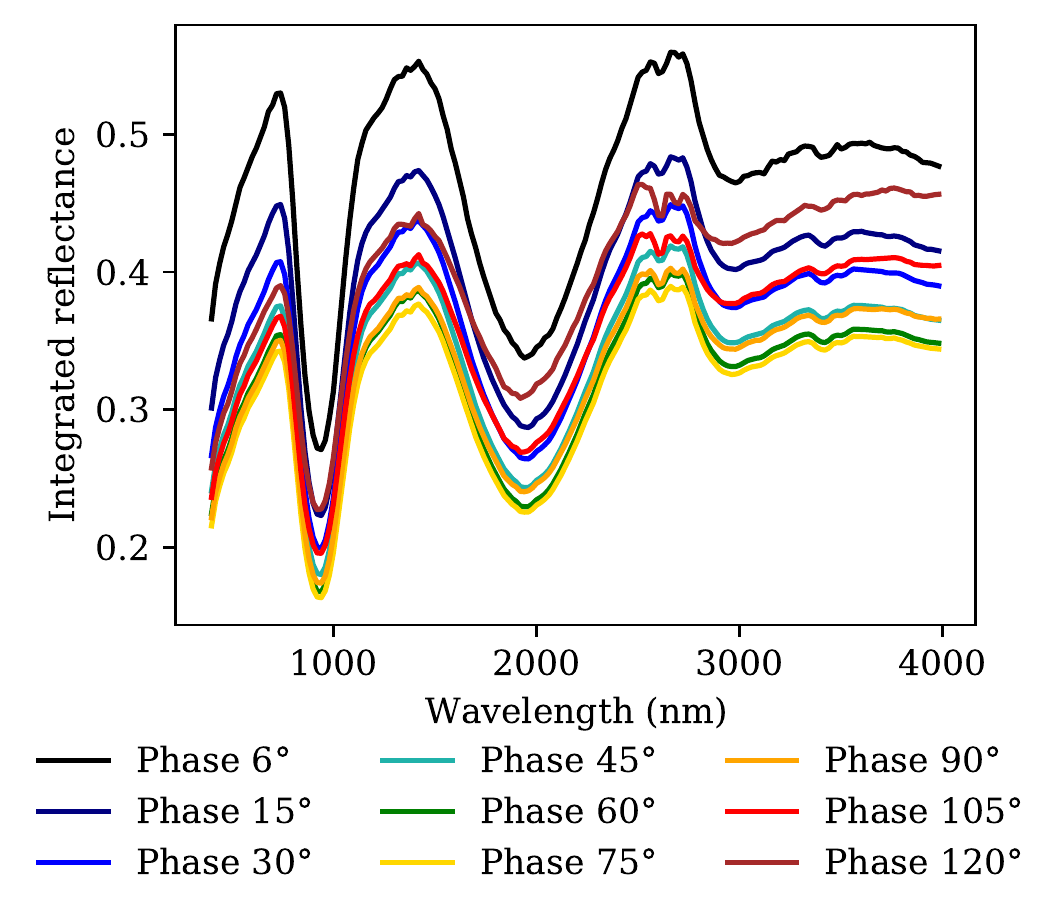}
\caption{Integrated reflectance spectra of Vesta resulting from the polar fly-by.}
\label{spectra FB polaire}
\end{center}
\end{figure}

Due to some limitations in the simulation program, the observation of Vesta with a phase angle of 135° was impossible. The polar fly-by is thus simulated for phase angles up to 120°. Figure \ref{spectra FB polaire} presents the integrated reflectance spectra of Vesta resulting from this simulation.

Similarly to the previous simulations, the unresolved spectra of Vesta calculated from the polar fly-by show a slightly red slope, as well as the pyroxene and hydration bands. With increasing phase angle, we can note that the reflectance of the simulated body first decreases from 0.530 at phase angle 6° to 0.342 at phase angle 75°. The reflectance then increases until reaching 0.390 at phase angle 120°. Figure \ref{spectra FB polaire} also shows that the spectral slope of the simulated Vesta increases with increasing phase angle.\\

\begin{figure}[H]
\begin{center}
\includegraphics[width = 0.6\textwidth]{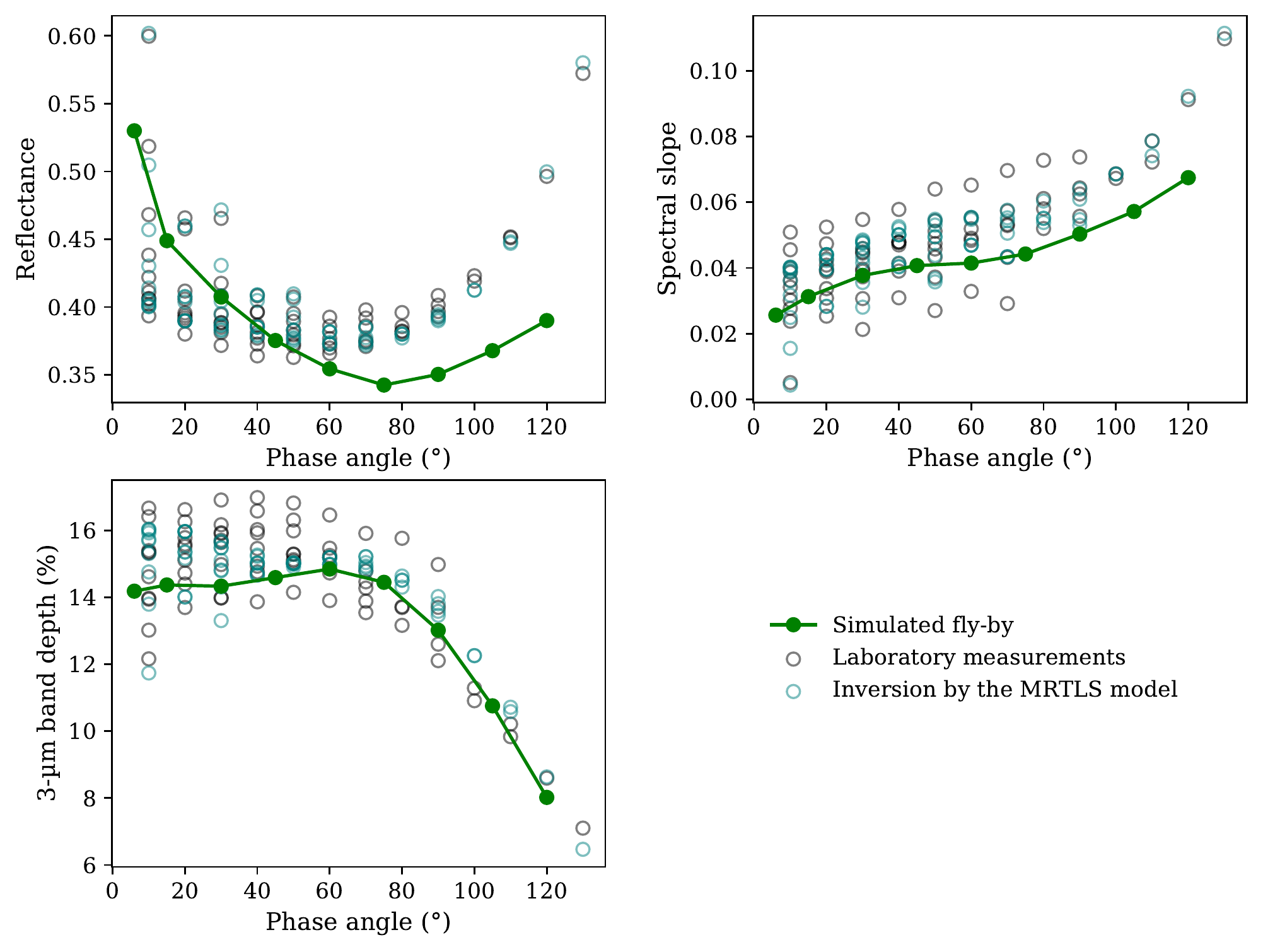}
\caption{Spectral phase curves resulting from the simulated polar fly-by around Vesta homogeneously covered with the howardite powder. The observations are compared to the spectral parameters derived from the reference surface.}
\label{PC FBP Vesta}
\end{center}
\end{figure}

Figure \ref{PC FBP Vesta} compares the spectral parameters of the reference surface to the phases curves of the simulated Vesta observed during a polar fly-by.\\

The spectral phase curves derived from the observation show similar behaviours compared to the reference surface. Similar to the results of the previous fly-by along the equator, the observation and reference photometric phase curves differ for phase angles wider than 75°. However, the concave shape shown by the observed photometric phase curve reduces the differences at the widest simulated phase angle. The unresolved observation of the simulated Vesta presents a reflectance of 0.39 at phase angle 120°, while the reference surface presents a reflectance value of 0.57. The evolution of the spectral slope of the simulated body is similar to the previous fly-by,  with, however, lower differences between the phase curves for phase angles wider than 90°. The simulated Vesta returns a spectral slope of 0.067 at phase angle 120°, for 0.110 measured on the reference surface. Finally, the evolution with increasing phase angle of the band depth on the simulated Vesta shows the same convex shape than the reference surface.\\

Similar to the results obtained with the equatorial fly-by, the phase curves obtained on the polar fly-by differ from these of the reference surface. The simulated body is brighter than the reference surface alone for phase angles lower than 45°, its spectral slope is lower and absorption bands are shallower than the reference on the whole angular range.\\

\section{Discussion on the effect of shape and topography}
\hspace*{0.5cm}We here compare the results from all the simulated scenarii and suggest reasons for the observed differences in the phase curves. We first analyse the effect of shape on the reflectance spectroscopy of a planetary body with the comparison between the simulated homogeneous spheres and their reference surfaces. We then compare the phase curves of the simulated spot-pointings and fly-bys to assess the effect of topography. Finally, we discuss the potential bias of comparing unresolved observations and laboratory spectroscopy acquired under a single geometrical configuration.

\subsection{Comparison between the reference surfaces and simulated spheres: The effect of shape}
\hspace*{0.5cm}It has been shown previously on figure \ref{spectro sphere homogenes} that the phase curve of a spherical body of homogeneous composition differs from the results obtained on its surface itself, while staying in the range of values of each of its spectral parameters. The global shape of the sphere induces local variations of incidence and emergence angles on the surface, leading to different geometrical configurations, thus spectral variations. Moving away from the subsolar point, the local incidence angle changes from nadir to grazing illumination at the limbs, and this effect is similar for the emergence angle. Both reference surfaces show an increase of reflectance if illuminated and observed at grazing angles (figure \ref{BRDF compar model}). The reflectance values higher than the references measured on the simulated spheres can be due to the wide area on the spheres showing grazing local angles, thus increasing the integrated reflectance. The variations due to the spherical shape are responsible for the higher reflectance value, lower spectral slope and shallower absorption bands of the unresolved simulated body compared to the reference surface.\\

\subsection{Comparison between the different simulated observations: The effect of topography}
\begin{figure}[H]
\begin{center}
\includegraphics[width = 0.6\textwidth]{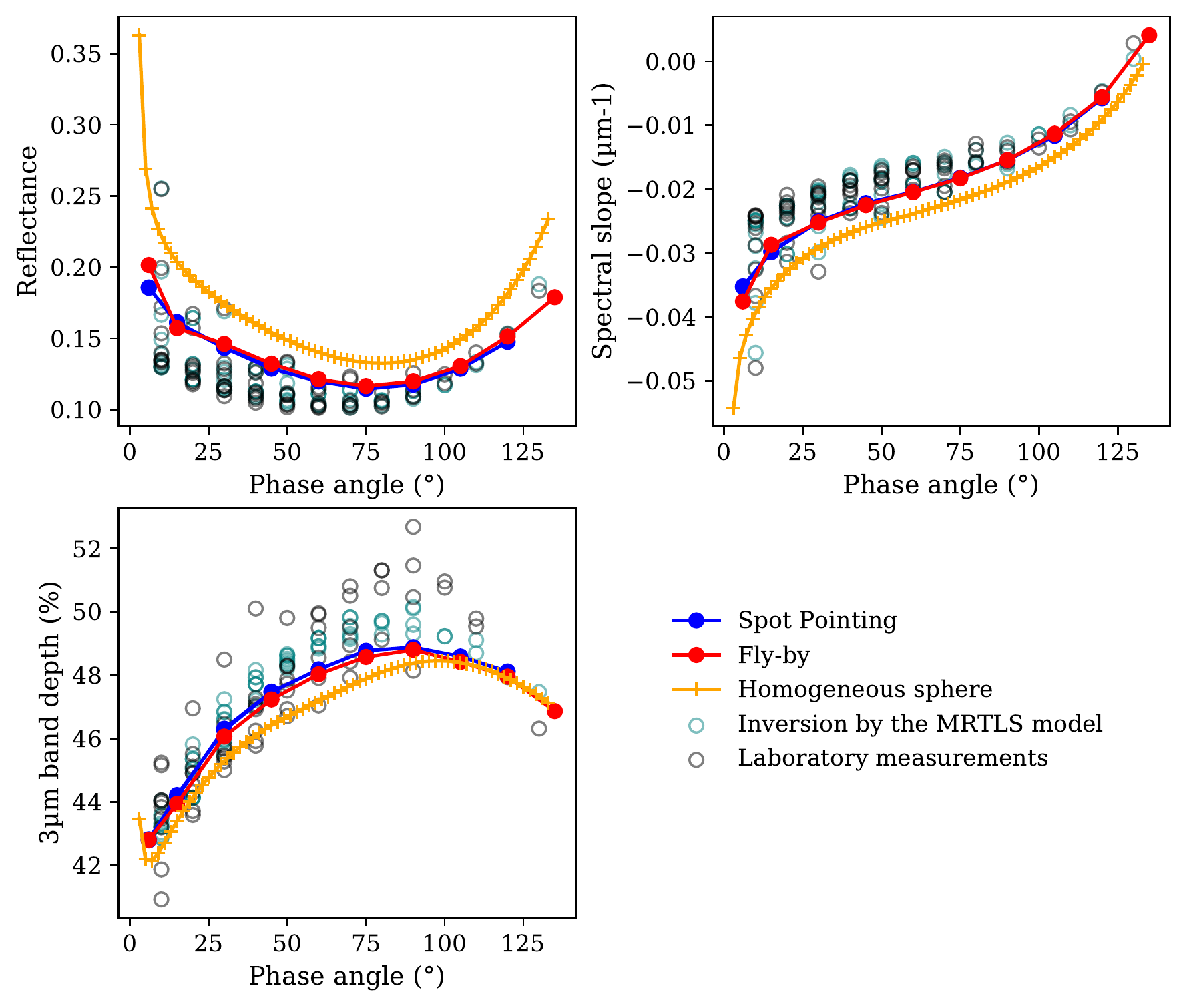}
\caption{Spectral phase curves resulting from the simulated spherical body (orange line and markers), spot-pointing (blue line and markers) and fly-by (red line and markers) around Ceres, compared to the spectral parameters of the reference surface calculated on the measured spectra (black circles) and resulting from the inversion by the MRTLS model (blue circles).}
\label{PC compar SP FB Ceres}
\end{center}
\end{figure}

\hspace*{0.5cm}We now compare for each target the phase curves resulting from the homogeneous spheres, spot-pointing and fly-by simulations. Figure \ref{PC compar SP FB Ceres} compares the phase curves of the simulated Ceres.\\

The phase curves resulting from the simulations of Ceres differ from the results obtained with the homogeneous spherical body covered with the Ceres analogue. The simulated sphere is brighter and shows a bluer slope than than the simulated Ceres over the whole geometrical range studied here. The spherical body also presents a 3-µm absorption band fainter than Ceres for phases angles lower than 100°, the phase curve of the sphere then matches those of the small body. We can make the hypothesis that these differences come from the small amount of shadows casted on the surface by the topography. The spot pointing and fly-by simulations show almost identical photometric and spectral phases curves. Only the observations under a phase angle of 6° present a noticeable difference, due to the low amount of active pixels for the fly-by. Identical behaviours with respect to the reference surface are observed for the simulated spot-pointing and fly-by.\\

\begin{figure}[H]
\begin{center}
\includegraphics[width = 0.5\textwidth]{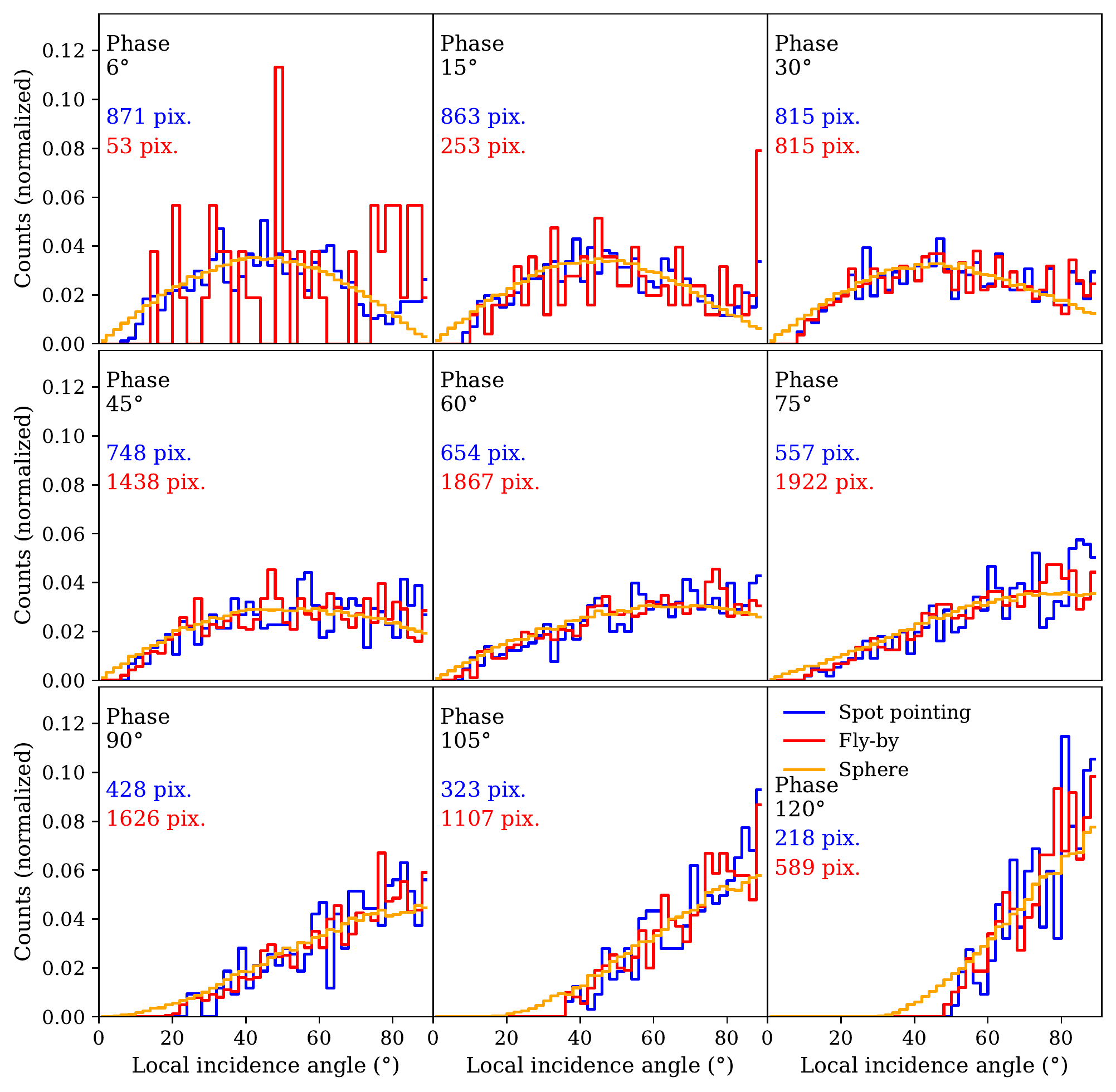}
\caption{Histograms of the local incidence angles resulting from the spot-pointing (blue) and fly-by (red) around Ceres, and from the simulation of a spherical body (orange). Each panel presents a phase angle. The counts have been normalized to the area under the histogram, so to the total number of pixels illuminated and observed on the images. This number of pixels is indicated on each panel for both the spot-pointing (blue) and fly-by (red).}
\label{histogram compar inci FB SP Ceres}
\end{center}
\end{figure}

\begin{figure}[H]
\begin{center}
\includegraphics[width = 0.5\textwidth]{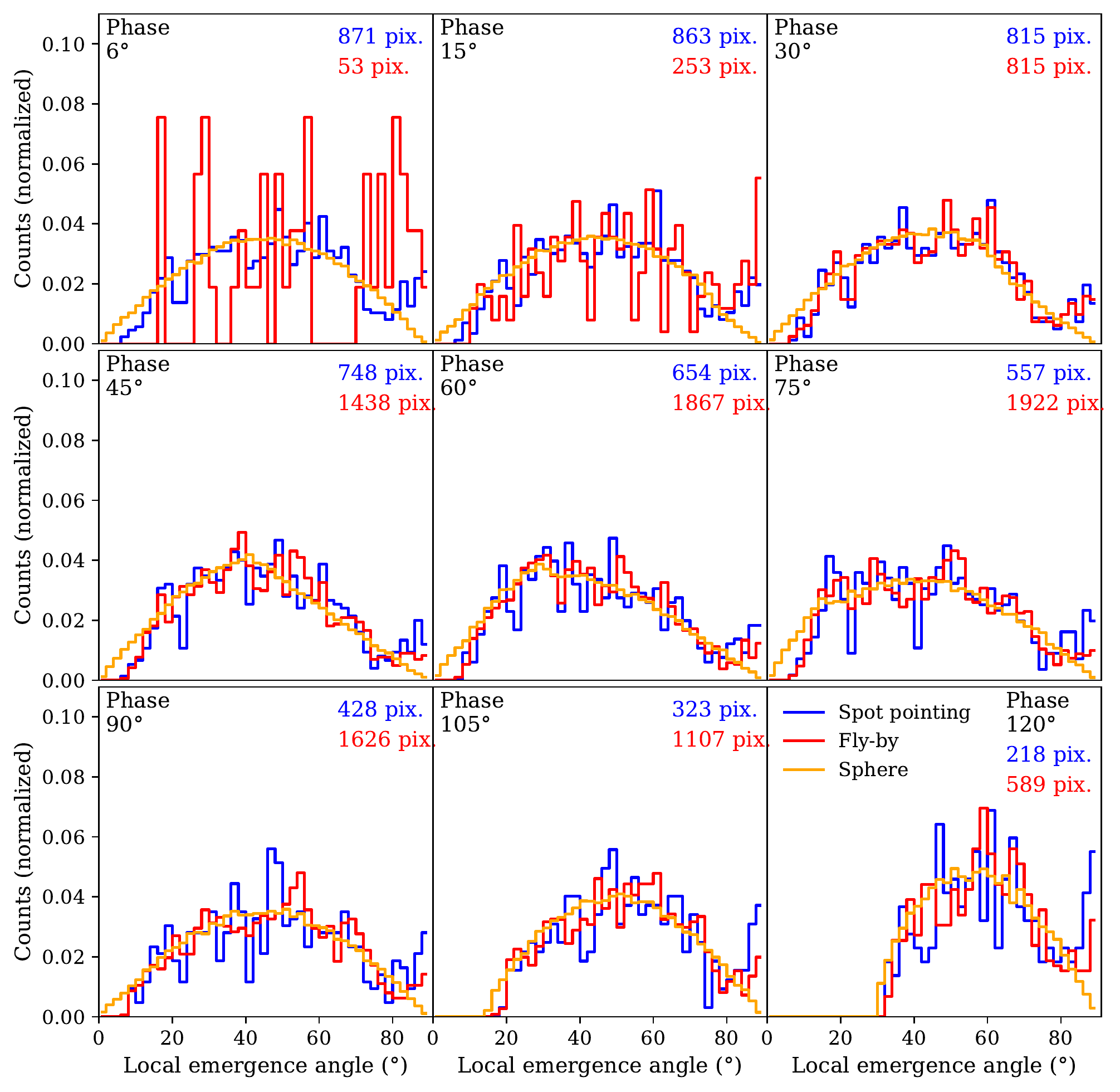}
\caption{Histograms of the local emergence angles resulting from the spot-pointing (blue) and fly-by (red) around Ceres, and from the simulation of a spherical body (orange). Each panel presents a phase angle. The counts have been normalized to the area under the histogram, so to the total number of pixels illuminated and observed on the images. This number of pixels is indicated on each panel for both the spot-pointing (blue) and fly-by (red). }
\label{histogram compar emer FB SP Ceres}
\end{center}
\end{figure}

The shape of the planetary body and the topography of its surface generates varying directions of illumination and observation on each facet of the shape model. These incidence and emergence angles are then averaged on all facets included in a single pixel, e.g. the local incidence and emergence angles. Differences in the distribution of these local angles will results in photometric variations, even with a constant phase angle. We now compare the histograms of the local incidence and emergence angles resulting from the simulated spot-pointing and fly-by around Ceres. Figure \ref{histogram compar inci FB SP Ceres} and  \ref{histogram compar emer FB SP Ceres} presents the distributions of the local incidence and emergence angles respectively. \\

In the case of Ceres, we observe that the distributions of the local incidence and emergence angles are similar on both simulations, and follow the distribution of geometries obtained on the spherical body. The differences of distribution are due to the sampling, i.e. varying with the number of active pixels in each observation, and to the observed area shifting on the simulated body between observations during the fly-by. As the observed topography changes, the distribution of local incidence and emergence angles should change as well. Because of the spherical shape of Ceres, no variations in shape is observed with increasing phase angle either on the spot-pointing or fly-by simulation. The great similarity between the angular distributions in both simulations and on the sphere implies that the variations in topography and observed area have negligible effects in this case. One can suppose that size scale between the diameter of Ceres and its surface topography is too important and smoothen the geometric variations due to the topography.\\

\begin{figure}[H]
\begin{center}
\includegraphics[width = 0.6\textwidth]{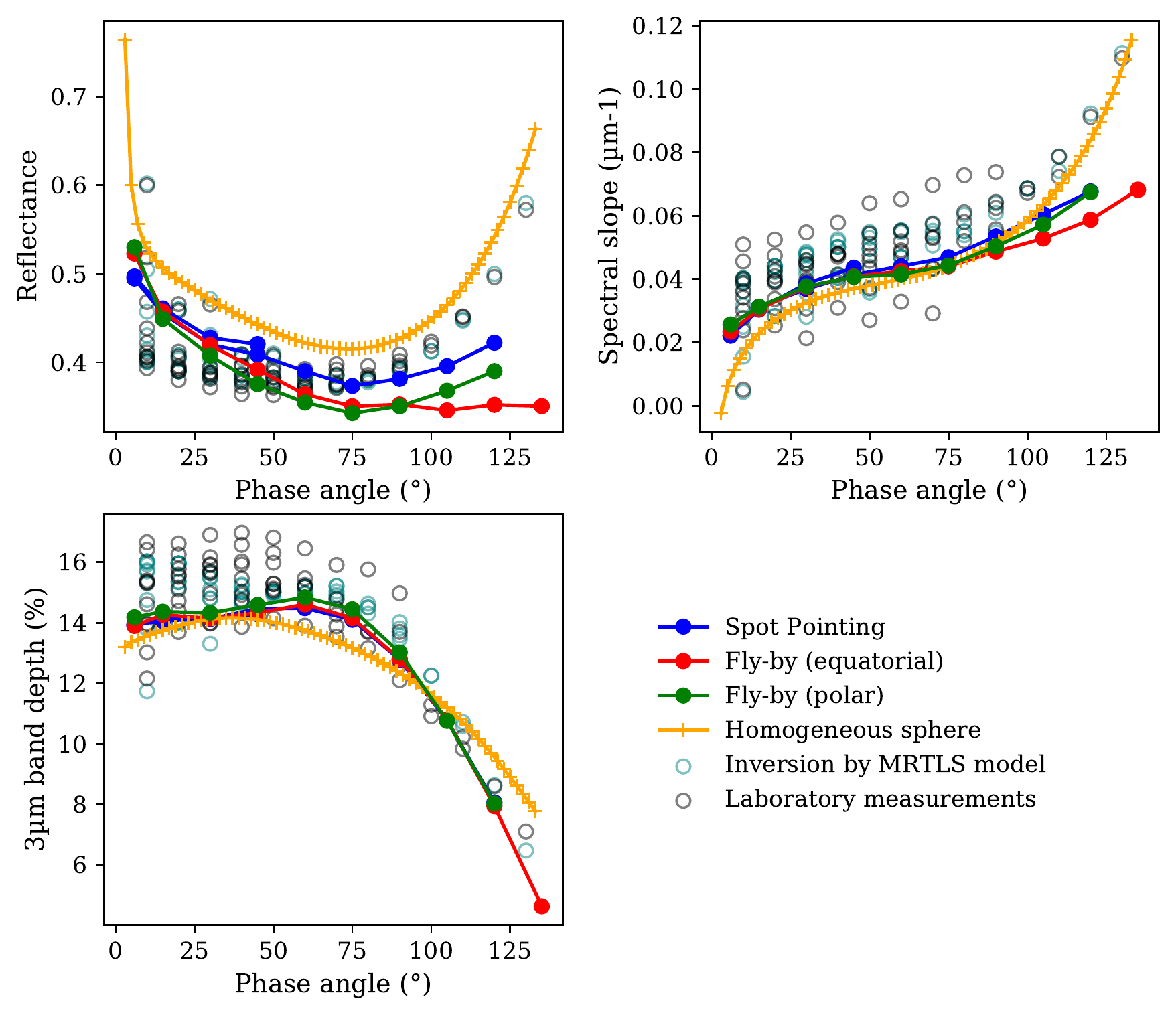}
\caption{Spectral phase curves resulting from the simulated homogeneous sphere (orange line and markers), spot-pointing (blue line and markers), equatorial fly-by (red line and markers) and polar fly-by (green line and markers) around Vesta, compared to the spectral parameters of the reference surface calculated on the measured spectra (black circles) and resulting from the inversion by the MRTLS model (blue circles).}
\label{PC compar SP FB Vesta}
\end{center}
\end{figure}

We now analyse the differences between the simulated scenarios observing Vesta and the homogeneous sphere. We remind that the spot pointing simulation led to variations in the illumination and  observation conditions, the equatorial fly-by added a shift of the observed area, and the polar fly-by induced a different observed area and projected shape. Figure \ref{PC compar SP FB Vesta} compares the phase curves of the simulated Vesta obtained during the spot-pointing, equatorial and polar fly-by.\\

It is interesting to note that, at low phase angle, the differences between the phase curves of the reference surface and simulated scenarii are similar to the differences observed for a spherical body. The phase curves begin to differ from phase angles between 60° for the fly-bys or 75° for the spot pointing simulation. One can suggest that at low phase angle, the shadows casted on the surface by the topography are at the lowest, implying that most observed spectral variations are due to the shape of the body. At widest phase angle, the shadows casted on the surface induce a stronger effect of the topography compared to the shape itself, leading to drastic differences between the phase curves.\\

\begin{figure}[H]
\begin{center}
\includegraphics[width = 0.6\textwidth]{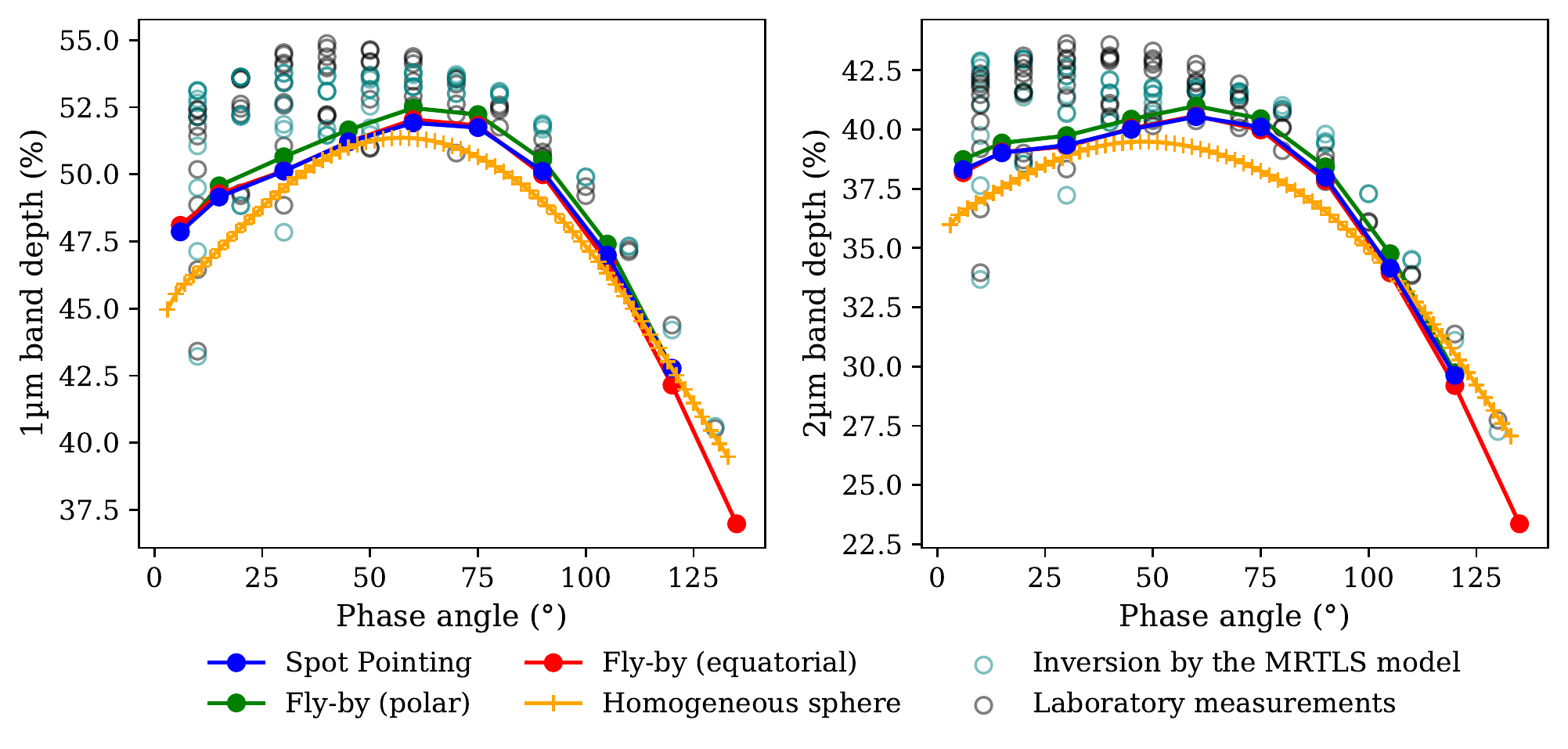}
\caption{Evolution with increasing phase angle of the amplitude of the absorption bands detected around 1µm (left) and 2µm (right) on the howardite and the simulated Vesta observations. The values are calculated using the observations from the previous simulations of the homogeneous sphere (orange line and markers), spot-pointing (blue line and markers), equatorial fly-by (red line and markers) and polar fly-by (green line and markers). The values are compared to the band depths of the reference surface calculated on the laboratory measurements (black circles) and spectra resulting from the inversion by the MRTLS model (blue circle).}
\label{PC band 1 2}
\end{center}
\end{figure}

We observe from figure \ref{PC compar SP FB Vesta} that, unlike the simulations of Ceres, the three simulated scenarii return different results. The spot pointing simulation returns the smallest photometric differences with the reference surface. It is interesting to note that the phase curves resulting from the spot pointing and the polar fly-by simulations present a similar shape, inflexing at 75°, while only the equatorial fly-by results in a phase curve drastically different from the reference surface. The spectral slope phase curve resulting from the spot pointing and equatorial fly-by simulations are almost identical, showing however a slight difference of 0.004 at phase angle 90°. The slope phase curve resulting from the equatorial fly-by differs from the other at phase angles wider than 75° until reaching 0.059 at phase angle 120°, for 0.067 resulting from the spot pointing and polar fly-by. Finally, the phase curves showing the evolution of the band depth with increasing phase angle do not show any significant dependency with the simulation scenario. The phase curves resulting from all observations of the simulated Vesta differ from the phase curves obtained with the simulated spherical body. The sphere presents a higher reflectance than all simulated values over the whole range of phase angles. The spectral slopes obtained on the sphere are lower than the values resulting from the observation scenarii until 75° of phase angle. The simulated sphere is then redder than the observation spectra until the widest phase angle investigated here. Finally, the amplitude of the 3-µm band increases until 37° of phase angle, then decreases until reaching 7.77$\%$ at the widest phase angle simulated. In the case of the observations, the inflexion presented by the phase curves is met at a phase angle of 60°. The band presented by the spherical body is also fainter than its counterpart on the observations, until a phase angle of 100° where the latest decreases until reaching 4.63$\%$ at phase angle 135° in the case of the equatorial fly-by.\\
\hspace*{0.5cm}We can suppose that this effect is due to the shadows casted on the surface with increasing phase angle. The general amount of reflectance is lowered by the increasing shadowed surface, while the spectral slope and band depth are not impacted by this effect. The photometric phase curves displayed in this work consider the reflectance in the visible range, while the spectral slope is calculated between the visible and near-infrared range, and the band depth is calculated in the near-infrared range. We can thus suppose that the effects due to the shape and topography are stronger at the shorter wavelengths. We thus analyzed the evolution of the amplitudes of the 1-µm and 2-µm bands detected on the howardite in the same simulations as presented above. The results are compared to the reference surface and shown on figure \ref{PC band 1 2}. The band depths are calculated using the same formula presented in section 3.1, considering a linear continuum, between 740nm and 1180nm centered at 940nm for the 1-µm band, and between 1440nm and 2700nm centered at 1940nm for the 2-µm band.\\

The phase curves both of the 1-µm and 2-µm band for the Vesta and the simulated spherical body experiments, for which howardite uniformly covers the surface, show similar behaviour to what has been previously observed for the 3-µm band. The phase curves of the simulated sphere present the concave shape similar to the evolution of the 3-µm band, with amplitudes fainter than what has been measured on the reference surface. For the simulations of Vesta, the evolution of the amplitude of the bands with increasing phase angle follows the reference surface, while however being shallower than most of the values measured on the howardite. We can thus conclude from figures \ref{PC compar SP FB Vesta} and \ref{PC band 1 2} that the effect of shape and topography does not significantly depends on the observed wavelength, at least considering the spectral range studied here. Textural effects, such as an increase of porosity for example, induce spectral variations depending on the studied wavelengths, as they act on a size scale similar to the wavelength \citep{Stefan_Ceres, hapke_2008}. In our case, the size scales of the topography and small body shape are significantly greater than the wavelengths, and so the spectral effects they induce do not depend on the considered wavelength.\\ 

\begin{figure}[H]
\begin{center}
\includegraphics[width = 0.5\textwidth]{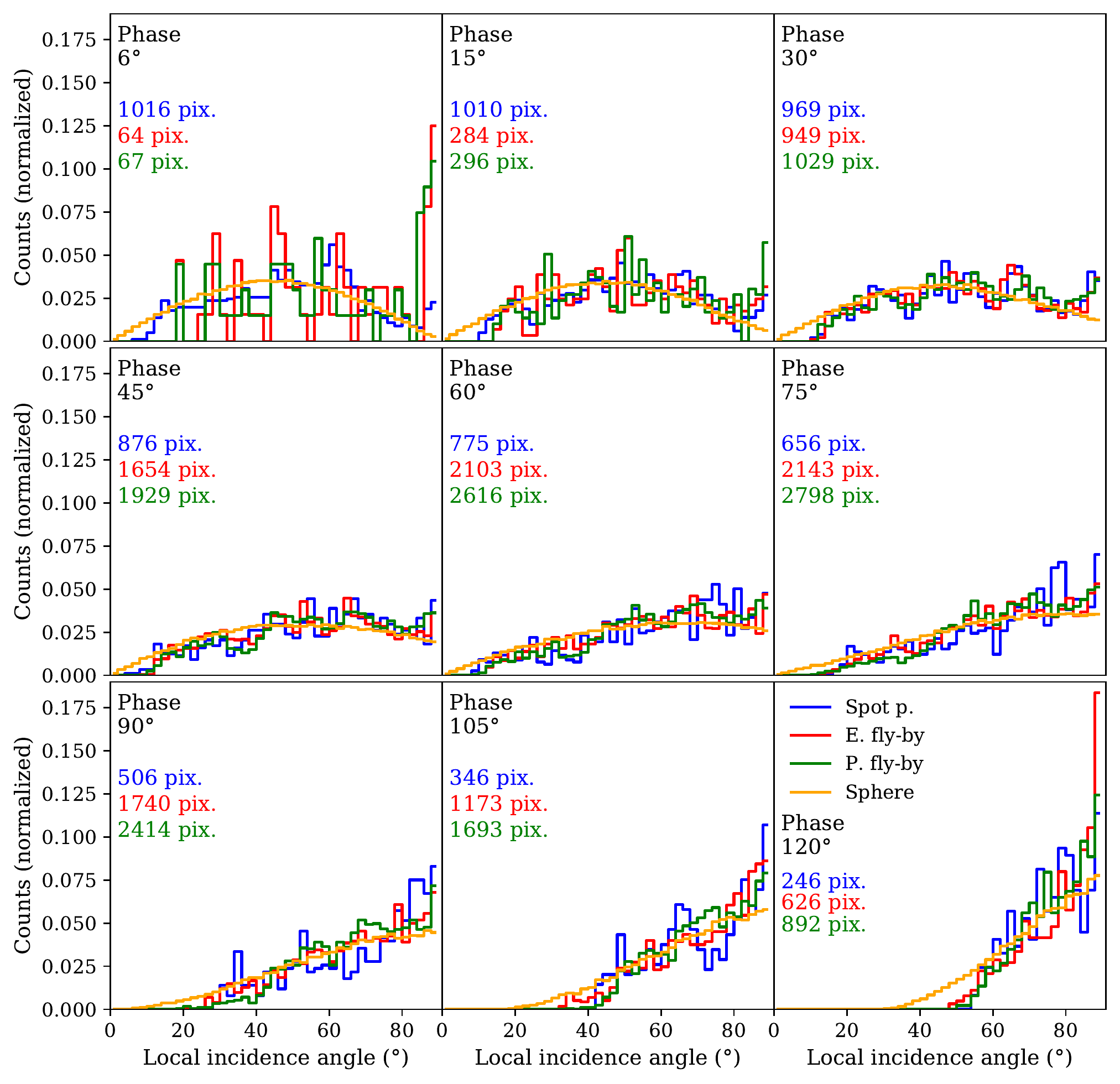}
\caption{Histograms of the local incidence angles resulting from the spot-pointing (blue), equatorial fly-by (red) and polar fly-by (green) around Vesta, and from the simulation of a spherical body (orange). Each panel presents a phase angle. The counts have been normalized to the area under the histogram, so the total number of pixels illuminated and observed on the images. This number of pixels is indicated on each panel for the spot-pointing (blue), equatorial fly-by (red) and polar fly-by (green).}
\label{histo incidence Vesta}
\end{center}
\end{figure}

\begin{figure}[H]
\begin{center}
\includegraphics[width = 0.5\textwidth]{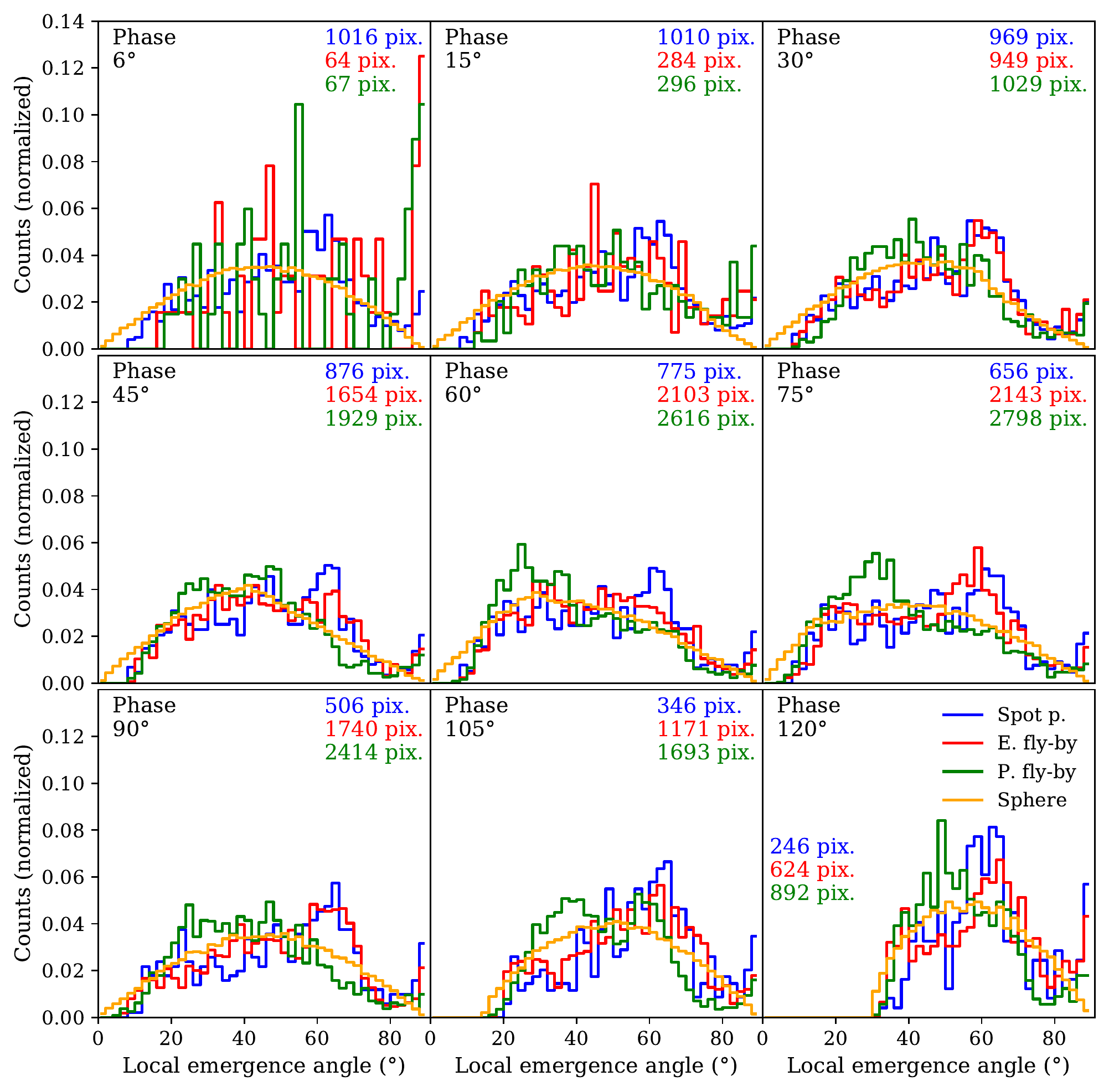}
\caption{Histograms of the local emergence angles resulting from the spot-pointing (blue), equatorial fly-by (red) and polar fly-by (green) around Vesta, and from the simulation of a spherical body (orange). Each panel presents a phase angle. The counts have been normalized to the area under the histogram, so the total number of pixels illuminated and observed on the images. This number of pixels is indicated on each panel for the spot-pointing (blue), equatorial fly-by (red) and polar fly-by (green). }
\label{histo emergence Vesta}
\end{center}
\end{figure}

Figures \ref{histo incidence Vesta} and \ref{histo emergence Vesta} present the distributions of the local incidence and emergence angles on the surface of Vesta observed during the spot-pointing and both fly-by experiments. The correspondant histograms for the sphere experiments are also added. \\

We observe from figure \ref{histo incidence Vesta} that the distributions of the local incidence angles are similar on the all scenarii. Moreover, the evolution of the distributions with increasing phase angles is similar to what was observed in the case of Ceres (figure \ref{histogram compar inci FB SP Ceres}). At low phase angle, the majority of the facets shows a low incidence angle. The distribution then tilts until the majority of the facets shows a near grazing local incidence angle, at the widest phase angle simulated here. Stronger differences are met when comparing the distributions of the local emergence angles between all simulations. As the spot pointing and the two fly-bys do not observe the same portion of the surface, the distributions of the local emergence, each departing significantly from the sphere, trace the various topographies observed, and the changing projected shape of the small body. These variations are seen when comparing the distributions of the local incidence angles, but the differences are fainter than with the local emergence angles. We can note on figure \ref{histo emergence Vesta} that the distribution of local emergence angles resulting from the spot pointing simulation show a peak of population between 25 and 29°, on the complete angular range studied here. This is also shown by the equatorial fly-by for phase angles wider than 90°. \\

In the case of Vesta, the non-spherical shape and noticeable topography of the surface induce non-negligible variations in the distributions of the local incidence and emergence angles, depending on the observation scenario. This thus implies the strong differences observed between the phase curves resulting from each simulated observation, and between the small body and its reference surface.

\subsection{The biases in laboratory measurements}
\hspace*{0.5cm}We showed in this study that the phase curves of simulated bodies could differ from the measurements on the reference surfaces acquired under several geometrical configurations. However, bidirectional reflectance spectroscopy can be limited or impossible in the laboratory. The reflectance spectrum is thus generally measured under a single geometry. Table \ref{compar phase 30} compares the spectral parameters of the spectra obtained for each simulation at phase angle 30° with the spectroscopy of the reference surface acquired in the laboratory with a nadir illumination and an emergence angle of 30°.\\

\begin{table}[H]
\begin{center}
\begin{tabular}{lll}
\hline
 & Analogue & Howardite \\
\hline
\multicolumn{3}{c}{Laboratory measurements}\\
\hline
Reflectance & 0.1138 & 0.3851 \\
Slope ($\mu m^{-1}$) & -0.0213 & 0.0445 \\
BD ($\%$) & 45.28 & 15.64 \\
\hline
\multicolumn{3}{c}{Inversion by MRTLS model}\\
\hline
Reflectance & 0.1163 & 0.3866 \\
Slope ($\mu m^{-1}$) & -0.0204 & 0.0476 \\
BD ($\%$) & 45.95 & 15.47 \\
\hline
\multicolumn{3}{c}{Homogeneous sphere}\\
\hline
Reflectance & 0.1726 & 0.4687 \\
Slope ($\mu m^{-1}$)& -0.0288 & 0.0329 \\
BD ($\%$)& 45.39 & 13.41 \\
\hline
\hline
 & (1)Ceres & (4)Vesta \\
\hline
\multicolumn{3}{c}{Spot Pointing}\\
\hline
Reflectance & 0.1433 & 0.4199 \\
Slope ($\mu m^{-1}$)& -0.0251 & 0.0369 \\
BD ($\%$)& 46.27 & 14.13 \\
\hline
\multicolumn{3}{c}{Equatorial fly-by}\\
\hline
Reflectance & 0.1461 & 0.4194 \\
Slope ($\mu m^{-1}$)& -0.0252 & 0.0375 \\
BD ($\%$)& 46.06 & 14.15 \\
\hline
\multicolumn{3}{c}{Polar fly-by}\\
\hline
Reflectance &  & 0.4073 \\
Slope ($\mu m^{-1}$)&  & 0.0377 \\
BD ($\%$)&  & 14.33 \\
\hline
\end{tabular}
\caption{Comparison between the spectral parameters derived from the simulated observations and reference surfaces at phase angle 30°. The laboratory measurements and inverted spectra are here presented in the nominal laboratory geometry, with a nadir illumination and emergence angle of 30°. We considered the simulations using the surface resulting from the inversion by the MRTLS model.}
\label{compar phase 30}
\end{center}
\end{table}

The inversion by the MRTLS model induces negligable deviations of the calculated spectral parameters from the reference surface. For the Ceres analogue, the inversion added 2.20$\%$ relative to the reflectance value, increases the spectral slope of 4.22$\%$ relative and increases the band depth by 1.48$\%$ relative. For the howardite, the inversion added 0.39$\%$ relative to the reflectance value at 740nm, increases the spectral slope of 6.97$\%$ relative and decreases the band depth by 1.09$\%$ relative. Our results presented in figure \ref{spectra_spheres} and reminded in table \ref{compar phase 30} show that the application of the reference surface on a spherical shape induces noticable variations of the spectral parameters. For the Ceres analogue and howardite respectively, the unresolved observations of the simulated body present higher reflectance value of 48.41$\%$ and 21.24$\%$ relative, bluer spectral slope of 41.18$\%$ and 30.88$\%$ relative and absorption bands fainter by 1.22$\%$ and 13.31$\%$ relative.\\

We now compare the spectral parameters obtained on the simulated observations. In the case of Ceres homogeneously covered with the analogue, the unresolved spot-pointing observation resulted in an increase of reflectance of 23.22$\%$ relative, and decrease of the spectral slope of 23.04$\%$ relative, while the absorption band depth is not impacted, with an increase of 0.70$\%$ relative. The fly-by simulation returns similar deviations of 25.62$\%$, 23.53$\%$ and 0.24$\%$ respectively for the reflectance, spectral slope and band depth. The phase 30° was used as a reference point to which the simulated object was observe in identical illumination and observation conditions between the various experiments. We noted previously that the phase curves of the simulated Ceres were similar between the spot pointing and fly-by simulation and concluded on the low effect on the reflectance value of the surface topography in the case of Ceres. However, we can note from table \ref{compar phase 30} that the spectral parameters derived from the simulated Ceres and homogeneous spherical body are different. Considering the homogeneous sphere as reference, the observations of Ceres differ from 16.97 and 15.35$\%$ relative for the reflectance value, 12.83 and 12.50$\%$ relative for the spectral slope and 1.94 and 1.48$\%$ relative for the band depth, for the spot pointing and fly-by respectively. The results suggest that the surface topography of Ceres induces changes in the local incidence and emergence angles acting on the spectral parameters when considered under a single geometrical configuration, but are too shallow compared to the size to induce casted shadows at high phase angle.\\
\hspace*{0.5cm}The simulated spot pointing and equatorial fly-by of Vesta resulted in a difference with the reference surface of 8.61$\%$ and 8.48$\%$ relative for the reflectance value, 22.48$\%$ and 21.22$\%$ relative for the spectral slope and 8.66$\%$ and 8.35$\%$ relative for the 3-µm band depth. The polar fly-by returned deviations from the reference surface of 5.35$\%$, 20.80$\%$ and 7.37$\%$ relative for the reflectance value, spectral slope and 3-µm band depth respectively. In the case of Vesta, the differences of spectral parameters with the spherical body homogeneously covered with the howardite are due to the local angles induced by the non-spherical shape and non-negligible surface topography of Vesta.\\

Our results show that when comparing unresolved observations of planetary body and laboratory spectroscopy acquired under a single geometry, the value of reflectance and the spectral slope are the most impacted parameters. The unresolved small body returns a higher reflectance value and bluer slope than its surface alone. These differences are increased if the unresolved body is not observed under the same phase angle as the laboratory measurement, as shown by all phase curves simulated in this study. In case of Earth-based observations, asteroids from the Main Belt cannot be observed under phase angles wider than 30° \citep{cellino_2014, rivkin_2019, usui_2019}, comets and Near-Earth asteroids are observed at phase angles up to 90° \citep{levasseur_1996, sanchez_2012}, but the Trans-Neptunian Objects can only be observed near opposition as they are the most distant objects in the Solar System \citep{rousselot_2005}. According to the phase curves resulting from our simulations, if the target is observed under a phase angle lower than the laboratory measurement, it will present a greater reflectance value, lower spectral slope and shallower absorption band. To better compare the unresolved reflectance spectrum of a planetary body with laboratory measurements, it is important to recreate the geometrical configuration of the observation in the laboratory. This would suppress the effects on the spectral parameters due to the phase curve of the target, leaving only the effect of shape and surface topography. If the adjustment of the geometrical configuration is not possible for instrumental reasons or limitations by the sample, one would have to keep in mind the spectral effects due to the different geometries.\\
\hspace*{0.5cm}We used in this study available reference surfaces consistent with the considered shape models. One can expect greater differences between the unresolved observations of the simulated body and its surface itself if the surface presents a stronger spectral dependency with the geometrical configuration. Moreover, our simulations presented here only consider a simulated body homogeneously covered with the same surface. Patches of various compositions and textures applied on the simulated body will increase the difference between the unresolved observation and laboratory measurement. 

\section*{Conclusion}
\hspace*{0.5cm}We measured in the laboratory the bidirectional reflectance spectroscopy of a powder of howardite and a sublimation residue of a Ceres analogue. The obtained spectra were inverted using the MRTLS and Hapke models to calculate the reflectance of the surfaces under any geometrical configuration. Comparison between the models show that both accurately result in similar bidirectional behaviour of the surface, though spectra resulting from the inversion by the Hapke model are noisier than the spectra resulting from the inversion by the MRTLS model. We then applied the bidirectional spectroscopic behaviour of the reference surfaces on shape models of spherical bodies and of the small bodies (1)Ceres and (4)Vesta, simulated several unresolved observation scenarii, and compared the obtained phase curves to the laboratory measurements on the reference surfaces.\\
\hspace*{0.5cm}The reflectance spectra of a spherical body tends to be brighter, bluer and showing fainter absorption features compared to the reference surfaces. These differences from the reference surfaces are due the high area on the spherical surface presenting near grazing local incidence and emergence angles and the strong backscattering behaviour of the reference surfaces used here.\\
\hspace*{0.5cm}The simulated spot pointing and fly-by of Ceres resulted in phase curves similar to one another, but slightly different from the spherical body covered with the same surface. We concluded that the spherical shape of Ceres coupled with its large size soften the effect of the surface topography, creating however variations of local incidence and emergence angles strong enough to generate a different phase curve from a perfectly spherical body, but not deep enough to create shadows with varying geometrical configuration.\\

We simulated the spot pointing, equatorial fly-by and polar fly-by observations of Vesta. The unresolved reflectance spectra resulted in phase curves strongly different from one another. Moreover, the spectral parameters derived from the observations differ from the laboratory measurements for phase angle wider than 60 or 75°. We showed that the reflectance value is the most impacted parameter, while the phase curves of spectral slope and amplitude of the absorption bands approach and sometimes match the laboratory measurements. These variations are due to these of the local incidence and emergence angles induced by the shape of the simulated Vesta coupled with its varying surface topography.\\
\hspace*{0.5cm}Finally, we presented the differences in spectral parameters between unresolved spectroscopic observations of a planetary body with laboratory reflectance spectroscopy of the same surface acquired under a single standard geometry (nadir illumination and emergence angle of 30°). We showed that unregarding the composition, surface texture and geometry of observation, the shape and topography of the planetary body will impact its unresolved reflectance spectroscopy and generate noticeable differences with the reflectance spectroscopy of its surface alone.

\section*{Acknowledgments}
The authors would like to thank Dr. Stefan Schröder for his authorization to use the BRDF of the sublimation residue of the Ceres analogue. The instrument SHADOWS was founded by the OSUG@2020 Labex (Grant ANR10 LABX56), by ‘Europlanet 2020 RI’ within the European Union’s Horizon2020 research and innovation program (Grant N° 654208) and by the Centre National d’Etudes Spatiales (CNES). 

\section*{Appendix A: Bidirectional reflectance spectroscopy of the reference surfaces}
\hspace*{0.5cm}Two surfaces are used in reference in this study, and were analyzed with bidirectional reflectance spectroscopy. We present in this appendix the dependencies of the reflectance, spectral slope and amplitude of the absorption bands of the reference surfaces with the geometrical configuration. 

\subsection*{A.1. Howardite}
\begin{figure}[H]
\begin{center}
\includegraphics[width = 0.6\textwidth]{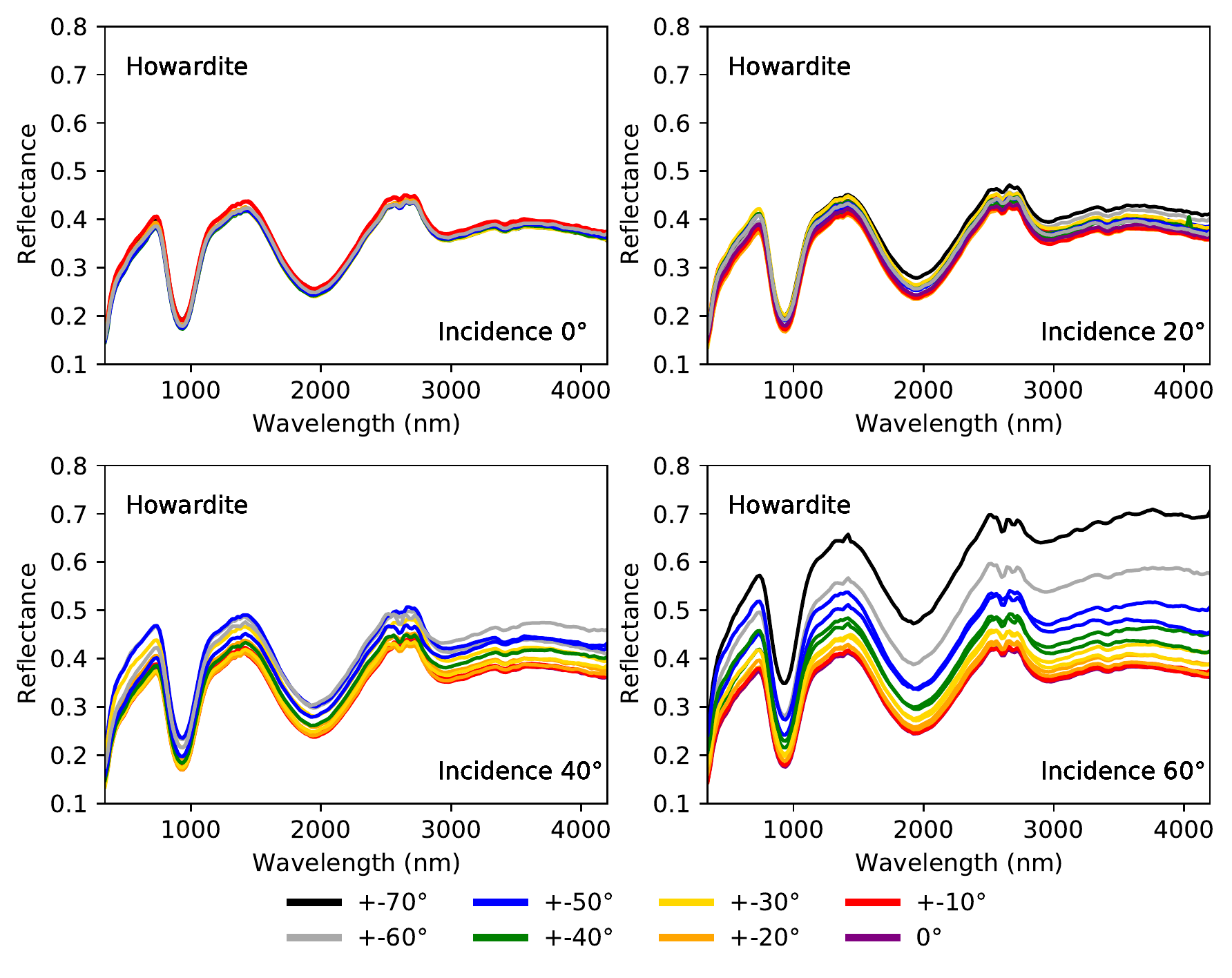}
\caption{Bidirectional reflectance spectra of the powder of howardite, acquired with an incidence angle of 0° (top left panel), 20° (top right panel), 40° (bottom left panel) and 60° (bottom right panel). The data are available in the GhoSST @ SSHADE database \citep{Expe_SSHADE}.}
\label{all spectra howardite}
\end{center}
\end{figure}

\hspace*{0.5cm}We first present the complete set of spectra acquired on the howardite. The spectra are shown in Figure \ref{all spectra howardite}.\\

\begin{figure}[H]
\begin{center}
\includegraphics[width = 0.6\textwidth]{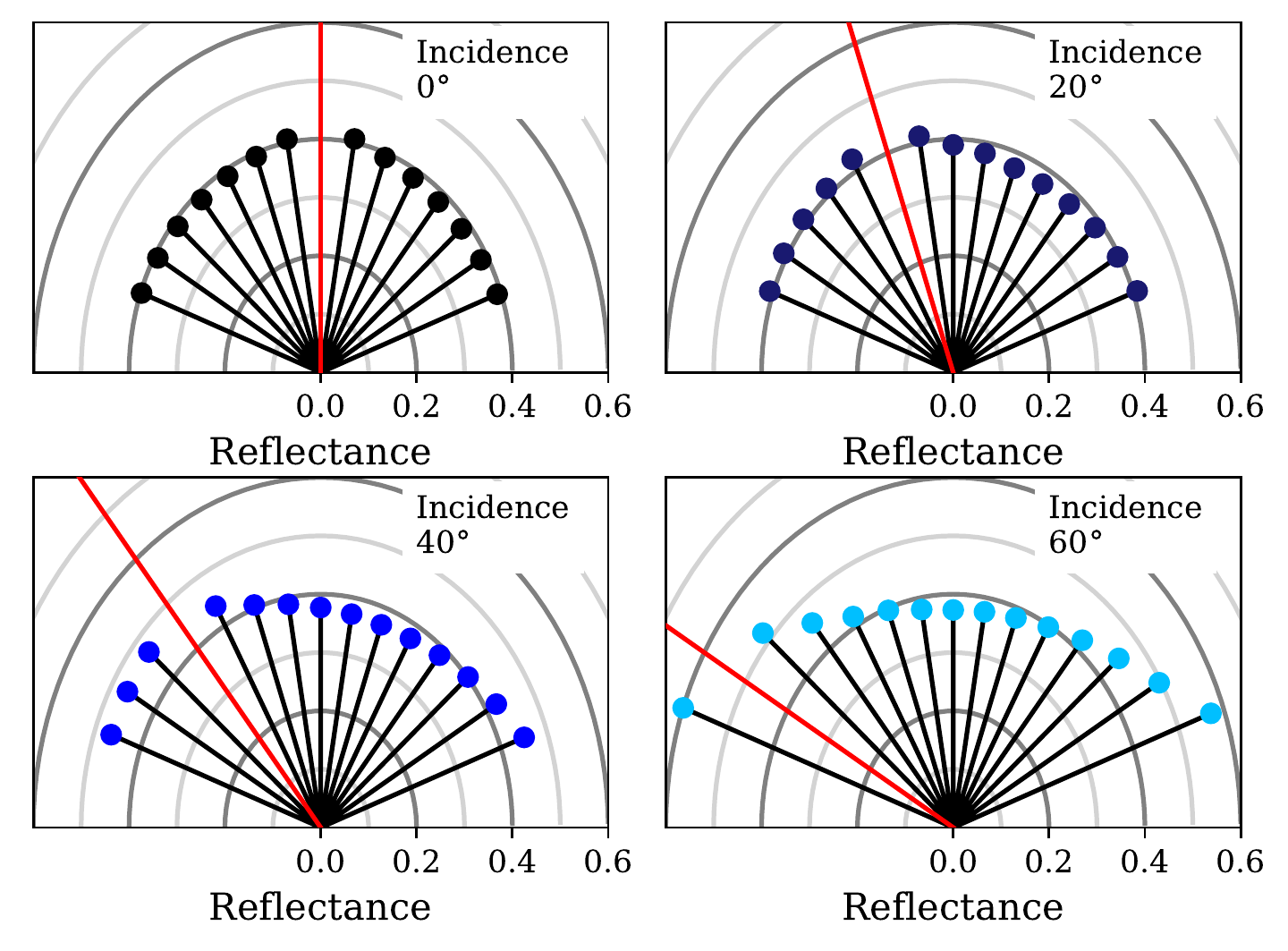}
\caption{Polar plot of the reflectance measured at 740 nm on the powder of howardite at incidence 0° (top left panel), 20° (top right panel), 40° (bottom left panel) and 60° (bottom right panel). The dots mark the reflectance measurements, and the red line represents the direction of illumination.}
\label{ref BRDF howardite}
\end{center}
\end{figure}

Figure \ref{all spectra howardite} shows that the bidirectional reflectance spectroscopy of the howardite is increasigly dependent more on the geometrical configuration with increasing incidence angle, i.e. towards grazing incidence. All spectral parameters, reflectance, spectral slope and absorption bands are now analyzed independently to highlight all effects of the geometrical configurations. Figure \ref{ref BRDF howardite} presents the variations of the reflectance value at a given wavelength with the geometry.\\

The reflectance of the howardite does not show significant variations with the geometrical configuration under nadir illumination (0°). Under grazing incidence (60°), the measured reflectance shows signs of backscattering and forward scattering, inducing an increase of reflectance at low phase angle and around the specular direction respectively. At incidence 60°, the reflectance has been measured at 0.373 at emergence 0°, when 0.599 is measured around the opposition at a phase angle of 10° and 0.496 in the specular direction. The bidirectional reflectance function strongly depends on the surface texture \citep{ pommerol_2008a, Potin_mukundpura, deangelis_2019}, and in this case, the sample is a fine powder with a narrow grain size distribution. The shadowing of one particle over another is reduced, which implies a relatively low shadow hiding opposition effect (SHOE, \cite{hapke_1986}) responsible for the increase of reflectance at low phase angle.\\

\begin{figure}[H]
\begin{center}
\includegraphics[width = 0.45\textwidth]{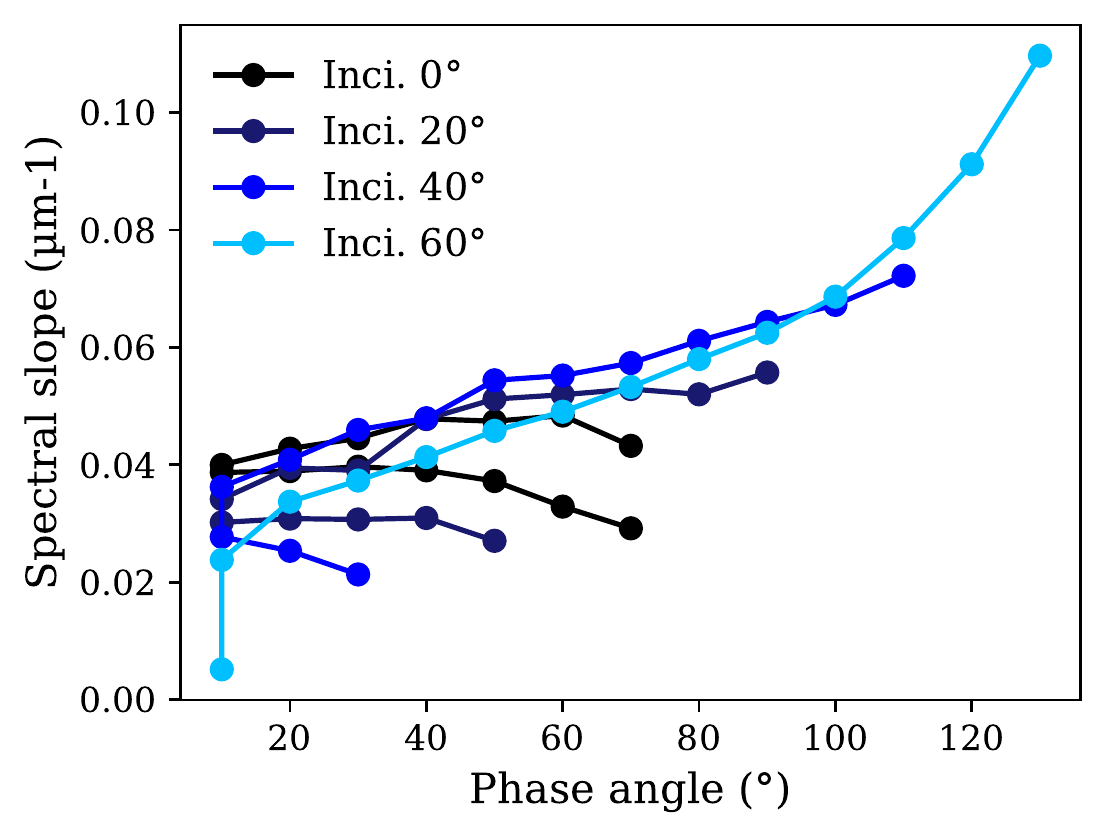}
\caption{Evolution of the spectral slope with increasing phase angle for the various incidence angles considered here.}
\label{ref SLOPE howardite}
\end{center}
\end{figure}

Figure \ref{ref SLOPE howardite} presents the variations of the calculated spectral slope with the geometrical configuration, resulting from the reflectance measurement on the howardite.\\

Interestingly, the spectral slope derived from the measured spectra does not depend only on the phase angle, but also on the illumination angle. Figure \ref{ref SLOPE howardite} shows that the slope of the spectra acquired with an incidence angle of 0° and 20° generally decreases with increasing phase angle. However, the spectral slope of the spectra acquired at 40° and 60° present the effect of phase reddening, so the increase of spectral slope with increasing phase angle. This effect was already detected in previous studies \citep{Potin_mukundpura}.\\

\begin{figure}[H]
\begin{center}
\includegraphics[width = 0.6\textwidth]{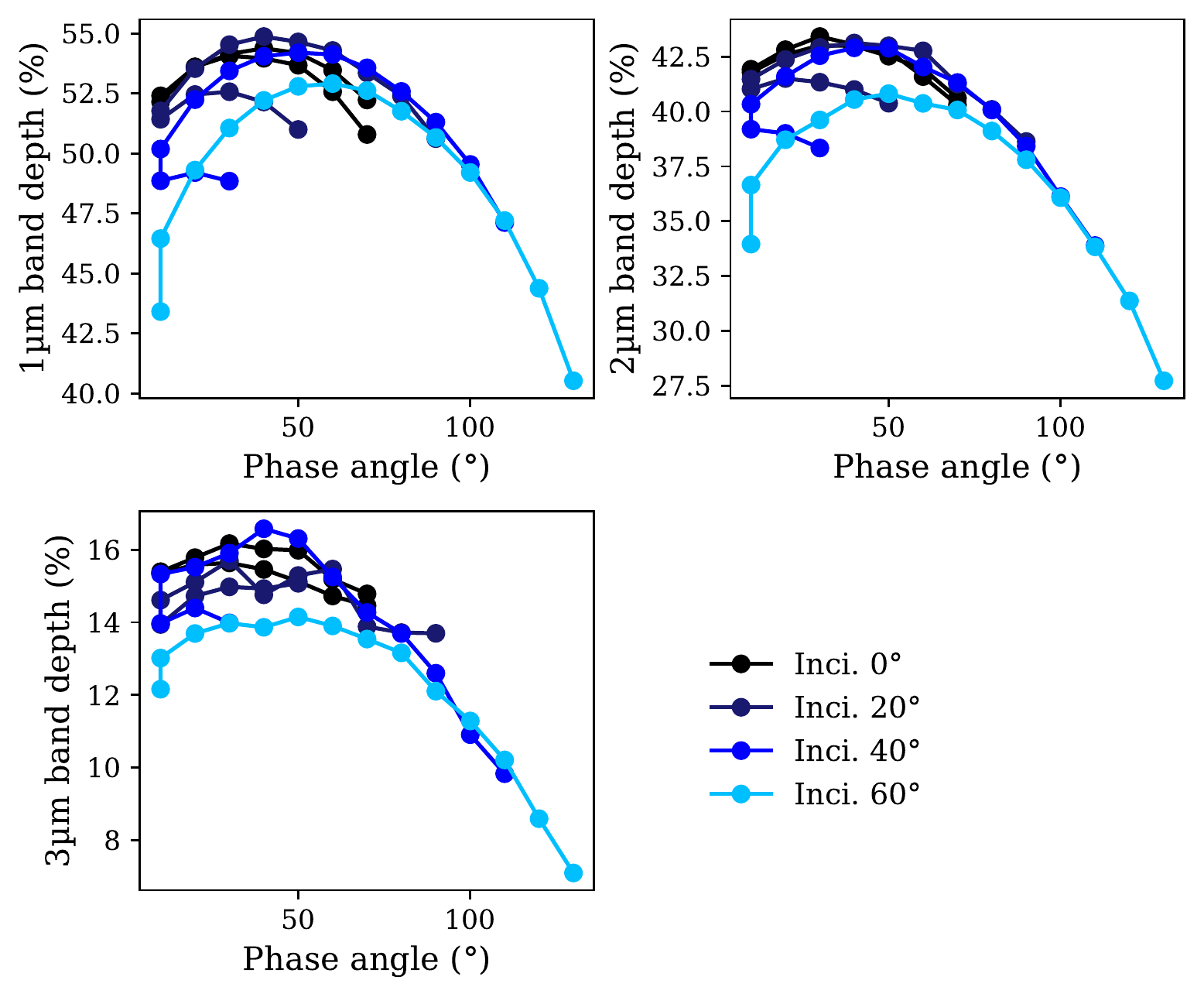}
\caption{Evolution of the amplitude of the 1-µm (top left panel), 2-µm (top right panel), and 3-µm band (bottom left panel) with increasing phase angle for the various incidence angles considered here.}
\label{ref BD howardite}
\end{center}
\end{figure}

Figure \ref{ref BD howardite} presents the variations of the amplitudes of the absorption bands with the geometrical configuration.\\

The three absorption bands detected on the powder of howardite show the same evolution with increasing phase angle, under all considered incidence angles. All curves show a deepening of the absorption bands until the phase angle reaches around 40°, then a decrease of amplitude. This effect can be explained by an increase of the relative number of photons scattered out of the powder before penetrating the sample and further interacting with the grains \citep{pommerol_2008b}. Figure \ref{ref BD howardite} also shows that the amplitude of the absorption band depends on the illumination angle. For the three bands considered here, the spectra acquired at incidence 60° presents the shallowest bands, with 46.45, 36.67 and 13.02$\%$ calculated for the 1-µm, 2-µm and 3-µm band respectively at low phase angle. In the same configuration, the spectra acquired under a nadir illumination returned amplitudes of 52.14, 41.78 and 15.34$\%$.\\

The reflectance spectroscopy of the powder of howardite shows a clear dependency of the reflectance, spectral slope and band depth with both the directions of illumination and observation.

\subsection*{A.2. Ceres analogue}
\begin{figure}[H]
\begin{center}
\includegraphics[width = 0.6\textwidth]{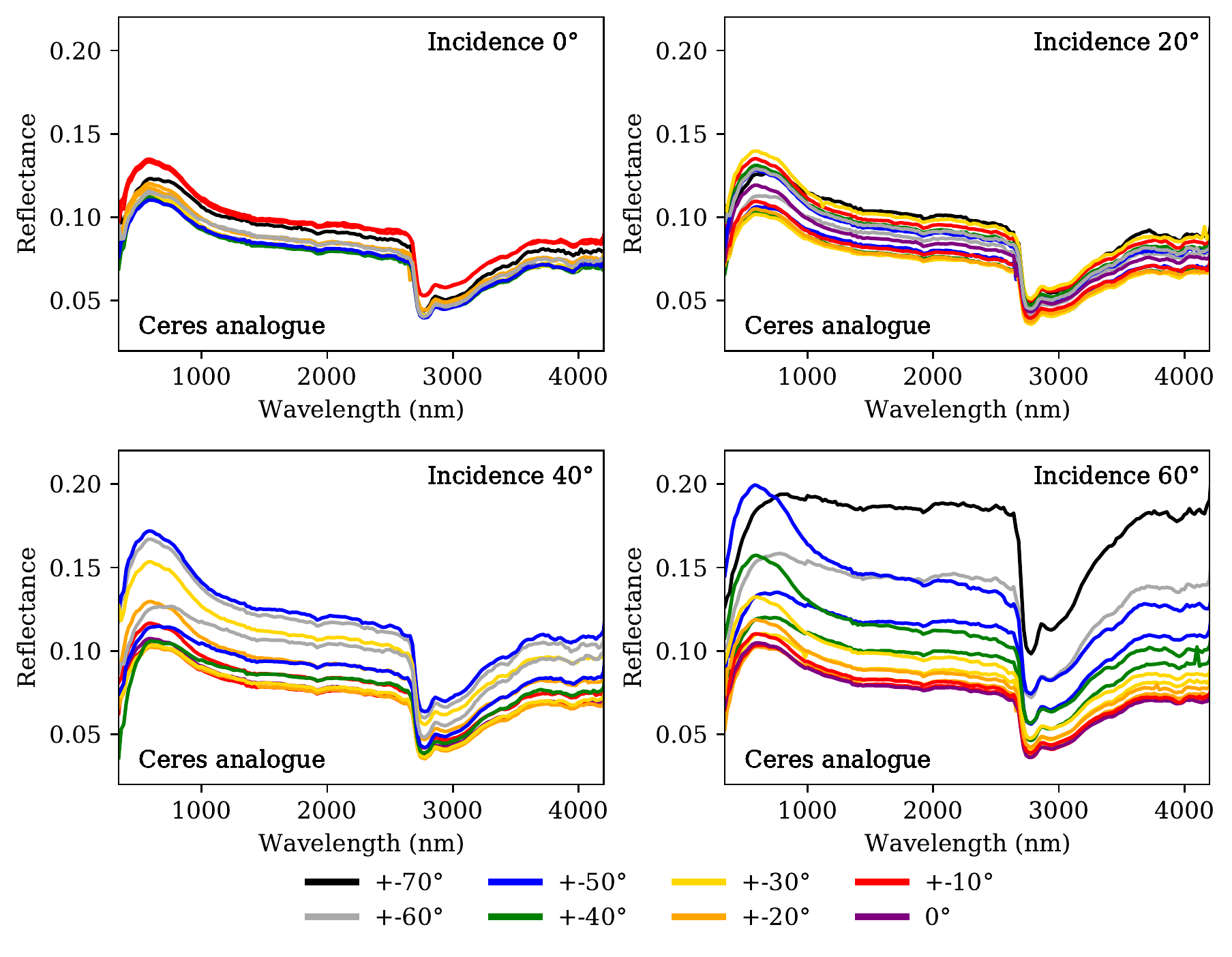}
\caption{Bidirectional reflectance spectra of the Ceres analogue, acquired with an incidence angle of 0° (top left panel), 20° (top right panel), 40° (bottom left panel) and 60° (bottom right panel). The data are available in the GhoSST @ SSHADE database \citep{Expe_SSHADE}.}
\label{ref spectres Ceres}
\end{center}
\end{figure}

\hspace*{0.5cm}We now analyze the bidirectional behaviour of the reflectance acquired in the laboratory on the sublimation residue of the Ceres analogue. The spectra are shown in figure \ref{ref spectres Ceres}.\\

\begin{figure}[H]
\begin{center}
\includegraphics[width = 0.6\textwidth]{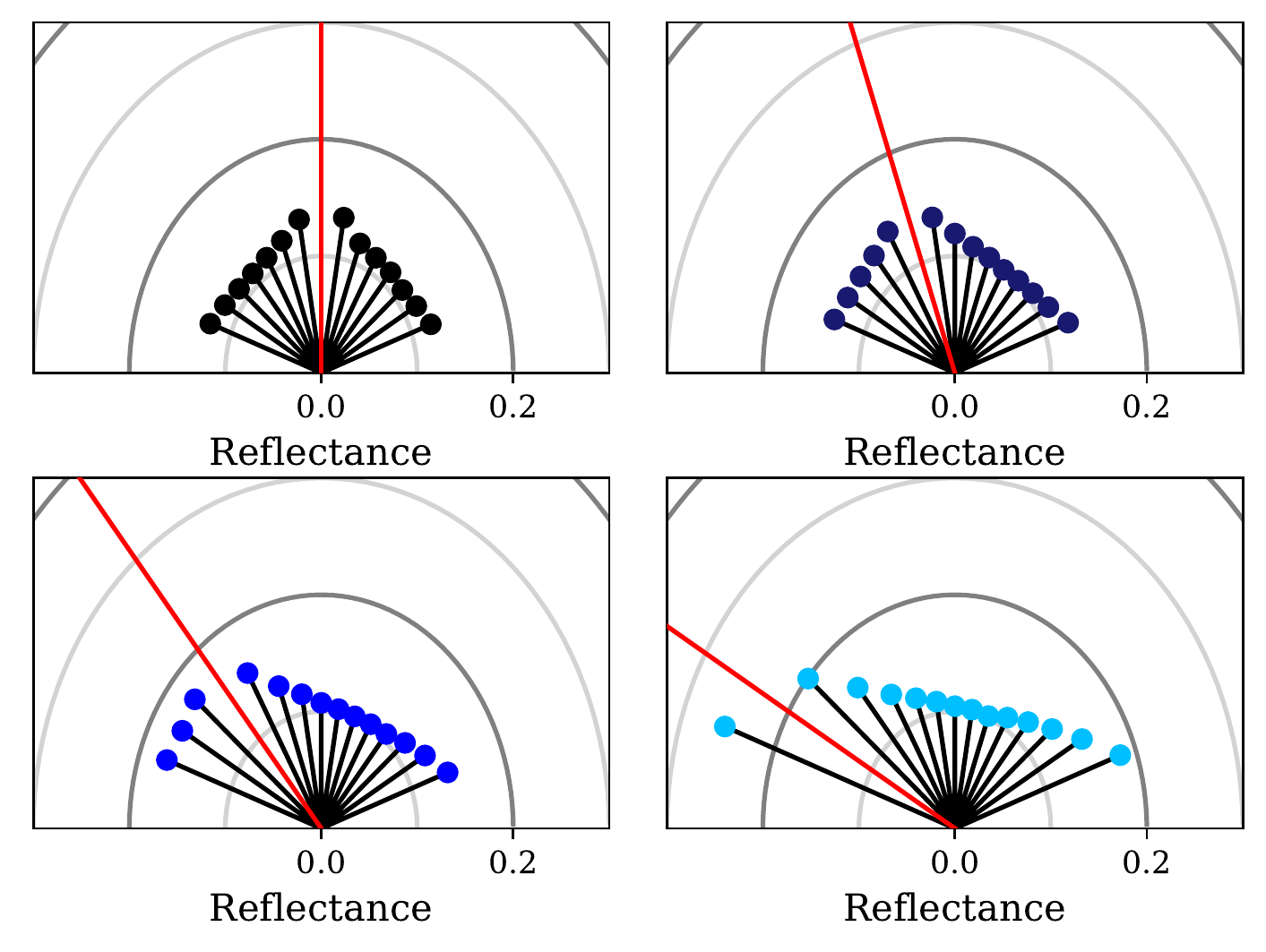}
\caption{Polar plot of the reflectance measured at 580 nm on the sublimation residue of the Ceres analogue at incidence 0° (top left panel), 20° (top right panel), 40° (bottom left panel) and 60° (bottom right panel). The dots mark the reflectance measurements, and the red line represents the direction of illumination.}
\label{ref BRDF Ceres}
\end{center}
\end{figure}

This sample shows similar dependency with the geometrical configuration to the sample of howardite. The greatest variations between spectra are observed under grazing illumination. The reflectance, spectral slope and absorption bands are now analyzed independently. Figure \ref{ref BRDF Ceres} presents the variations of the reflectance value at a given wavelength with the geometry.\\

The sample of Ceres analogue presents a strong backscattering, i.e. increase of reflectance at low phase angle and for all incidence angles considered here. At a phase angle of 10°, reflectance values of 0.135, 0.140, 0.172 and 0.255 have been measured at incidence 0°, 20°, 40° and 60° respectively, while 0.121, 0.125, 0.140 and 0.183 have been measured at the widest phase angle available for each illumination direction. The high porosity and foam-like texture of the sample increases the amount of shadowing of the grains on their surrounding, and so resulting in an important SHOE effect.\\

\begin{figure}[H]
\begin{center}
\includegraphics[width = 0.45\textwidth]{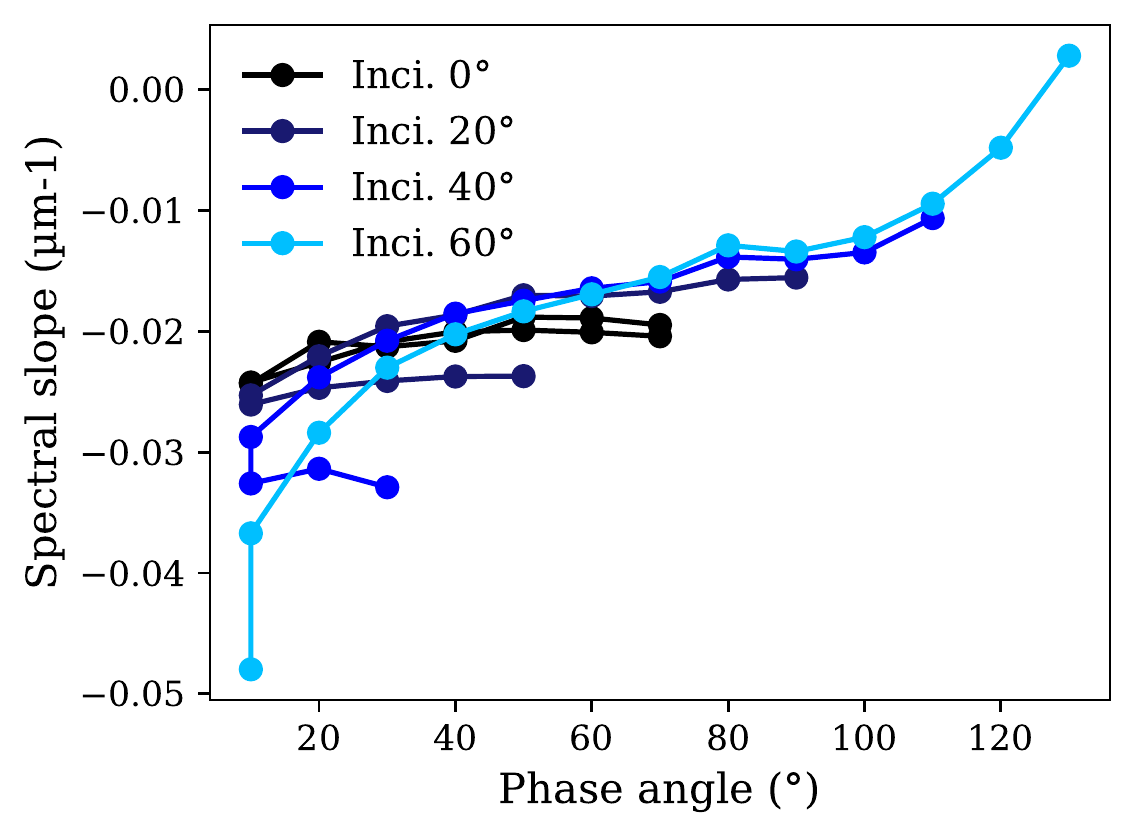}
\caption{Evolution of the spectral slope with increasing phase angle for the various incidence angles considered here.}
\label{ref slope Ceres}
\end{center}
\end{figure}

Figure \ref{ref slope Ceres} presents the variations of the calculated spectral slope with the geometrical configuration.\\

The spectral slope derived from the spectra of the Ceres analogue shows the dependency highlighted previously, so on both the directions of illumination and observation. The phase reddening is appearant for spectra acquired at illumination 20°, 40° and 60°. \\

\begin{figure}[H]
\begin{center}
\includegraphics[width = 0.45\textwidth]{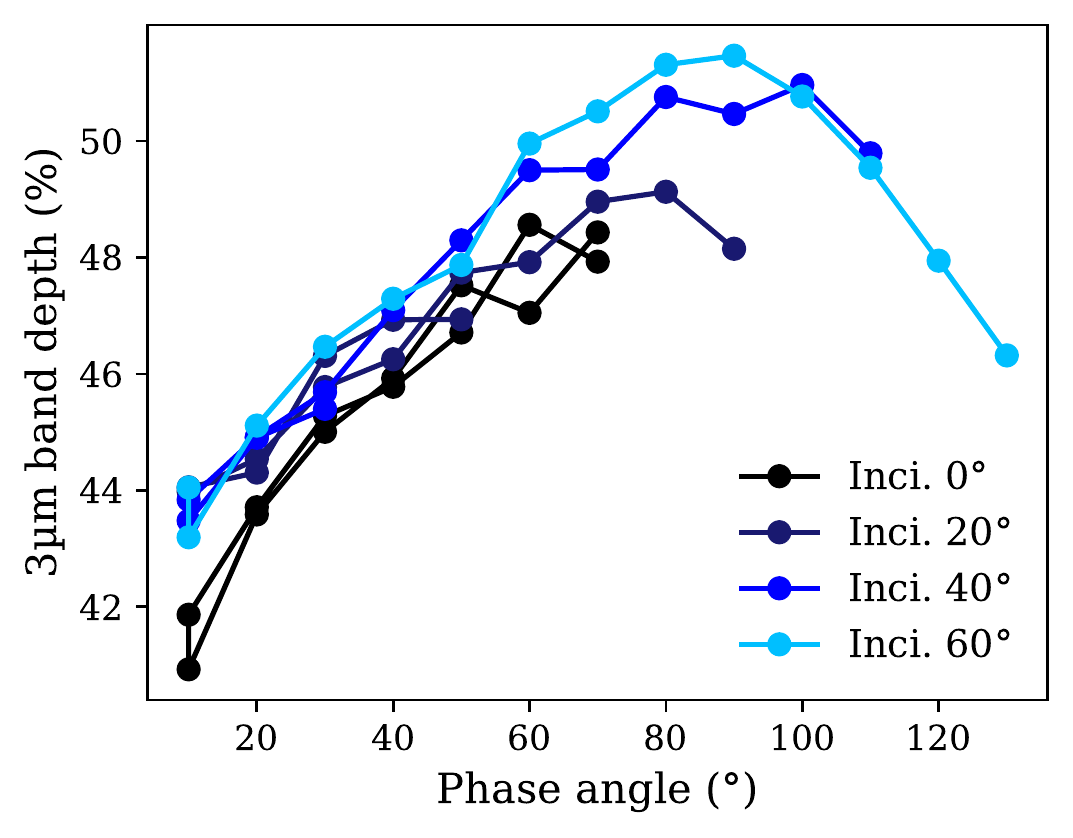}
\caption{Evolution of the amplitude of the 3-µm band with increasing phase angle for the various incidence angles considered here.}
\label{ref BD Ceres}
\end{center}
\end{figure}

Figure \ref{ref BD Ceres} shows the variation of the amplitude of the absorption band around 3µm with the geometrical configuration. \\

The 3-µm band detected on this sample shows an deepening for phase angles up to 90°, then a decrease of amplitude. Unlike the results from the howardite, the spectra acquired on the Ceres analogue show deeper absorption bands under grazing illumination than under nadir. At phase angle 10°, 41.8$\%$ has been measured under nadir illumination, for 44.0$\%$ under illumination 60°. One can suppose that this effect is due to the foam-like texture of the surface, allowing the photons to penetrate deeper in the sample and interacting with the grains under grazing geometries.

\section*{Appendix B: Images of the simulated observations}
\hspace*{0.5cm}We present here the various images resulting from the spot-pointing and fly-bys of the simulated Ceres and Vesta, showing the radiance of the bodies at 400nm.

\begin{figure}[H]
\begin{center}
\includegraphics[width = 0.45\textwidth]{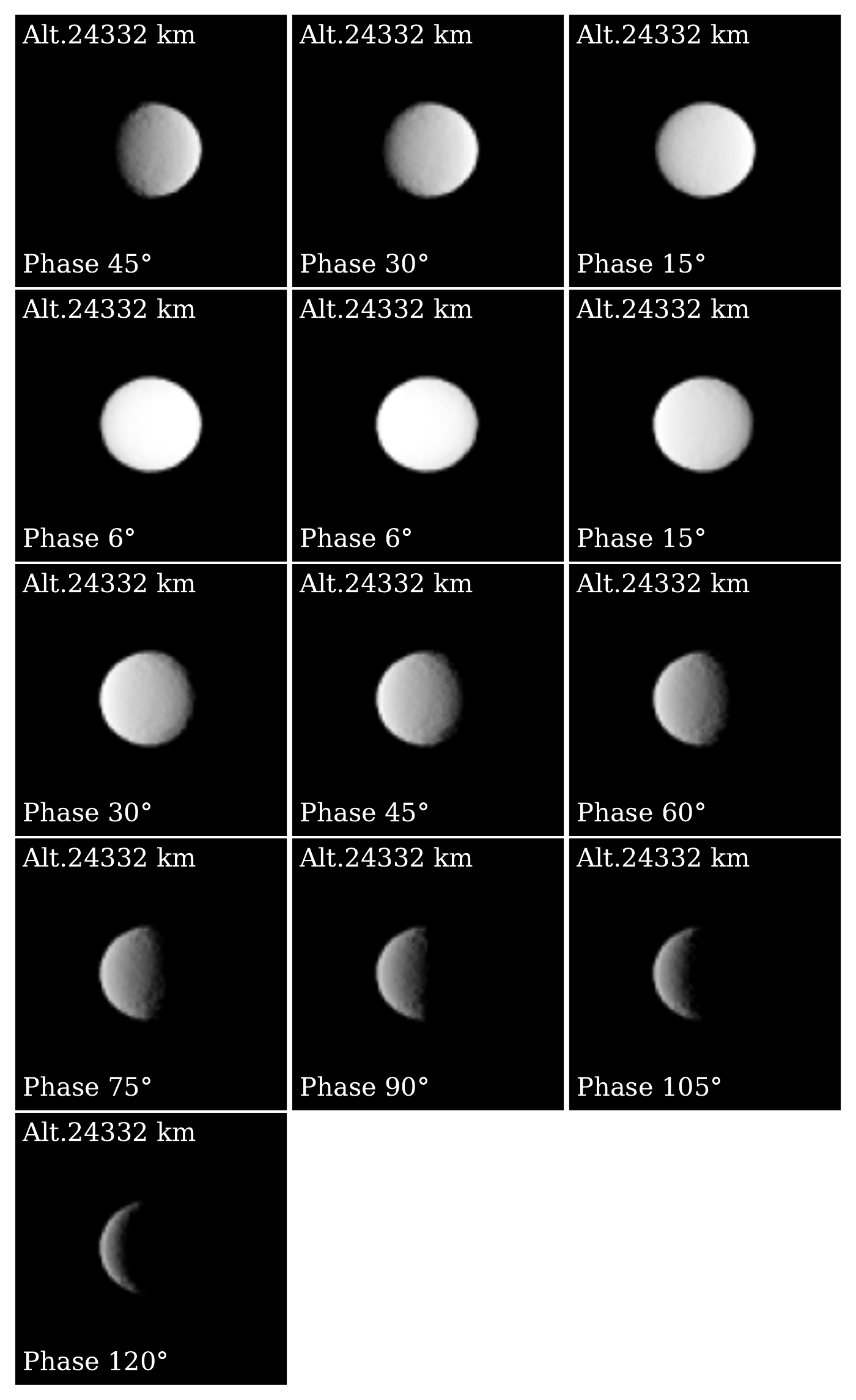}
\caption{Simulated observations of Ceres resulting from the spot-pointing.}
\label{obs sp_Ceres}
\end{center}
\end{figure}

\begin{figure}[H]
\begin{center}
\includegraphics[width = 0.45\textwidth]{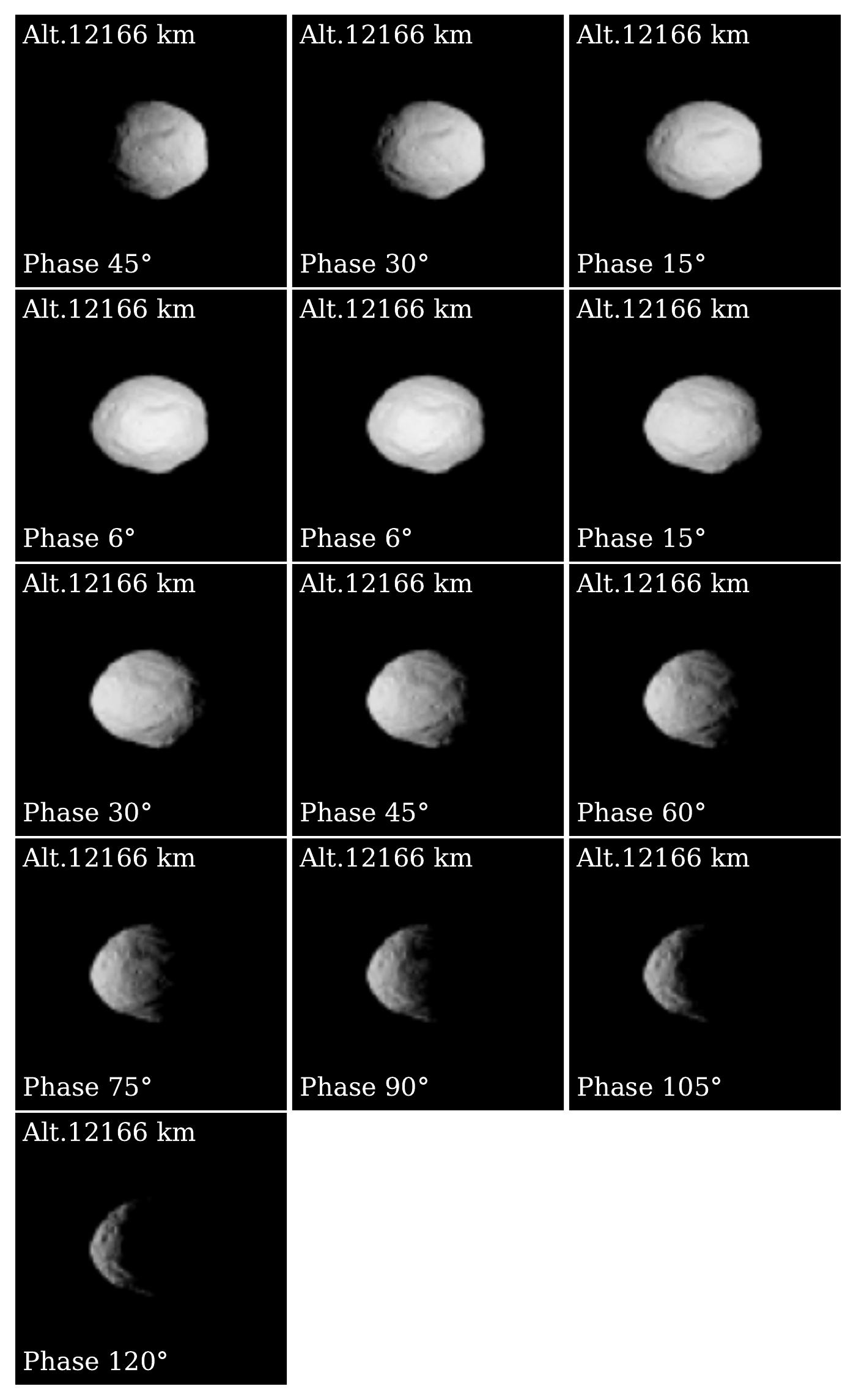}
\caption{Simulated observations of Vesta resulting from the spot-pointing.}
\label{obs sp_Vesta}
\end{center}
\end{figure}

\begin{figure}[H]
\begin{center}
\includegraphics[width = 0.45\textwidth]{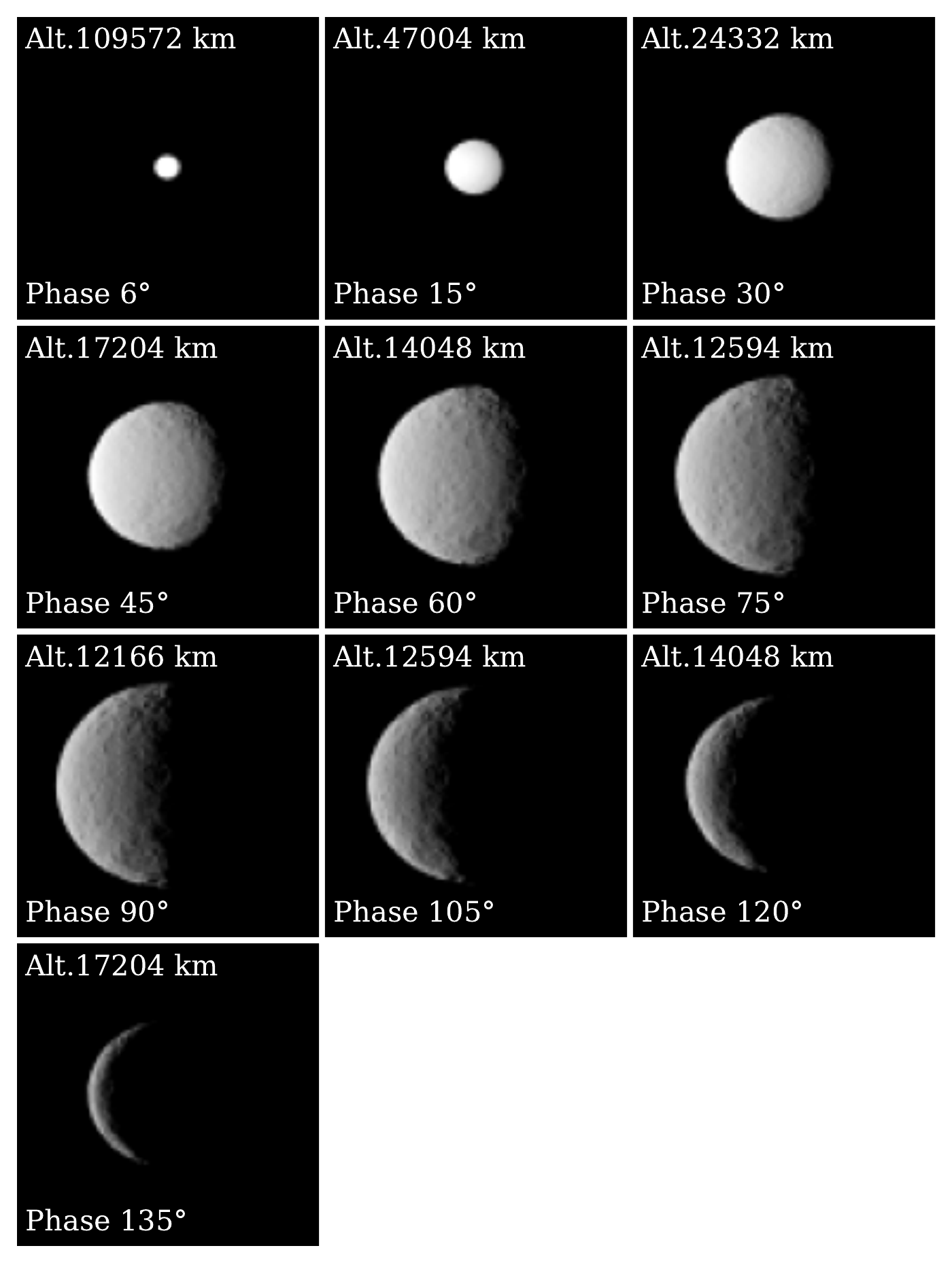}
\caption{Simulated observations of Ceres resulting from the equatorial fly-by.}
\label{obs fb_Ceres}
\end{center}
\end{figure}

\begin{figure}[H]
\begin{center}
\includegraphics[width = 0.45\textwidth]{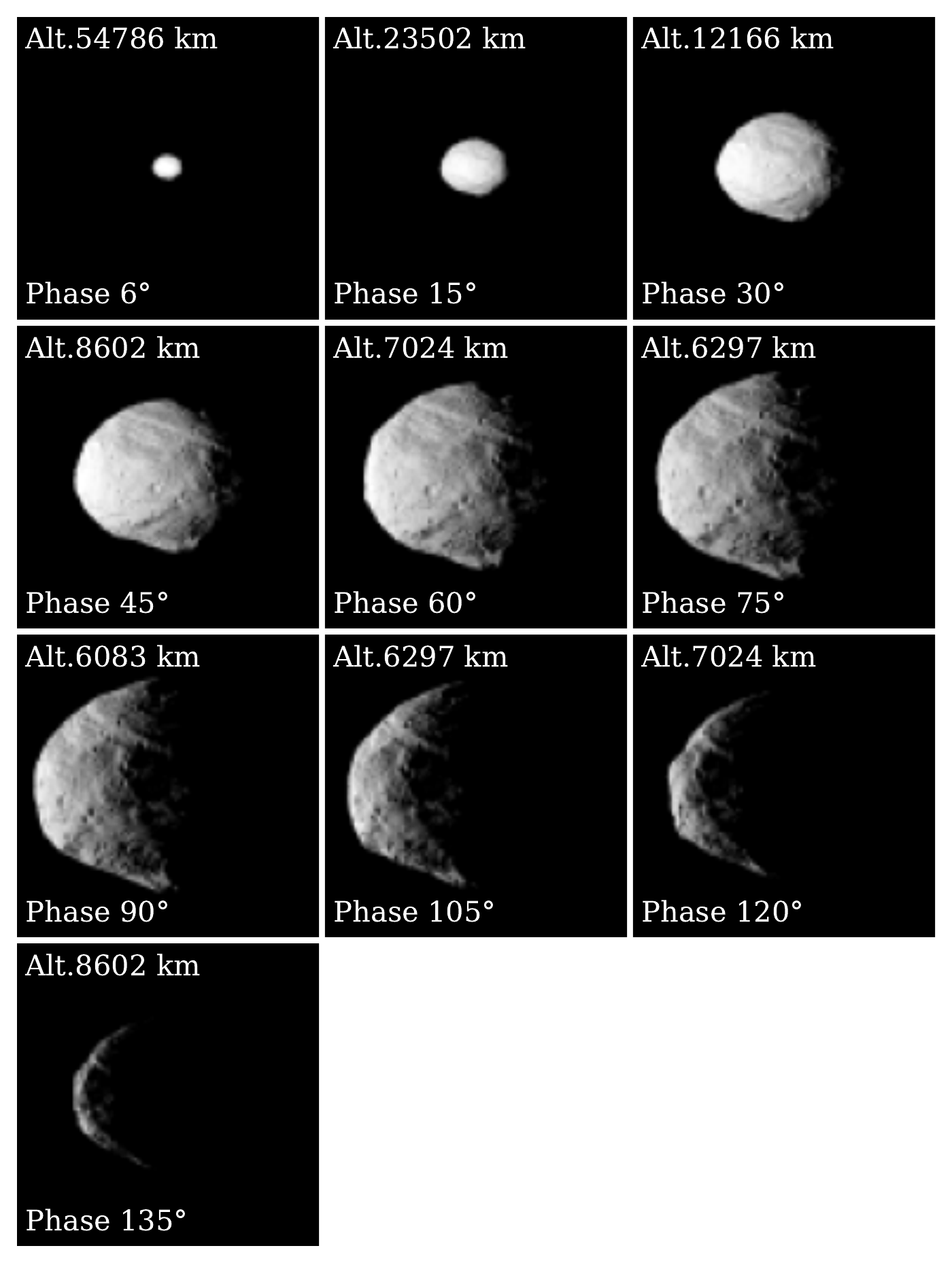}
\caption{Simulated observations of Vesta resulting from the equatorial fly-by.}
\label{obs fb_Vesta}
\end{center}
\end{figure}

\begin{figure}[H]
\begin{center}
\includegraphics[width = 0.45\textwidth]{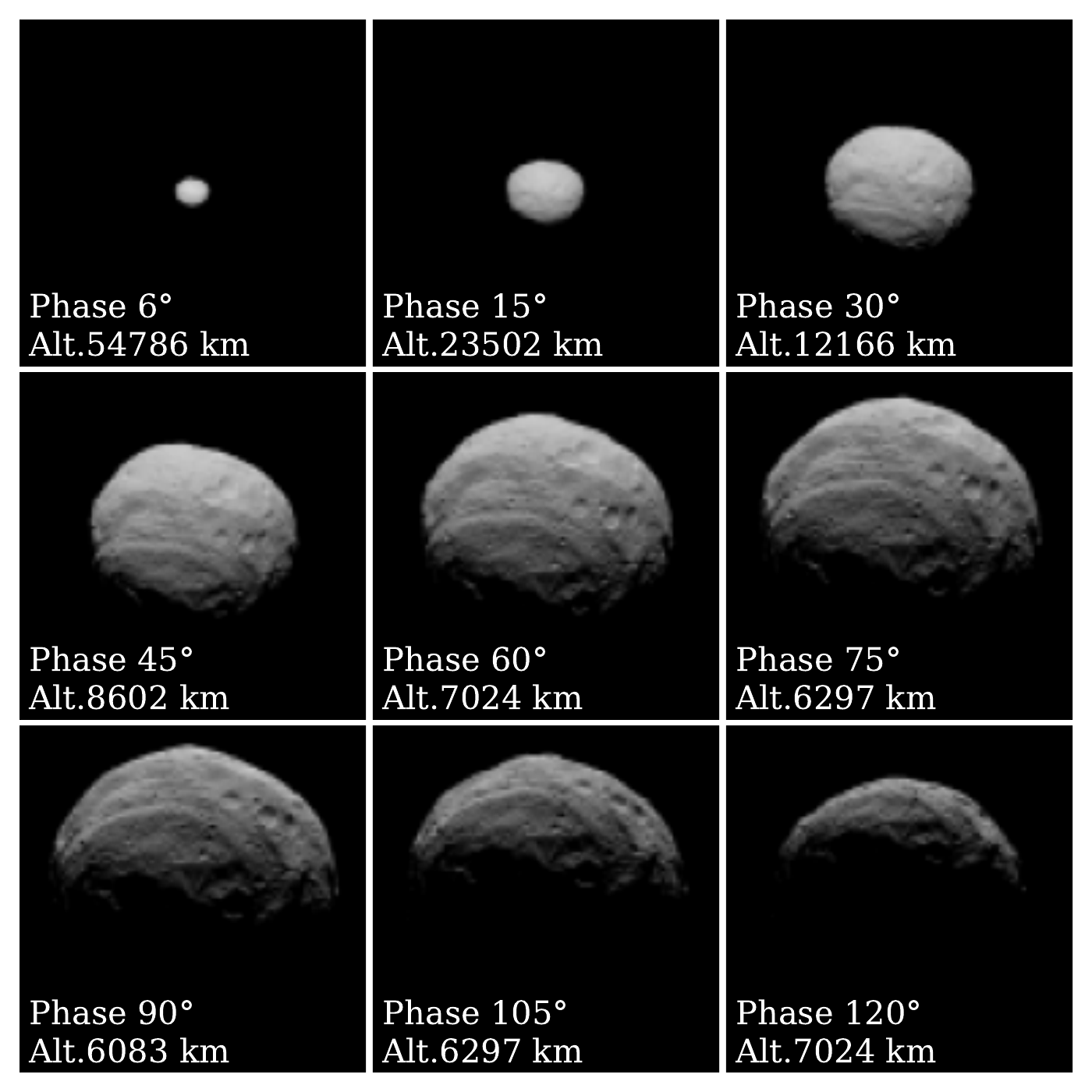}
\caption{Simulated observations of Vesta resulting from the polar fly-by.}
\label{obs fb_Vesta polar}
\end{center}
\end{figure}

\bibliography{biblio_surface_simu}

\end{document}